\documentclass [10pt,twocolumn,oneside,a4paper] {article}
\usepackage{epsf,epsfig}
\newcommand{\nn}{\nonumber}
\newcommand{\ec}{electron capture }
\newcommand{\bt}{$\beta$ }

\newcommand{\btm}{$\beta^{-}$ }
\newcommand{\gt}{Gamow-Teller }
\newcommand{\wi}{weak interaction }

\newcommand{\lt}{log(T) }

\newcommand{\ry}{$\rho Y_{e}$ }
\def\be{\begin{equation}}
\def\ee{\end{equation}}
\def\bea{\begin{eqnarray}}
\def\eea{\end{eqnarray}}
\def\bdm{\begin{displaymath}}
\def\edm{\end{displaymath}}
\begin{document}
\onecolumn
\title {\textbf{Stellar Weak Interaction Rates and Energy Losses for $\mathbf{fp}$-Shell Nuclei Calculated in the Proton-Neutron Quasiparticle Random-Phase Approximation (I).\\ (A = 40 to 60)} }
\author{\textbf{Jameel-Un Nabi}  \and
\textbf {Hans Volker Klapdor-Kleingrothaus} \\Max-Planck-Institut f\"ur Kernphysik, 69029 Heidelberg, Germany}
\normalsize
\maketitle
\begin{abstract}
Nuclear weak interaction rates for $fp$-shell nuclei in stellar matter and the associated energy losses are calculated using a modified form of proton-neutron quasiparticle RPA model with separable Gamow-Teller forces. The stellar weak rates are calculated over a wide range of densities (10$\leq$ $\rho Y_{e}$ (gcm$^{-3}$) $ \leq$ 10$^{11}$) and temperatures (10$^{7}$ $\leq$ T(K) $\leq$ 30 $\times$ 10$^{9}$). This is the first ever extensive compilation of weak interaction rates in stellar matter calculated over a wide temperature-density grid and over a larger mass range. The calculated capture and decay rates take into consideration the latest experimental energy levels and $ft$ value compilations. We take into consideration for the first time the effect of particle emission processes from excited states in the calculations of stellar rates which considerably affect our calculated rates. We have calculated stellar weak interaction rates for a total of 619  nuclei in the mass range A = 40 to 100. These also include many important neutron-rich nuclei which play an important role in the evolution process of stellar collapse. In a series of papers, starting from this one, we will be presenting our results on an abbreviated scale of temperature and density. This paper contains the stellar weak rates in the mass range 40 to 60.
\end{abstract}
\clearpage
\twocolumn
\section {\normalsize INTRODUCTION}   
The weak interaction has several crucial effects in the course of development of a star. It initiates the gravitational collapse of the core of a massive star triggering a supernova explosion, plays a key role in neutronisation of the core material via electron capture by free protons and by nuclei and effects the formation of heavy elements above iron via the r-process at the final stage of the supernova explosion (including the so-called cosmochronometers which provide information about the age of the Galaxy and of the universe). The weak interaction also largely determines the mass of the core, and thus the strength and fate of the shock wave formed by the supernova explosion.

The $\beta$ decay properties of neutron-rich nuclei are a prerequisite for a better understanding of the r-process. The r-process theory was put forward by \cite{Bur57,Cam57} more than four decades ago. Both the element distribution on the r-path, and the resulting final distribution of stable elements are highly sensitive to the $\beta$ decay properties of the neutron-rich nuclei involved in the process. This was also stressed by Klapdor \cite{Kla83} (see also \cite{Cla68} and references therein).There are about 6000 nuclei between the $\beta$ stability line and the neutron drip line. Most of these nuclei cannot be produced in terrestrial laboratories and one has to rely on theoretical extrapolations in respect of beta decay properties. The microscopic calculations of weak interaction rates \cite{Kla84,Sta90a,Sta89,Hir93} led to a better understanding of the r-process. However there was a need to go to domains of high temperature and density scales where the weak interaction rates are of decisive importance in studies of the stellar and galactic evolution processes and nucleosynthesis calculations.

During the collapse to a neutron star or to a black hole, observation of the neutrino emission provides the only possibility for direct diagnosis of the events in the interior of the star. The neutrino spectrum with its flavors, its luminosities and its time dependence reflects the entire internal evolution and thus provides a simultaneous test for theories of stellar collapse and of the formation of neutron stars and black holes. The formation of neutron stars is one of the few instances where the weak interaction plays a crucial role in a macroscopic event in nature. It is generally assumed that massive stars in the region M $\geq$ 8M$_{\odot}$ (M$_{\odot}$ denotes the solar mass) end in supernova explosions (of Type II).

If the mass of the Fe core exceeds the so-called Chandrasekhar limit \cite{Cha39}
\begin{equation}
M_{Ch}=1.45(2Y_{e})^{2}M_{\odot}
\end{equation}
(Y$_{e}$ is the number of electrons present per nucleon) the pressure of the relativistic degenerate electron gas can no longer resist the force of gravity, and the core becomes unstable. The reason for the collapse is the photodisintegration of iron group nuclei and the onset of electron capture by free protons and nuclei \cite{Bro82} which is made possible by the increase in the Fermi energy of the degenerate electron gas. The electron capture rates determine the initial dynamics of the collapse and also, via Eq.~(1), the size of the collapsing core--and therewith, the fate of the shock wave released later. These weak rates, energetically forbidden under earth conditions, increase by many orders of magnitude, due to the thermal population of parent excited states which also contribute to the overall rates.

Prior to this work, the most extensive calculations of stellar weak rates over a wide range of temperature (10$^{7}$ $\leq$ T(K) $\leq$ 10$^{11}$) and density (10$\leq$ $\rho Y_{e}$ (gcm$^{-3}$) $\leq$ 10$^{11}$) was done by Fuller, Fowler and Newman \cite{Ful82} (hereafter refered to as FFN). They calculated stellar electron and positron emission rates and continuum electron and positron capture rates, as well as the associated neutrino energy loss rates for 226 nuclei with masses between A = 21 and 60. Measured nuclear level information and matrix elements available at that time were used and unmeasured matrix elements for allowed transitions were assigned an average value of $log ft = $5. To complete the FFN rate estimate, the Gamow-Teller contribution to the rate was parameterized on the basis of the independent particle model and supplemented by a contribution simulating low-lying transitions. The FFN rates were then updated and extended to heavier nuclei \cite{Auf94}. These authors also considered the quenching of the Gamow-Teller strength by reducing the independent particle estimate for the Gamow-Teller resonance contribution by a common factor of two. However there was a need for an improved theoretical description of the problem \cite{Auf91}. The calculation of the weak interaction mediated reactions under the astrophysically reliant conditions requires the knowledge of not only the total strength in a given direction ($\beta^{-}$-decay or electron capture) but also that of the strength distributed in nuclear excitation energy. Attempts have been made to obtain these distributions theoretically from shell model calculations. However, the number of shell model basis states can get very large for mid-$fp$ nuclei, even at low energies. Hence the direct method of obtaining Gamow-Teller Giant Resonance distribution from shell model calculations using full $0\hbar\omega$ basis is computationally a formidable task even for a few nuclei. Several attempts to calculate the Gamow-Teller distribution from direct shell model calculations using a truncated shell model basis space have also been made (eg. \cite{Auf91}) which because of the above reasons, despite substantial computational efforts, are as yet approximate. Shell model Monte Carlo technique \cite{Joh92,Koo97} was then used to further refine the calculations in the $fp$-shell. Recently KB3 interaction \cite{Pov81} was used to perform calculations of electron capture and $\beta$ decay rates for a few nuclei by shell model diagonalization in the $fp$-shell \cite{Lan98,Lan98a,Mar98}. These authors calculated the supernova electron capture and beta-decay rates for a few nuclei. In any case the calculation of the weak interaction mediated reaction rates for a moderately large number of nuclei required in astrophysical situations requires a straightforward and computationally manageable approach.

Nuclei with A $>$ 60 were in the beginning not included in the stellar evolution calculations because nuclei with A $>$ 60 were not thought to become abundant enough to be important. It was shown later that these large nuclei also have important effects on the physics of core collapse \cite{Auf90}. These authors emphasized that calculations of presupernova evolution generate cores that are so neutron-rich ($Y_{e}$ reaches 0.42 and lower) that nuclei more massive than A~=~60 must be considered, thus requiring an extension of the FFN rate tables. Soon work was started on calculation of $\beta$ decay for nuclei with A $>$ 60 \cite{Kar94}. These authors reported calculation of $\beta$ decay for 11 nuclei using an average beta strength function for typical presupernova matter density ($\rho = 3 \times 10^{7}$ to $3 \times 10^{9}$ g ~cm$^{-3}$) and temperature ($T = (2$ to $5) \times 10^{9}$K). Later a search was made for important weak interaction nuclei in presupernova evolution \cite{Auf94}. One of the difficulties faced by these authors in performing their study was the lack of enough $\beta$ decay and electron capture rates for nuclei with A $>$ 60. The authors then had to made an estimate of these rates and approximate rates were derived. Since then there was a need to have microscopic stellar rates for $fp$-shell nuclei.

This work is the first ever extensive calculation of stellar weak rates in the $fp$- and $fpg$-shell nuclei ranging from A = 40 to 100. A total of 619 nuclei were considered for the calculation of stellar weak rates. These include also proton-rich and neutron-rich nuclei. The $\beta$ decay of neutron-rich nuclei prevents the decrease of the overall electron fraction in the stellar core \cite{Kar94}. Nuclei with A $>$ 60 may make significant contributions to the electron capture and $\beta$ decay rates in the last stages of stellar evolution \cite{Auf90}. In this work twelve different weak rates have been calculated for each parent nucleus. These include $e^{\pm}$-capture rates, $\beta^{\pm}$-decay rates, neutrino (anti-neutrino) energy loss rates, energies of beta-delayed proton (neutron) and the probabilities of these beta-delayed particle emission processes.  The good comparison of pn-QRPA to experimental data \cite{Sta90,Hir93} encouraged us to use pn-QRPA theory to calculate stellar weak rates. Further the pn-QRPA theory is supposed to work even better for $fp$-shell nuclei where the number of nucleon is fairly large.

Section 2 presents the formalism of our rate calculations for various nuclear processes. The references used for accumulating experimental data and their incorporation to our calculation is treated in Section~3. Results and discussions are introduced in Section~4. Here we also compare our results to earlier compilations. We discuss the reliability of the calculated rates using the pn-QRPA model in Section~5. Section~6 finally summarizes our results. 

\section{FORMALISM}
\subsection{Assumptions}
The following main assumptions are made:

(1) Only allowed \gt and superallowed Fermi transitions are calculated. It is assumed that contributions from forbidden transitions are relatively negligible.

(2) It is assumed that the temperature is high enough to ionize the atoms completely. The electrons are not bound anymore to the nucleus and obey the Fermi-Dirac distribution (see also the discussions in \cite{Ful82}). At high temperatures (kT $>$ 1~MeV), positrons appear via electron-positron pair creation, and the positrons follow the same energy distribution function as the electrons.
 
(3) Neutrinos and antineutrinos escape freely from the interior of the star. Therefore, there are no (anti)neutrinos which block the emission of these particles in the capture or decay processes. Also, (anti)neutrino capture is not taken into account.

(4) The distortion of electron (positron) wavefunction due to the Coulomb interaction with a nucleus is represented by the Fermi function in phase space integrals.

(5) Particle emissions from excited states are taken into account. This was neglected in the previous calculations of weak interaction rates and is found to considerably affect the capture and decay rates. The effects of separation energy of protons ($S_{p}$) and neutrons ($S_{n}$) on the stellar rates are considered. All excited states, with energy greater than $S_{p}$ or $S_{n}$, are assumed to decay to the ground state by emission of protons or neutrons, respectively. Due to uncertainties in the calculation of energy levels and the effect of the Coulomb barrier for the case of proton emission, it is assumed that particles are emitted at excited energies 1~MeV higher (called $E_{crit}$ throughout this section) than the particle decay channel.

(6) For the parent nuclei, the cut-off excitation energy is set at $E_{crit}$. All excited states in daughter nuclei, with excitation energy ($E_{j} \leq E_{crit}$) decay directly to the ground state through $\gamma$ transitions. All excited energy states ($E_{j} > E_{crit}$) decay by emitting particles (protons or neutrons) with increasing kinetic energies. It is further assumed that either protons (if $S_{p} < S_{n}$) or neutrons (if $S_{n} < S_{p}$) are emitted from these excited states. The $\beta$ decay of a possible isomer is not taken into account.
\subsection{Weak Decay in Stellar Matter}
The weak decay rate from the $\mathit{i}$th state of the parent to the $\mathit{j}$th state of the daughter nucleus is given by \footnote{Throughout subsection 2.2 natural units $(\hbar=c=m_{e}=1)$ are adopted, unless otherwise stated, where $m_{e}$ is the electron mass.}
\begin{equation}
\lambda_{ij} =ln2 \frac{f_{ij}(T,\rho,E_{f})}{(ft)_{ij}},
\end{equation}
where $(ft)_{ij}$ is related to the reduced transition probability $B_{ij}$ of the nuclear transition by
\be
(ft)_{ij}=D/B_{ij}.
\ee
The $D$ appearing in Eq.~(3) is a compound expression of physical constants,
\be
D=\frac{2ln2\hbar^{7}\pi^{3}}{g_{V}^{2}m_{e}^{5}c^{4}},
\ee
and,
\be
B_{ij}=B(F)_{ij}+(g_{A}/g_{V})^2 B(GT)_{ij},
\ee
where B(F) and B(GT) are reduced transition probabilities of the Fermi and ~Gamow-Teller transitions respectively,
\be
B(F)_{ij} = \frac{1}{2J_{i}+1} \mid<j \parallel \sum_{k}t_{\pm}^{k} \parallel i> \mid ^{2},
\ee 
\be
B(GT)_{ij} = \frac{1}{2J_{i}+1} \mid <j \parallel \sum_{k}t_{\pm}^{k}\vec{\sigma}^{k} \parallel i> \mid ^{2},
\ee

In Eq.~(7), $\vec{\sigma}^{k}$ is the spin operator and $t_{\pm}^{k}$ stands for the isospin raising and lowering operator. The value of D=6295 s is adopted and the ratio of the axial-vector $(g_{A})$ to the vector $(g_{V})$ coupling constant is taken as 1.254. 

\subsubsection{Phase space integrals}
The phase space integral $(f_{ij})$ is an integral over total energy,
\be
f_{ij} = \int_{1}^{w_{m}} w \sqrt{w^{2}-1} (w_{m}-w)^{2} F(\pm Z,w) (1-G_{\mp}) dw,
\ee
for electron (\textit{upper signs}) or positron (\textit{lower signs}) emission, or 
\be
f_{ij} = \int_{w_{l}}^{\infty} w \sqrt{w^{2}-1} (w_{m}+w)^{2} F(\pm Z,w) G_{\mp} dw,
\ee
for continuum positron (\textit{lower signs}) or electron (\textit{upper signs}) capture.

In Eqs.~(8) and (9), $w$ is the total kinetic energy of the electron including its rest mass, $w_{l}$ is the total capture threshold energy (rest+kinetic) for positron (or electron) capture. One should note that if the corresponding electron (or positron) emission total energy, $w_{m}$, is greater than -1, then $w_{l}=1$, and if it is less than or equal to 1, then $w_{l}=\mid w_{m} \mid$. $w_{m}$ is the total $\beta$-decay energy,
\be
w_{m} = m_{p}-m_{d}+E_{i}-E_{j},
\ee
where $m_{p}$ and $E_{i}$ are mass and excitation energies of the parent nucleus, and $m_{d}$ and $E_{j}$ of the daughter nucleus, respectively.

$G_{+}$ and $G_{-}$ are the positron and electron distribution functions, respectively. Assuming that the electrons are not in a bound state, these are the Fermi-Dirac distribution functions,
\be
G_{-} = [exp (\frac{E-E_{f}}{kT})+1]^{-1},
\ee
\be
G_{+} = [exp (\frac{E+2+E_{f}}{kT})+1]^{-1}.
\ee
Here $E=(w-1)$ is the kinetic energy of the electrons, $E_{f}$ is the Fermi energy of the electrons, $T$ is the temperature, and $k$ is the Boltzmann constant.

In the calculations, the inhibition of the final neutrino phase space is never appreciable enough that neutrino (or anti-neutrino) distribution functions had to be taken into consideration. $F(\pm Z,w)$ are the Fermi functions and are calculated according to the procedure adopted by Gove and Martin \cite{Gov71}.

The number density of electrons associated with protons and nuclei is $\rho Y_{e} N_{A}$, where $\rho$ is the baryon density, $Y_{e}$ is the ratio of electron number to the baryon number, and $N_{A}$ is the Avogadro number.
\be
\rho Y_{e} = \frac{1}{\pi^{2}N_{A}}(\frac {m_{e}c}{\hbar})^{3} \int_{0}^{\infty} (G_{-}-G_{+}) p^{2}dp, 
\ee
where $p=(w^{2}-1)^{1/2}$ is the electron or positron momentum, and Eq.~(13) has the units of \textit{moles $cm^{-3}$}. This equation is used for an iterative calculation of Fermi energies for selected values of $\rho Y_{e}$ and $T$.

There is a finite probability of occupation of parent excited states in the stellar environment as a result of the high temperature in the interior of massive stars. Weak interaction rates then also have a finite contribution from these excited states. The occupation probability of a state $i$ is calculated on the assumption of thermal equilibrium,
\be
P_{i} = \frac {(2J_{i}+1)exp(-E_{i}/kT)}{\sum_{i=1}(2J_{i}+1)exp(-E_{i}/kT)},
\ee
where $J_{i}$ and $E_{i}$ are the angular momentum and excitation energy of the state $i$, respectively.

Unfortunately one cannot calculate the $J_{i}$'s in QRPA theory and hence Eq.~(14) is modified as
\be
P_{i} = \frac {exp(-E_{i}/kT)}{\sum_{i=1}exp(-E_{i}/kT)}.
\ee
This approximation is a compromise and can be justified when one takes into consideration the uncertainty in the calculation of $E_{i}$ which easily over-sheds the uncertainty in calculating the values of $J_{i}$ in the above Eq.~(14).

The rate per unit time per nucleus for any weak process is then given by
\be
\lambda = \sum_{ij}P_{i} \lambda_{ij}.
\ee
The summation over all initial and final states is carried out until satisfactory convergence in the rate calculations is achieved.
\subsubsection{Other calculated rates}
The neutrino energy loss rates are calculated using the same formalism except that the phase space integral is replaced by
\be
f_{ij}^{\nu} = \int_{1}^{w_{m}} w \sqrt{w^{2}-1} (w_{m}-w)^{3} F(\pm Z,w) (1-G_{\mp}) dw,
\ee
for electron (\textit{upper signs}) or positron (\textit{lower signs}) emission, or by
\be
f_{ij}^{\nu} = \int_{w_{l}}^{\infty} w \sqrt{w^{2}-1} (w_{m}+w)^{3} F(\pm ,w) G_{\mp} dw,
\ee
for continuum positron (\textit{lower signs}) or electron (\textit{upper signs}) capture.

 The $\lambda_{ij}$ in Eq.~(16) in this case is the sum of the \ec and positron decay rates, or the sum of the positron capture and electron decay rates, for the transition $i \rightarrow j$.

The proton energy rate from the daughter nucleus is calculated, whenever $S_{p}$ $<$ $S_{n}$, by
\be
\lambda^{p} = \sum_{ij}P_{i}\lambda_{ij}(E_{j}-E_{crit}),
\ee
for all $E_{j} > E_{crit}$, whereas for all $E_{j} \leq E_{crit}$ one calculates the $\gamma$ heating rate,
\be
\lambda^{\gamma} = \sum_{ij}P_{i}\lambda_{ij}E_{j}.
\ee
If on the other hand, $S_{n} < S_{p}$, then the neutron energy rate from the daughter nucleus is calculated using
\be
\lambda^{n} = \sum_{ij}P_{i}\lambda_{ij}(E_{j}-E_{crit}),
\ee
for all $E_{j} > E_{crit}$, and for all $E_{j} \leq E_{crit}$ the $\gamma$ heating rate is calculated as in Eq.~(20).  

The probability of $\beta$-delayed proton (neutron) emission is calculated by
\be
P^{p(n)} = \frac{\sum_{ij\prime}P_{i}\lambda_{ij\prime}}{\sum_{ij}P_{i}\lambda_{ij}},
\ee
where $j\prime$ are states in the daughter nucleus for which $E_{j\prime} > E_{crit}$. In Eqs.~[(19)-(22)] $\lambda_{ij(\prime)}$ is the sum of the electron capture and positron decay rates, or the sum of positron capture and electron decay rates, for the transition $i$ $\rightarrow$ $j(j\prime)$.

\subsection{\gt transitions}
The parent excited states can be constructed as phonon-correlated multi-quasi-particle states. An extension of the pn-QRPA model is straight forward to the \gt transitions from nuclear excited states. Low-lying excited states of a nucleus are obtained by one-proton (or one-neutron) excitations. They are described, in the quasi-particle (q.p.) picture, by adding two-proton (two-neutron) q.p.'s to the ground state \cite{Mut92}. The transition amplitudes between the multi-quasi-particle states can be reduced to those of single-quasi-particle states. Details of the formalism can be found in \cite{Nab99}.

In the calculation of the \gt transitions a quenching of the transitions was not explicitly taken into account. The quenching of the \gt strength cannot be a constant renormalization of the axial vector current \cite{Gro83,Kla84}. We did not choose a global quenching factor (as done in many shell model calculations) for the following reasons. In order to reproduce the \gt strength the \gt interaction strength parameters $\chi$ (for particle-hole interactions) and $\kappa$ (for particle-particle interactions) were adjusted in our calculations. These values were deduced from a fit to experimental 
half-lives and for every isotopic chain fixed values of $\chi$ and $\kappa$ allowed to deduce a locally best value of $\chi$ and $\kappa$, respectively. With the large model space (up to 7 major shells) considered in this QRPA calculation and the fine tuning of the \gt strength parameter, an extra quenching factor on one hand might slightly improve the experimental half-lives for some $\beta^{+}$ decays of very proton-rich  nuclei but the overall comparison with measured rates of  nuclei would remain more or less unaltered (see \cite{Sta90}). 
\subsection{Fermi transitions}
The Fermi operator is independent of space and spin [Eq.~(6)], and as a result, the Fermi strength is concentrated in a very narrow resonance centered around the isobaric analog state (IAS) for the ground and excited states (The IAS is generated by operating on the associated parent states with the isospin raising or lowering operator:
\bdm
T_{\pm}=\sum_{i}t_{\pm}(i), \nn
\edm 
where the sum is over the nucleons. This operator commutes with all parts of the nuclear Hamiltonian except for the Coulomb part). 
The superallowed Fermi transitions were assumed to be concentrated in the IAS of the parent state. The Fermi matrix element depends only on the nuclear isospin, T, and it's projection $T_{z}=(Z-N)/2$ for the parent and daughter nucleus. The energy of the IAS is calculated according to the prescription given in \cite[pages 111--112]{Gro90}
whereas the reduced transition probability is given by 
\be
B(F)=T(T+1)-T_{zi}T_{zf},  
\ee
where $T_{zi}$ and $T_{zf}$ are the third components of the isospin of initial and final analogue states.
\section{INCORPORATION OF EXPERIMENTAL DATA}
Table~(1) shows the references from which the latest experimental excitation energies and log $\mathit{ft}$ values are taken (NP $\rightarrow$ Nuclear Physics and NDS $\rightarrow$ Nuclear Data Sheets). For those nuclei where more than one references were given, the latest data was taken into account. 

The calculated excitation energies were replaced with the measured ones when they were within 0.5 MeV of each other. The log $\mathit{ft}$ value of this energy level was also then replaced by the measured one. Very low lying states were inserted in the calculations together with their log $\mathit{ft}$ values if the theory was missing them. Inverse and mirror transitions were also taken into consideration. If there appeared a level in experimental compilations without definite spin and parity assignment, no theoretical levels were replaced with experimental levels beyond this excitation energy nor were they inserted in the calculations.

The Q-value of each transition as well as $S_{p}$ and $S_{n}$ of each nucleus were derived using the experimental mass compilation of Audi et al. \cite{Aud95}. For nuclei where \cite{Aud95} failed to give mass defects M\"oller and Nix \cite{Moe81} and Myers and Swiatecki \cite{Mye96} were used to derive the corresponding energies necessary for the calculations. The two different theoretical mass formulae used reflect the sensitivity of stellar rates on input masses. Work is in progress to compare the results of using these different mass models on the rate calculations for the case of neutron-rich and proton-rich nuclei. In \cite{Nab99t} the stellar rates of neutron-rich nuclei using the mass model of M\"oller and Nix \cite{Moe81} are presented. Stellar rates using mass formula of Myers and Swiatecki \cite{Mye96}, for the case of neutron-rich and proton-rich nuclei, are given in this paper on an abbreviated scale of density. The complete rate table on magnetic tapes can be requested from the authors.   

\section{RESULTS AND DISCUSSION}
For nuclei in the mass region A = 40 to 60, some work has been done previously regarding the calculation of stellar weak rates. The calculations include the  work of FFN (which to-date is used in many simulations for example the KEPLER stellar evolution code \cite{Wea78}) and the recently reported shell model diagonalization approach calculations \cite{Lan98,Lan98a,Mar98}. The shell model diagonalization approach is used to calculate the stellar \ec and beta decay rates of a few nuclei. Our calculations are compared with the FFN and the shell model calculations (reported at the time of writing this paper) \cite{Lan98,Lan98a,Mar98}.
\subsection{Comparison with the FFN calculation}   
Ten different cases of \ec and beta-decay nuclei were considered for comparison with earlier calculations of FFN. The idea was to chose the most important weak interaction nuclei which play a key role in the evolution process of the cores. For these the nuclei in Table~25 and Table~26 of \cite{Auf94} were considered. In these tables the authors have presented a list of 90 \ec nuclei and 71 beta-decay nuclei averaged throughout the stellar trajectory for $0.40 \leq Y_{e} \leq 0.5$. These were the nuclei which, according to the calculations of the authors, affected $\dot{Y_{e}}$ (rate of change of $Y_{e}$) the most in the presupernova evolution. These lists also contain nuclei with A $>$ 60. The top 10 nuclei with masses 40 to 60 in both lists were chosen for comparison with this work. 

The QRPA rates are generally suppressed as compared to the rates of FFN. The QRPA rates are suppressed up to more than five orders of magnitude. There are two main reasons for this suppression. FFN does not take into effect the process of particle emission from excited states.
\begin{figure}[h!]
\epsfxsize=7.8cm
\epsffile{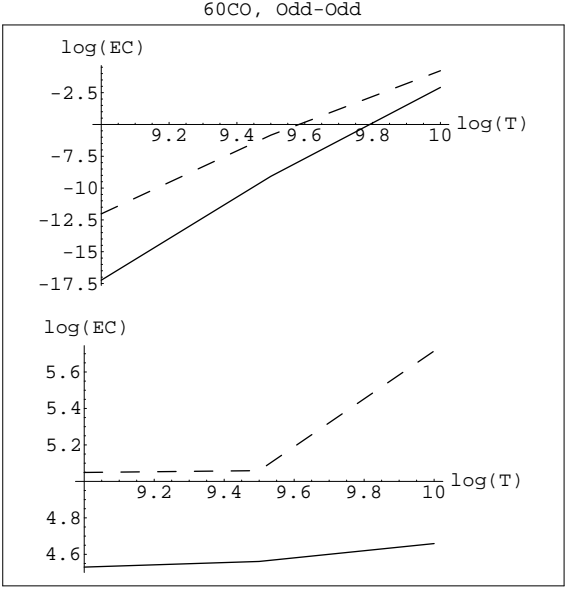}
\caption{ \footnotesize Comparison of the QRPA  electron capture (EC) rates (this work) with those of FFN. Solid lines represent the QRPA \ec rates calculated in this work, while broken lines represent the \ec rates of FFN. Log(T) is the log of temperature in units of Kelvin and log(EC) represents the log of electron capture rates in units of sec$^{-1}$. The upper graph is plotted at $\rho Y_{e}=$ 10$^{3}$ g cm$^{-3}$ and the lower graph is plotted at $\rho Y_{e}=$ 10$^{11}$ g cm$^{-3}$.}
\end{figure}
 The method for truncation of a state-density integral for calculation of 
the nuclear partition function was criticized by Tubbs and Koonin \cite{Tub79}. The
single particle level density, $g(\epsilon)$, which is used for the calculation of the 
partition function, was discussed in detail in \cite{Shl96}. In \cite{Shl96} it is argued that the no-truncation method, prescribed by \cite{Tub79}, is associated with the presence of very high angular momentum states. However in the \gt transitions, considered in our work, only low angular momentum states are considered. Thielemann and collaborators \cite{Thi98} also accept the consideration of energies close to the particle separation thresholds for the calculation of level densities for astrophysical purposes. A choice of particle threshold decay as the cutoff parameter for parent excitation energy seems to be a reasonable choice as also discussed earlier by Fowler, Engelbrecht and Woosley \cite{Fow78}. This work takes into consideration for the first time the particle emission processes, which constrain the parent excitation energies.  At any $E_{i}$, higher than the minimum of $S_{p}$ and $S_{n}$ (after accounting for the effective Coulomb barrier which prevents a proton from being promptly emitted, and, the uncertainty in calculation of energy levels), the nucleus will emit protons or neutrons (as the case may be) instead of continuing to undergo capture process. The $E_{i}$ considered in this work, as a result, are considerably lower as compared to those of FFN. Secondly for odd-A nuclei FFN places the centroid of the \gt strength at too low excitation energies (discussed also in \cite{Lan98,Lan98a,Mar98}). Their rates are thus somewhat overestimated.  
The top 10 \ec nuclei are, in order of decreasing importance, $^{60}$Co, $^{55}$Co, $^{56}$Ni, $^{57}$Co, $^{55}$Fe, $^{59}$Co, $^{54}$Mn, $^{53}$Mn, $^{54}$Fe and $^{56}$Fe in the mass range 40 to 60 \cite{Auf94}.

Fig.~1 shows the comparison of the QRPA \ec rates with those of FFN for the most important nucleus $^{60}$Co. The QRPA rate is suppressed by five orders of magnitude at \lt = 9 and \ry = 10$^{3}$ g cm$^{-3}$. The suppression factor decreases as density increases to \ry = 10$^{11}$ g cm$^{-3}$. FFN adopts the so-called Brink's hypothesis in their calculations. This hypothesis assumes that the \gt strength distribution on excited states is the same as for the ground state, only shifted by the excitation energy of the state. This hypothesis is used because no experimental data is available for the \gt strength distributions from excited states and FFN did not employ any microscopic theory to calculate the \gt strength functions from excited states.
\begin{figure}
\epsfxsize=7.8cm
\epsffile{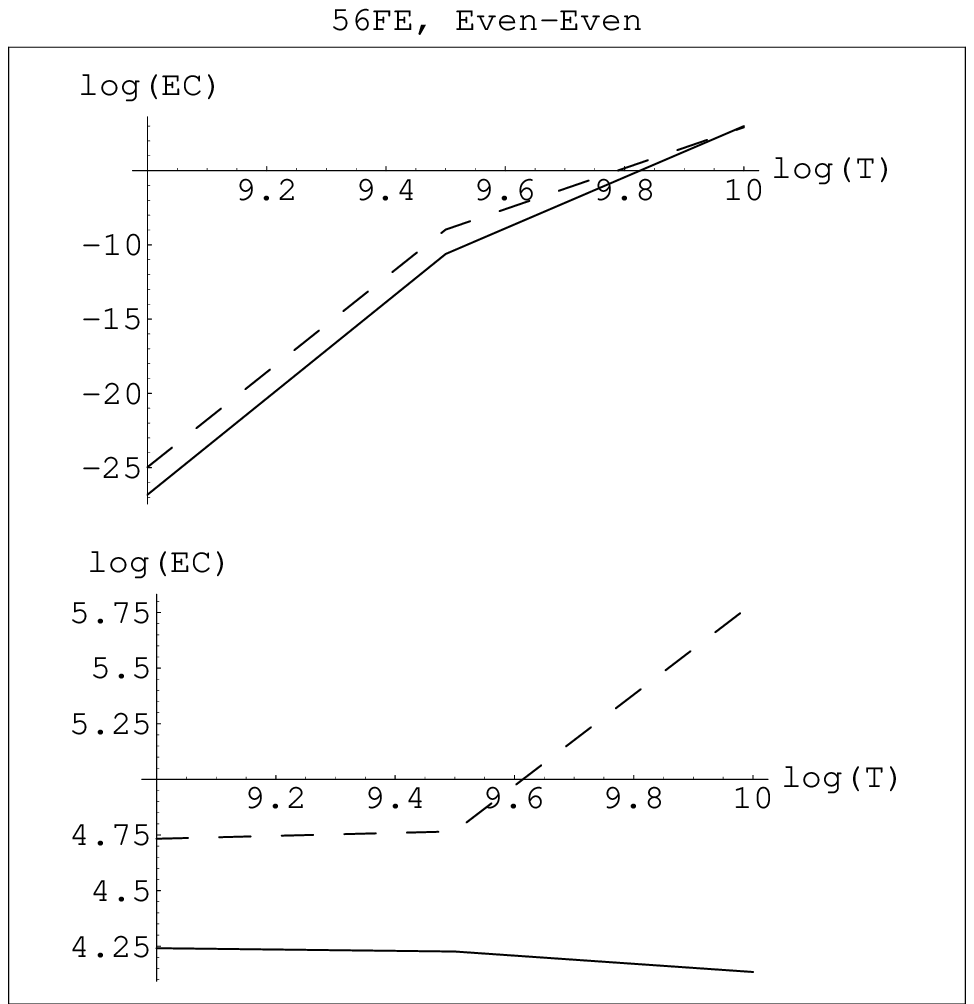}
\caption{ \footnotesize Same as Fig.~1 but for \ec by $^{56}$Fe.}
\end{figure}
\begin{figure}
\epsfxsize=7.8cm
\epsffile{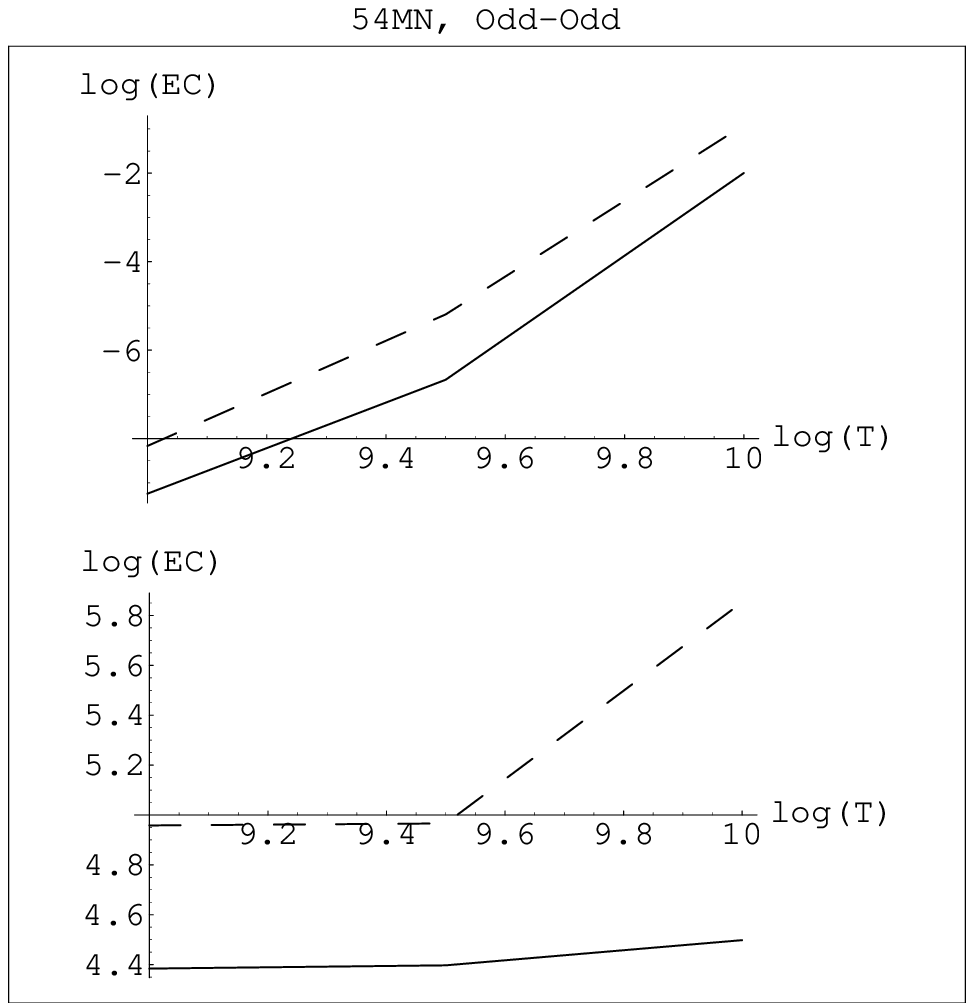}
\caption{ \footnotesize Same as Fig.~1 but for \ec by $^{54}$Mn.}
\end{figure}
 The pn-QRPA theory calculates stronger \gt strength distribution from these excited states compared to those assumed using Brink's hypothesis and hence the suppression is reduced to a factor three at \lt~=~9 and \ry = 10$^{11}$ g cm$^{-3}$. The FFN rates increase faster at higher temperatures as their parent excitation energies are not constrained, as mentioned above. 

Fig.~2, Fig.~3 and Fig.~4 show the comparison of \ec rates for $^{56}$Fe, $^{54}$Mn and $^{54}$Fe, respectively.
\begin{figure}
\epsfxsize=7.8cm
\epsffile{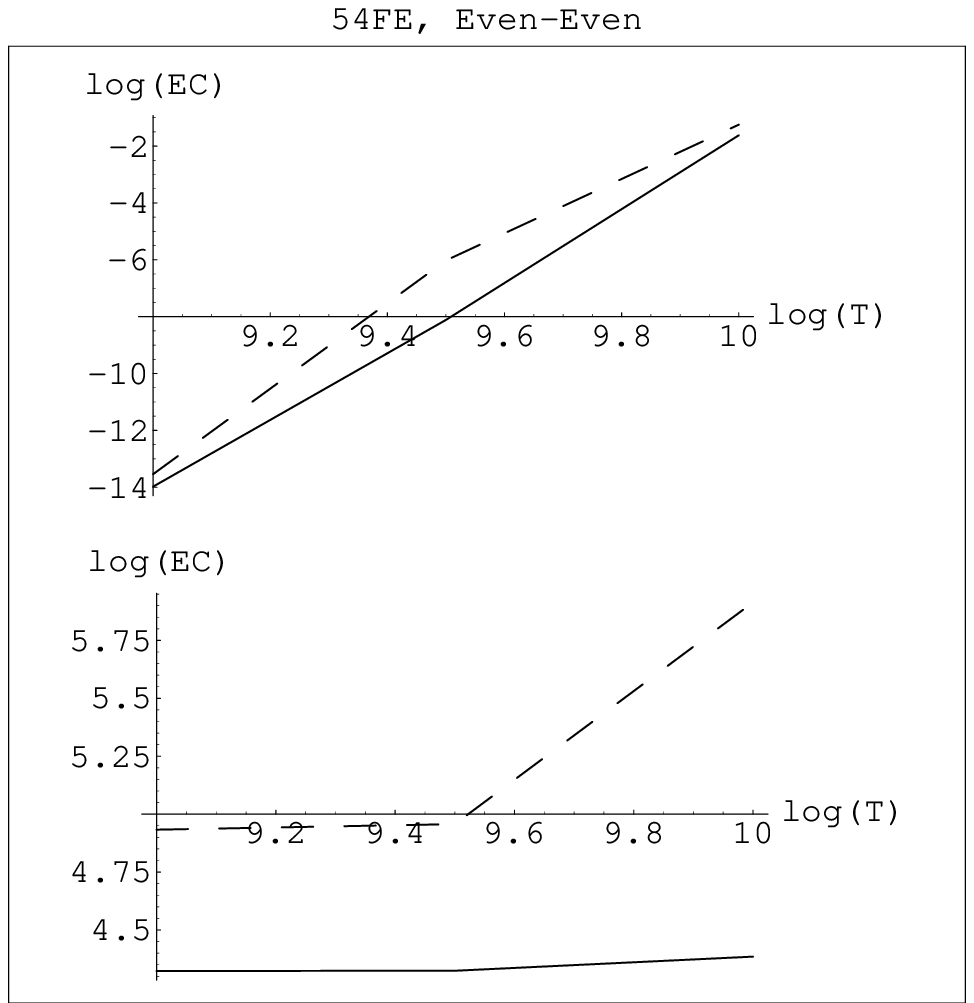}
\caption{ \footnotesize Same as Fig.~1 but for \ec by $^{54}$Fe.}
\end{figure}
These figures follow similar trends and one sees that the suppression factor at \lt = 9 and \ry = 10$^{3}$ g~cm$^{-3}$ keeps on decreasing from 2 orders of magnitude ($^{56}$Fe) to about a factor of three ($^{54}$Fe). The enhancement in the QRPA rates continues and for $^{57}$Co and $^{56}$Ni the agreement seems to be very good at \lt = 9 and \ry = 10$^{3}$ g~cm$^{-3}$ (Fig.~5 and Fig.~6, respectively). The trends are similar at  \ry = 10$^{11}$ g~cm$^{-3}$.  
\begin{figure}
\epsfxsize=7.8cm
\epsffile{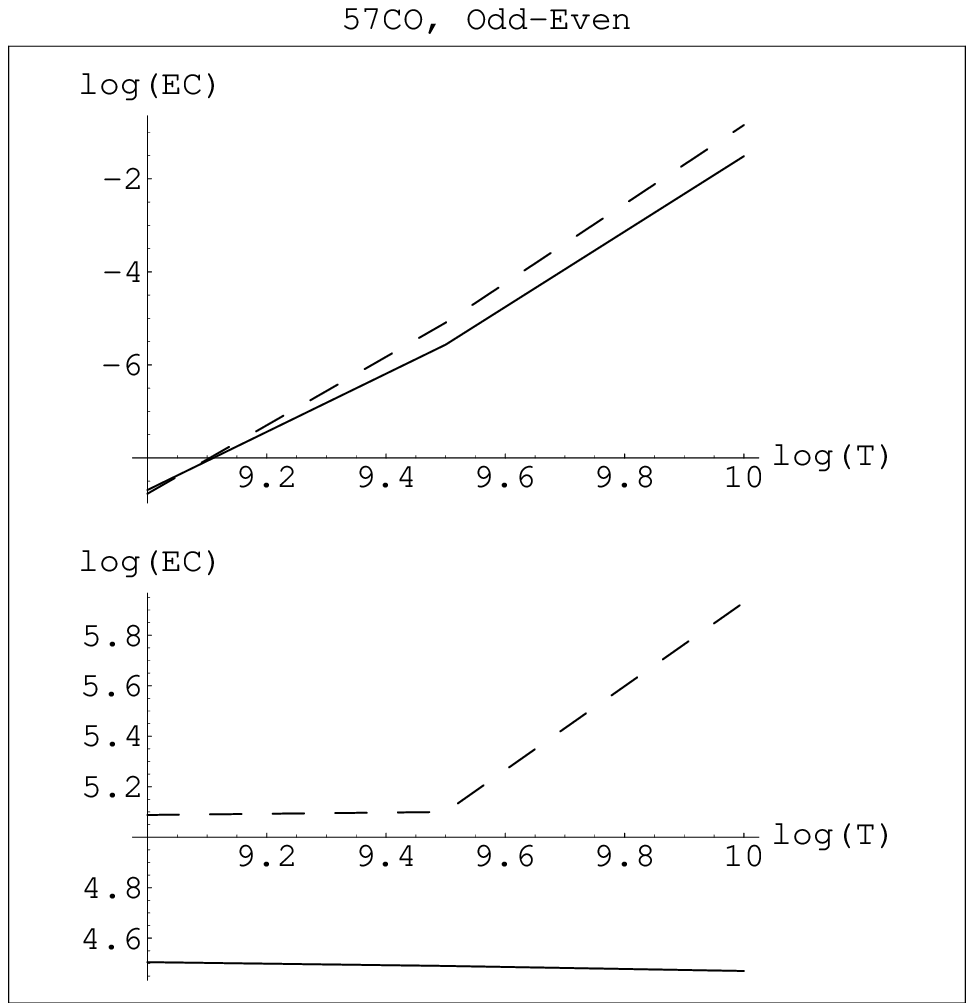}
\caption{ \footnotesize Same as Fig.~1 but for \ec by $^{57}$Co.}
\end{figure}
\begin{figure}
\epsfxsize=7.8cm
\epsffile{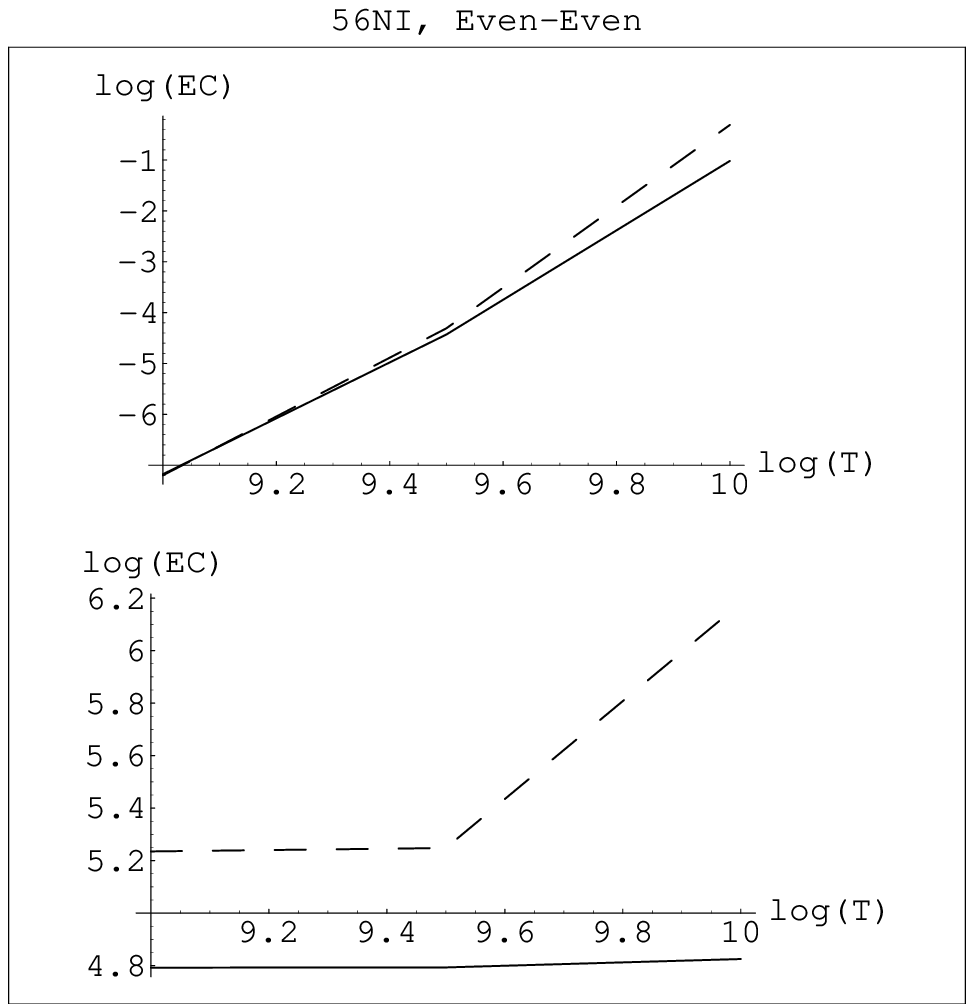}
\caption{ \footnotesize Same as Fig.~1 but for \ec by $^{56}$Ni.}
\end{figure}
This trend continues and one sees that for $^{59}$Co the QRPA \ec rate is enhanced by a factor of 5 in comparison to FFN at the lower density (Fig.~7). Clearly here the pn-QRPA theory calculates strong \gt distribution strength function from parent excited states. However one notes that at higher temperatures, \lt $>$ 9.5, the FFN rates are again enhanced in comparison to our rates. At high temperatures the probability of occupation of parent excited states ($E_{i}$) increases [Eq.~(14)] and, since FFN considers higher parent excited states, their rates are enhanced at higher temperatures.  
\begin{figure}
\epsfxsize=7.8cm
\epsffile{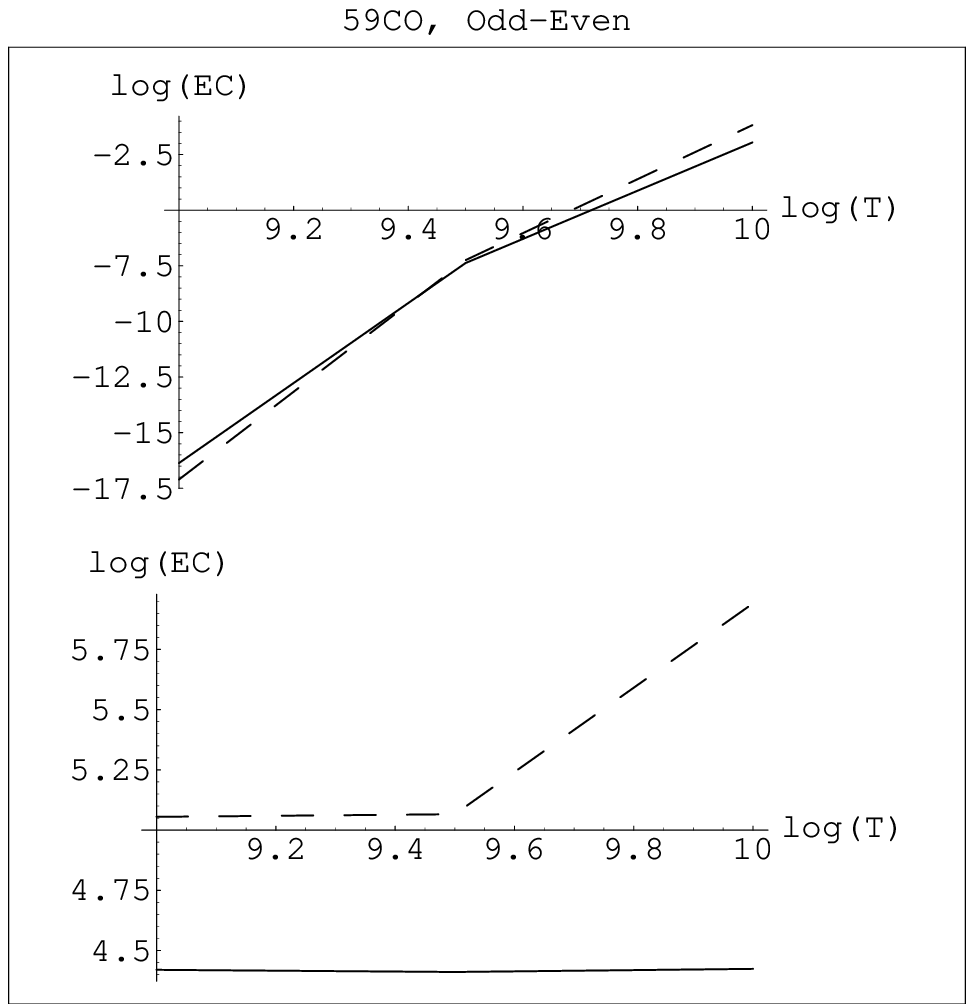}
\caption{ \footnotesize Same as Fig.~1 but for \ec by $^{59}$Co.}
\end{figure}
\begin{figure}
\epsfxsize=7.8cm
\epsffile{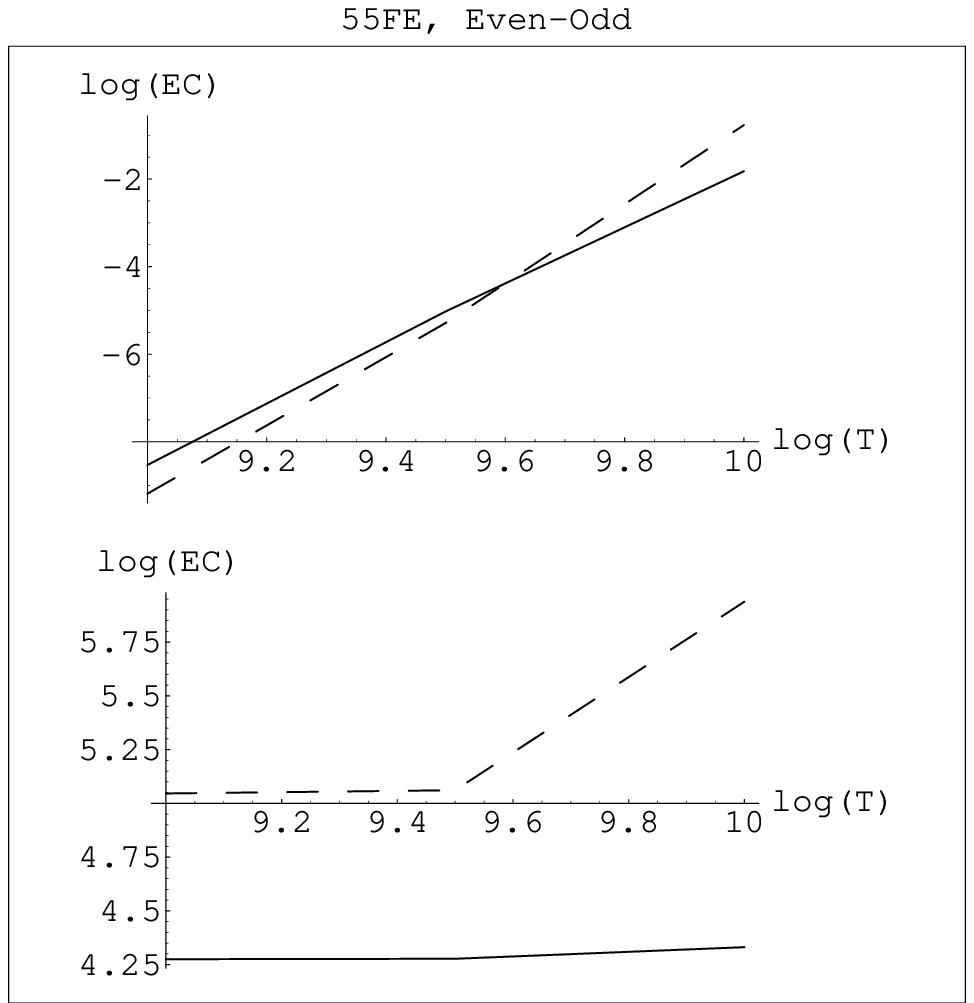}
\caption{ \footnotesize Same as Fig.~1 but for \ec by $^{55}$Fe.} 
\end{figure}
\begin{figure}
\epsfxsize=7.8cm
\epsffile{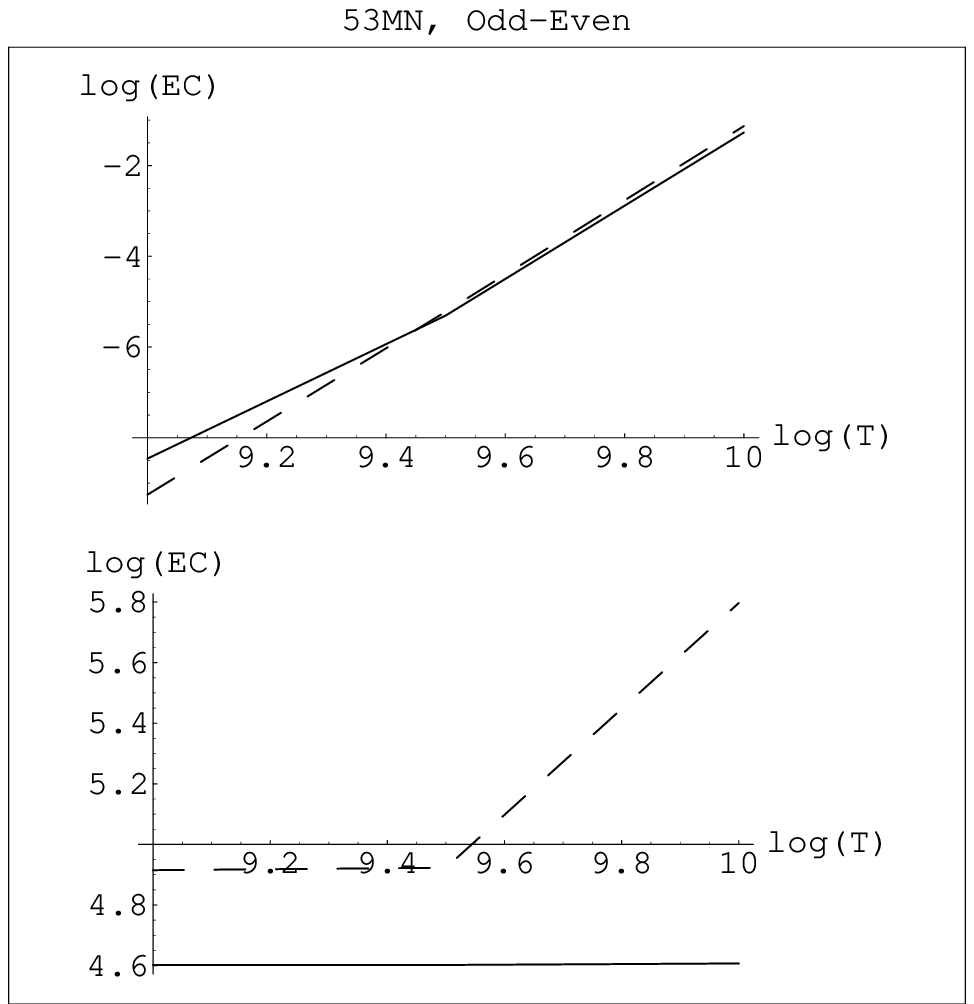}
\caption{ \footnotesize Same as Fig.~1 but for \ec by $^{53}$Mn.} 
\end{figure}
\begin{figure}
\epsfxsize=7.8cm
\epsffile{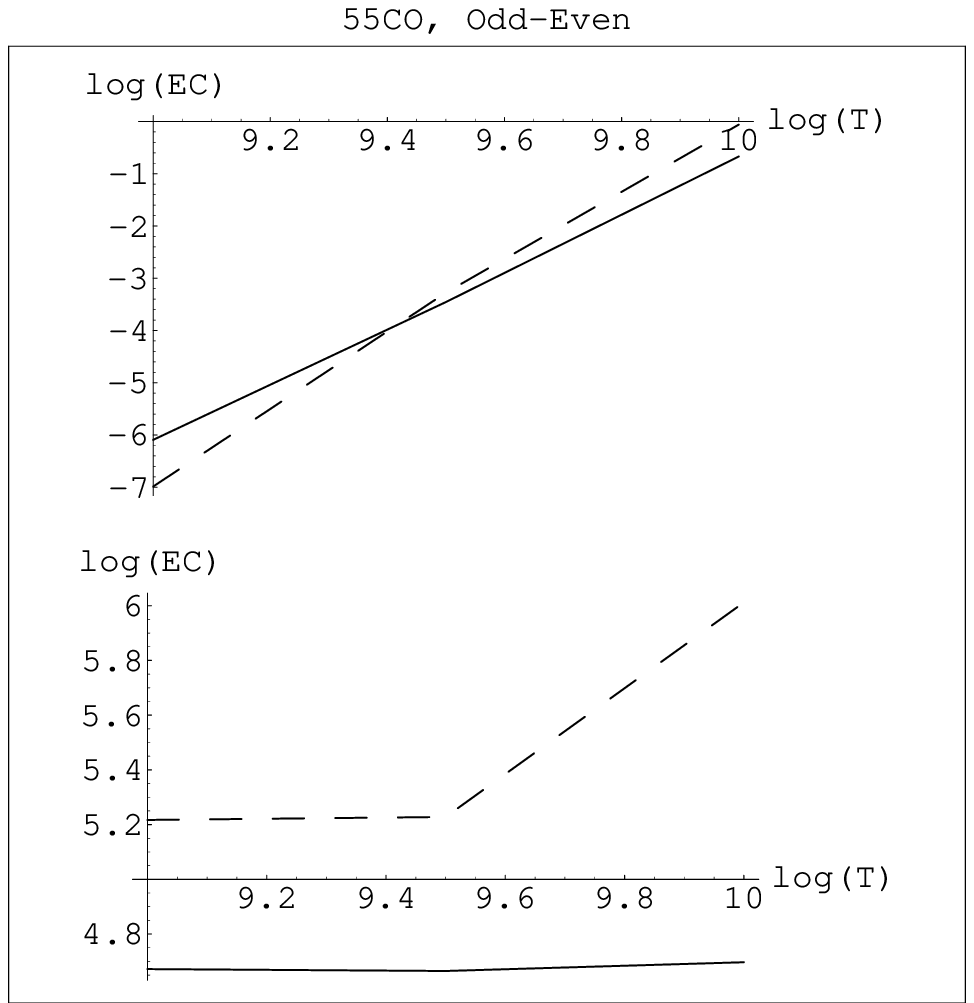}
\caption{ \footnotesize Same as Fig.~1 but for \ec by $^{55}$Co.} 
\end{figure}

Fig.~8, Fig.~9 and Fig.~10 show a comparison of \ec rates for the nuclei $^{55}$Fe, $^{53}$Mn and $^{55}$Co, respectively. The QRPA rates keep on enhancing at $\rho Y_{e}=$ 10$^{3}$ g cm$^{-3}$ and low temperatures. In Fig.~10 the QRPA rates are enhanced by up to an order of magnitude in comparison to the FFN rates. At high temperatures the effect of high $E_{i}$ considered by FFN dominates and hence FFN \ec rates exceed the QRPA rates. Comparison of the \ec rates at $\rho~Y_{e}~=$~10$^{11}$~g~cm$^{-3}$ remains more or less the same throughout the considered cases.
\begin{figure}[h!]
\epsfxsize=7.8cm
\epsffile{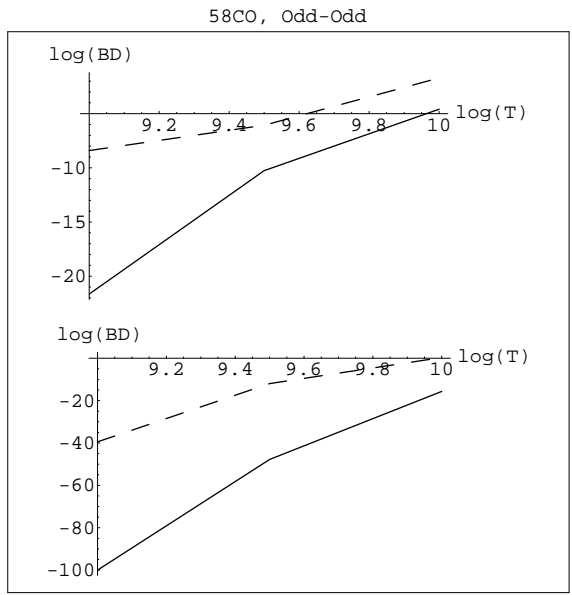}
\caption{ \footnotesize Comparison of the QRPA  beta-decay (BD) rates (this work) with those of FFN. Solid lines represent the QRPA beta-decay rates calculated in this work while broken lines represent the decay rates of FFN. Log(T) is the log of temperature in units of Kelvin and log(BD) represents the log of \btm rates in units of sec$^{-1}$. The upper graph is plotted at $\rho Y_{e}=$ 10$^{3}$ g cm$^{-3}$ and the lower graph is plotted at $\rho Y_{e}=$ 10$^{11}$ g cm$^{-3}$.}
\end{figure}

The top 10 beta-decay nuclei are, according to \cite{Auf94}, in order of decreasing importance, $^{56}$Co, $^{59}$Fe, $^{54}$Mn, $^{56}$Mn, $^{57}$Fe, $^{57}$Mn, $^{60}$Co, $^{54}$Cr, $^{58}$Co and $^{60}$Fe in the mass region 40 to 60. It is very interesting to note that the pn-QRPA theory reports vanishing decay rates for $^{56}$Co. This is in sharp contrast to Table 26 of \cite{Auf94} where $^{56}$Co is ranked as the most important beta-decay nuclei causing the largest change in $Y_{e}$. The Q-value for this decay is -2.136~MeV and $S_{p}$ is 5.849~MeV. One of the selection rules for the quasi-particle transition to occur in the pn-QRPA theory is that the transforming quasi-particles should belong to the same major oscillator shell, i.e the neutron which changes to a proton should belong to the same major oscillator shell (and same for proton). Even though the decay channel opens below the particle decay channel no decay rates are obtained for $^{56}$Co. In this case the above mentioned selection rule is satisfied only for high-lying major shells but then the parent excited state (which is the sum of the energies of constituent quasi-particles) is above the particle decay energy. $^{56}$Co can decay however to $^{56}$Ni via positron capture process  at higher temperatures. So the top 10 beta-decay nuclei taken for comparison with FFN are, in order of decreasing importance, $^{59}$Fe, $^{54}$Mn, $^{56}$Mn, $^{57}$Fe, $^{57}$Mn, $^{60}$Co, $^{54}$Cr, $^{58}$Co, $^{60}$Fe and $^{57}$Cr. 

The beta decay rates calculated in this work are also, in general, suppressed in comparison to the calculations of FFN by up to 13 orders of magnitude and more. 

Fig.~11 depicts the comparison of the QRPA beta-decay rates (this work) and those of FFN for the odd-odd cobalt isotope, $^{58}$Co. One notes that the QRPA rates are very much suppressed in comparison to FFN rates. In the figure the QRPA rate is suppressed by some 13 orders of magnitude at \lt = 9 and \ry = 10$^{3}$ g cm$^{-3}$. At \lt~=~9 and \ry = 10$^{11}$ g cm$^{-3}$, the QRPA rate is vanishing. Beta-decay rates are sensitive functions of $(E_{i}-E_{j})$. Even for positive Q-values where measured \gt transitions might also exist, the phase space $(=Q+E_{i}-E_{j})$ at high temperatures and densities might be zero if $E_{i}$ is considerably less than $E_{j}$. In this work $E_{i}$'s are constrained due to the particle emission process and no $E_{i}$ above particle decay channel (after accounting for the effective Coulomb barrier and the uncertainty in calculation of energy levels) are considered. Especially at high densities, where the chemical potential of electrons increases outside the nuclei (thus impeding the decay process), this phenomenon is very obvious. By a mere increment of 500~ keV in $E_{i}$ many of the QRPA rates at high densities can increase by some tens of orders of magnitude. At \ry = 10$^{11}$ g cm$^{-3}$, the phase space for calculation of beta-decay rates is negative resulting in vanishing decay rates till \lt = 9. Only at higher temperatures (\lt $>$ 9) the phase space becomes positive and then the QRPA rates tend to FFN rates.
\begin{figure}
\epsfxsize=7.8cm
\epsffile{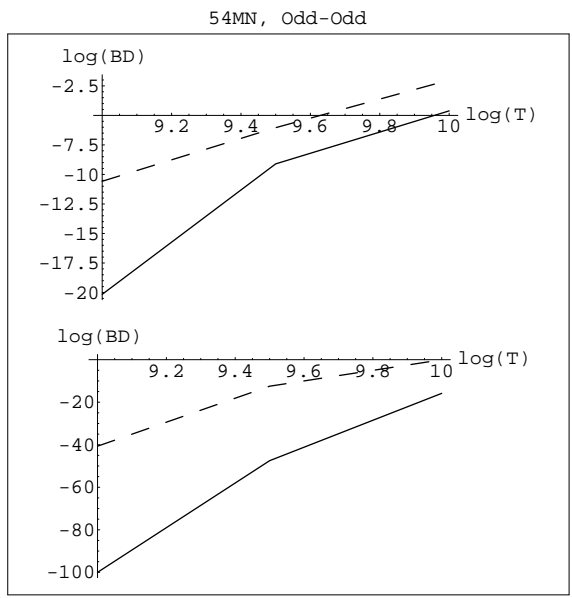}
\caption{ \footnotesize Same as Fig.~11 but for beta decay by $^{54}$Mn.}  
\end{figure}
\begin{figure}
\epsfxsize=7.8cm
\epsffile{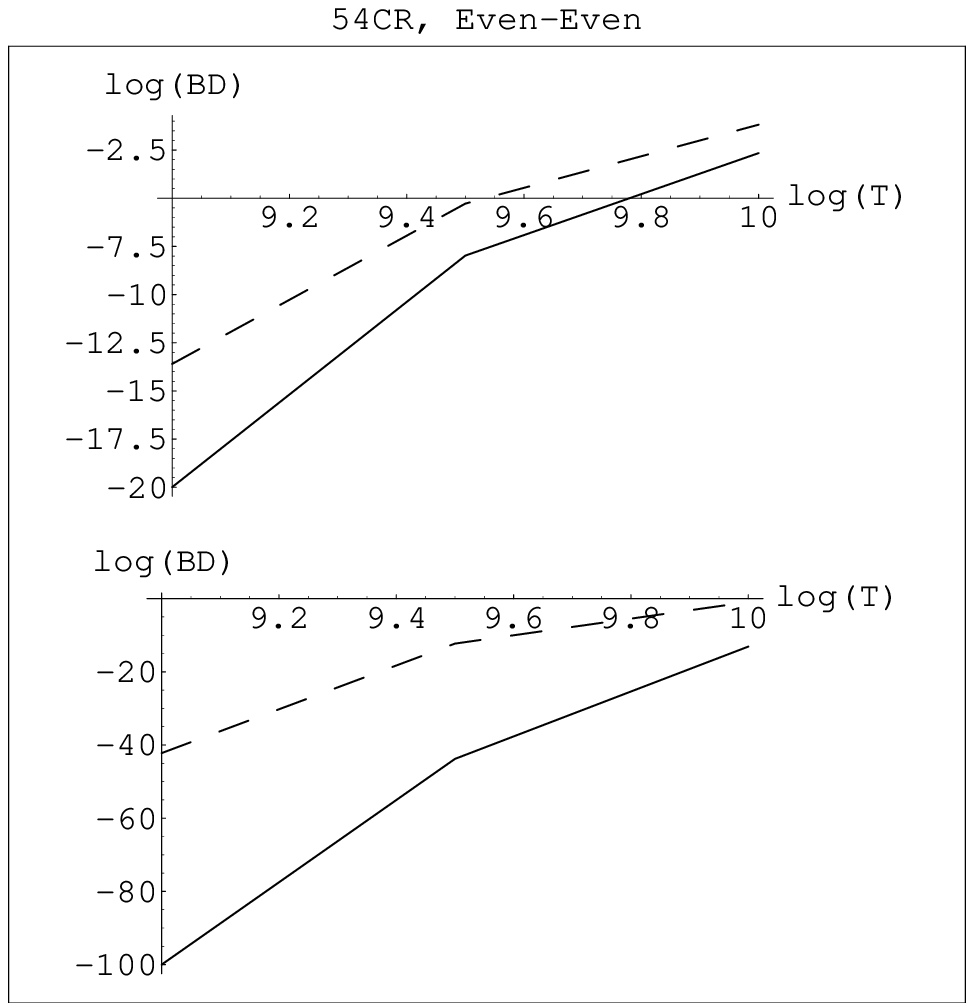}
\caption{ \footnotesize Same as Fig.~11 but for beta decay by $^{54}$Cr.}  
\end{figure}
\begin{figure}
\epsfxsize=7.8cm
\epsffile{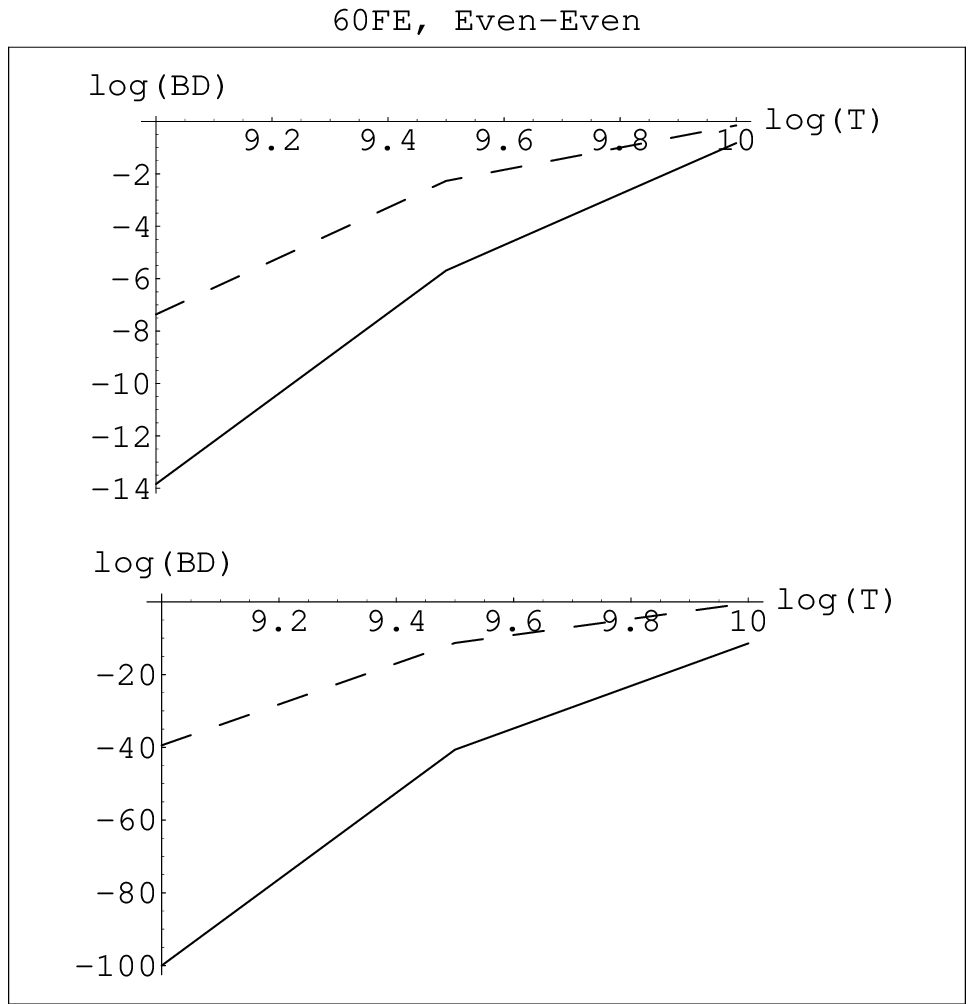}
\caption{ \footnotesize Same as Fig.~11 but for beta decay by $^{60}$Fe.}  
\end{figure}
\begin{figure}
\epsfxsize=7.8cm
\epsffile{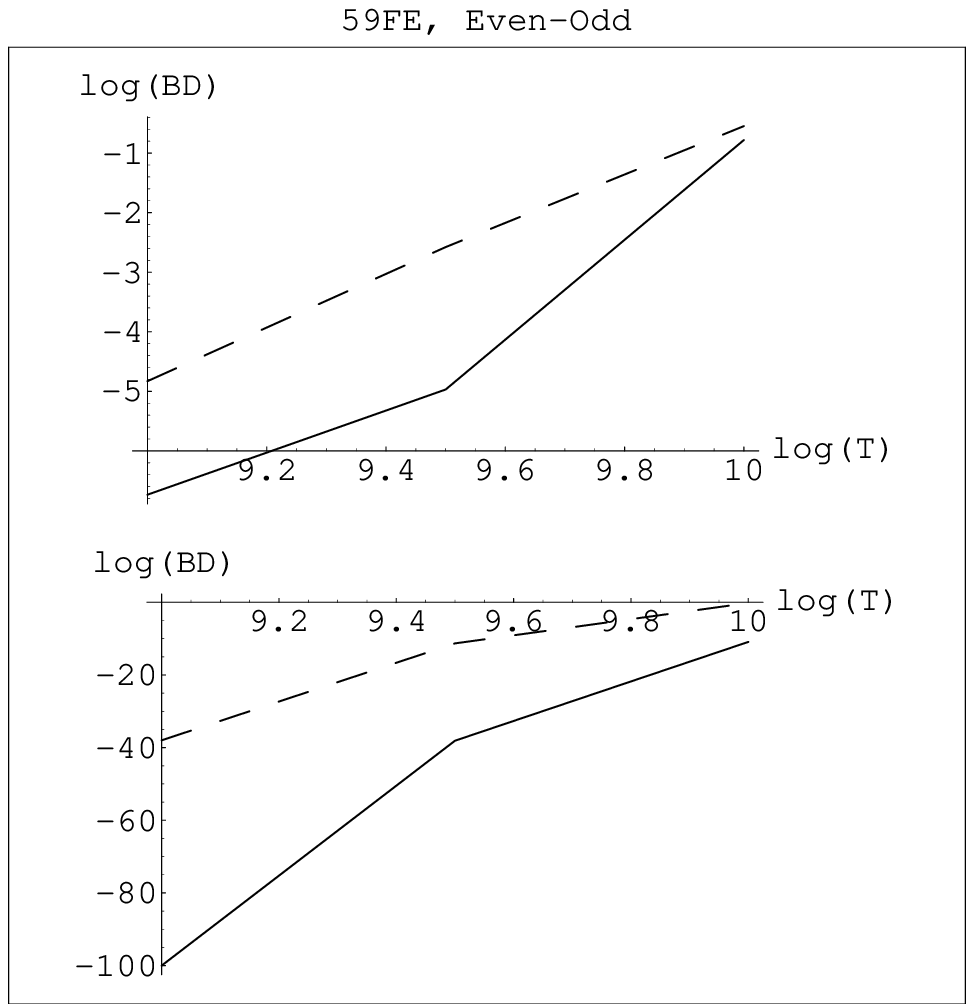}
\caption{ \footnotesize Same as Fig.~11 but for beta decay by $^{59}$Fe.}  
\end{figure}
\begin{figure}
\epsfxsize=7.8cm
\epsffile{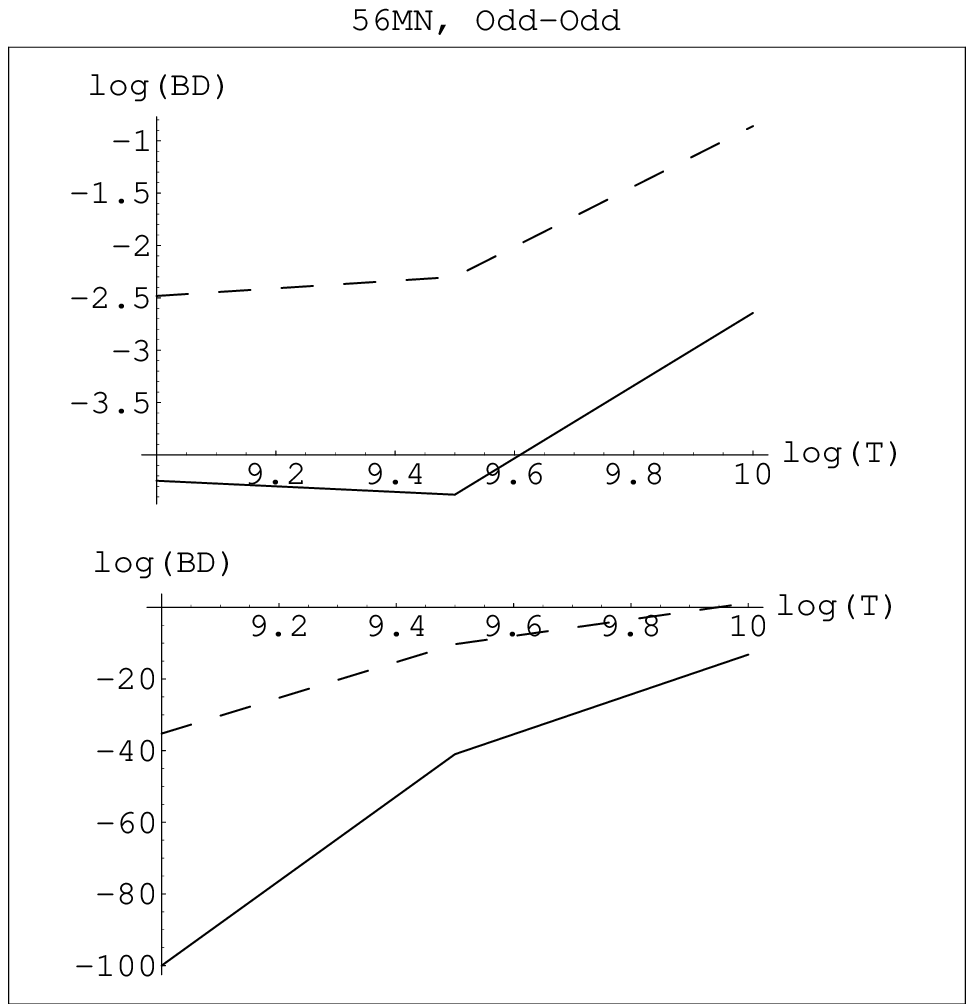}
\caption{ \footnotesize Same as Fig.~11 but for beta decay by $^{56}$Mn.}  
\end{figure}
\begin{figure}
\epsfxsize=7.8cm
\epsffile{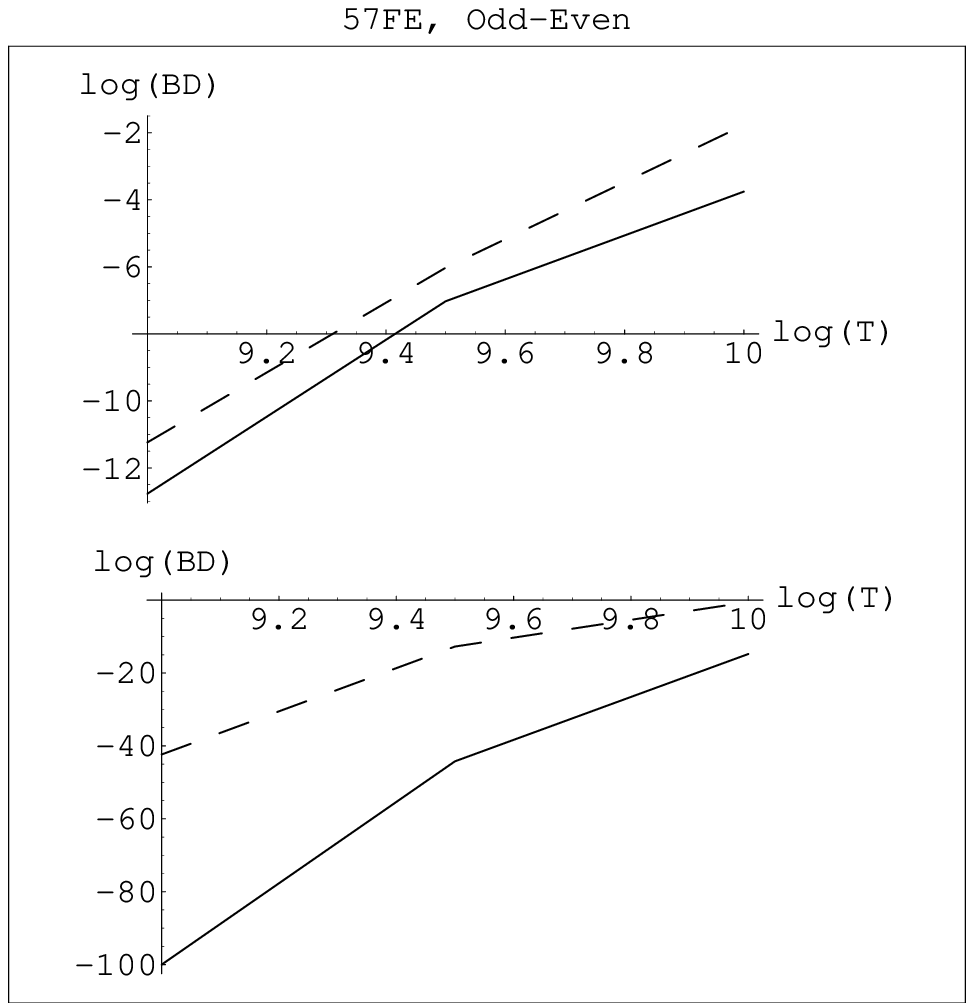}
\caption{ \footnotesize Same as Fig.~11 but for beta decay by $^{57}$Fe.}  
\end{figure}
\begin{figure}
\epsfxsize=7.8cm
\epsffile{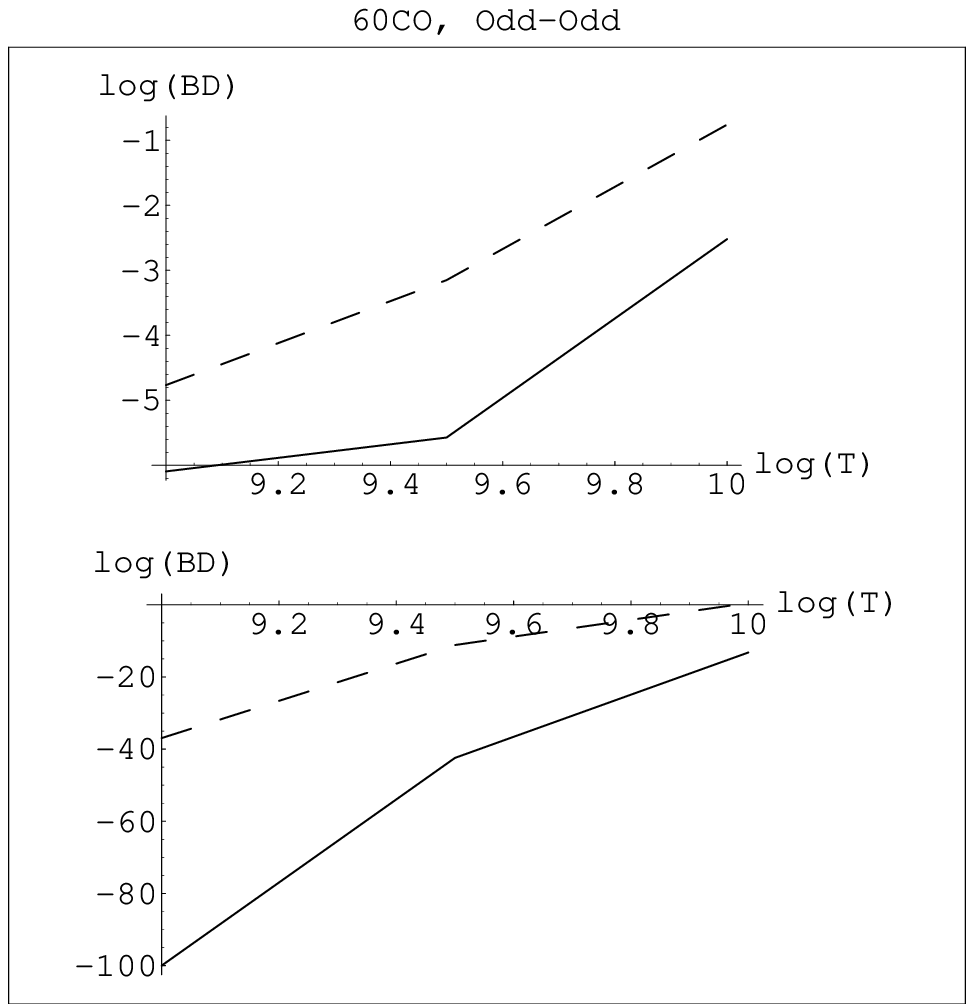}
\caption{ \footnotesize Same as Fig.~11 but for beta decay by $^{60}$Co.}  
\end{figure}
\begin{figure}
\epsfxsize=7.8cm
\epsffile{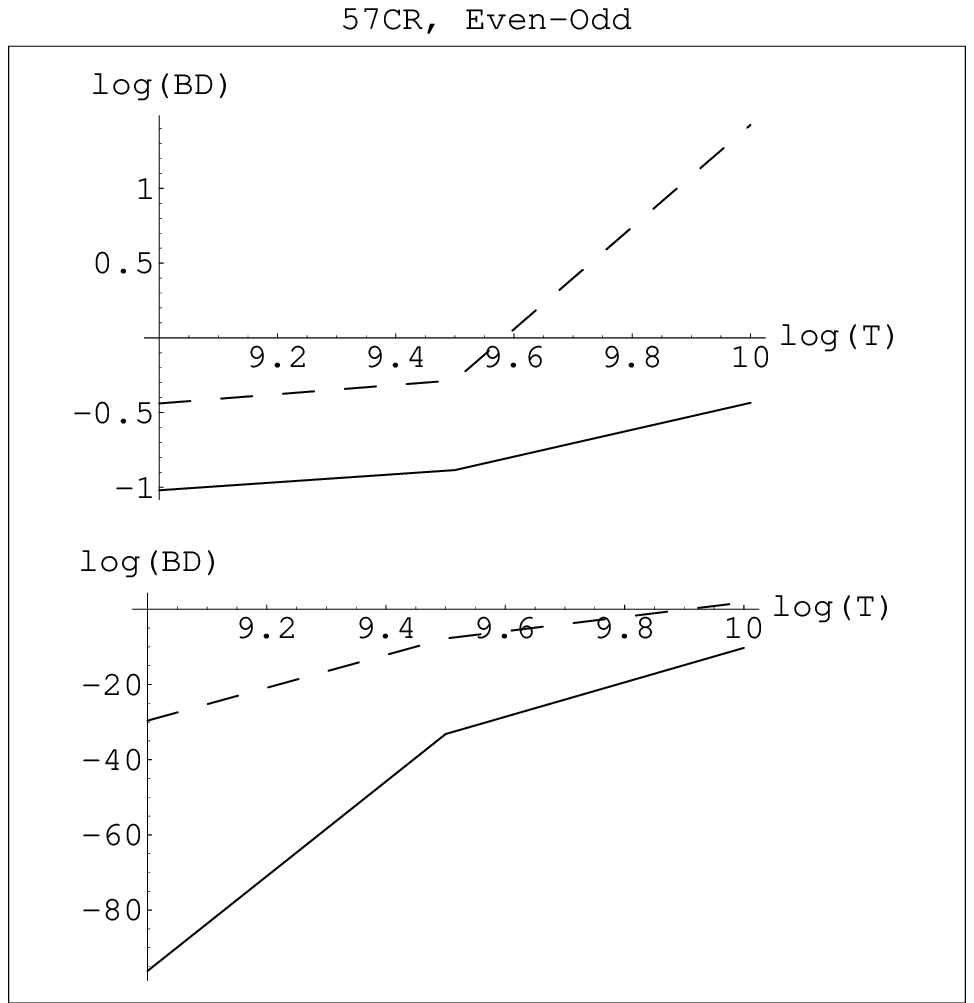}
\caption{ \footnotesize Same as Fig.~11 but for beta decay by $^{57}$Cr.}   
\end{figure}
\begin{figure}
\epsfxsize=7.8cm
\epsffile{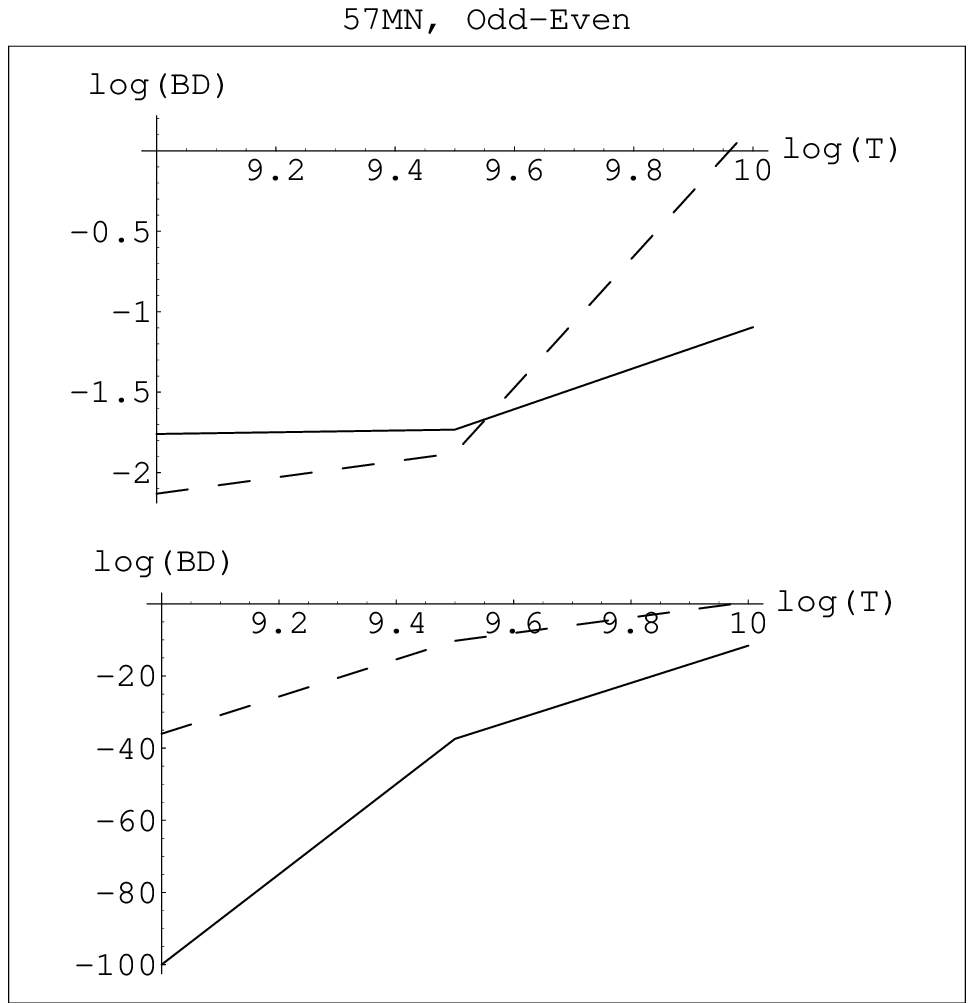}
\caption{ \footnotesize Same as Fig.~11 but for beta decay by $^{57}$Mn.}   
\end{figure}

Fig.~12, Fig.~13 and Fig.~14 depict the comparison of beta-decay rates for $^{54}$Mn, $^{54}$Cr and $^{60}$Fe, respectively. At lower density of  $\rho Y_{e} =$ 10$^{3}$ g~cm$^{-3}$, one notes that at \lt = 9 the suppression rate is reducing gradually. This trend continues and finally in Fig.~19 (representing  $^{57}$Cr) the QRPA rate is suppressed by a mere factor of 3 at $\rho Y_{e} =$ 10$^{3}$ g~cm$^{-3}$, \lt = 9. Fig.~20 (representing  $^{57}$Mn) then depicts a situation where the QRPA rates are enhanced in comparison with FFN rates at $\rho Y_{e} =$ 10$^{3}$ g~cm$^{-3}$ till \lt = 9.5. At higher temperatures FFN rates increase due to the phenomenon of considering higher $E_{i}$'s in their calculation. In all these figures one sees that the situation remains more or less the same at  $\rho Y_{e} =$ 10$^{11}$ g~cm$^{-3}$.       

The comparison of positron capture and positron decay rates is similar to the preceding comparison of electron capture and beta decay rates, respectively. In our table of rates, -~100 means that the rate is smaller than $10^{-100}$. It should be pointed out that the decay rates are sensitive functions of the difference of parent and daughter energies ($E_{i}-E_{j}$) and just by a mere addition of around 0.5 MeV, many of our -100 decay rates could rise to larger numbers. For all cases where experimental data were available our results are in very good agreement with the results of FFN. We differ when no experimental data are present. In all such cases FFN uses either some parameterization or some approximation for estimation of nuclear matrix elements whereas the nuclear matrix elements are calculated in the pn-QRPA theory in our compilation.

\subsection{Comparison with the shell model diagonalization approach calculation}
The shell model calculations describe well the nuclear structure and, especially for the light nuclei (eg., \cite{Mut91}), agree very well with the experiment. The dimension of the many-body space, however, grows exponentially as the number of particle increases, making it difficult to apply shell model techniques to heavier nuclei. A comparison of half-lives calculated by the pn-QRPA model for light nuclei with shell model calculations can be seen in Figure~8 of \cite{Sta90}. Recently the shell model diagonalization approach (SMDA) has been used to calculate \bt decay and \ec rates of some $fp$-shell nuclei \cite{Lan98,Lan98a,Mar98}. For the case of odd-odd nuclei, no experimental information exists about the \gt strength distribution. Comparison with the QRPA electron capture rates for odd-odd nuclei suggests that SMDA rates are suppressed. The density of states in odd-odd nuclei is quite high and doing calculations for a few $E_{i}$ is undesirable. Further the shell model gives very weak transition strengths to the low-lying states in the daughter nucleus. Due to the model used, they also had to consider very few $E_{i}$ in calculation of capture rates on odd-A and even-even nuclei (for $^{56}$Ni, they considered just the ground-state transition). Consequently their rates are generally suppressed in comparison to the earlier calculations of FFN and \cite{Auf94}. Similar conclusions can be drawn for the SMDA calculation of $\beta$-decay rates. Here they considered $E_{i}$ usually up to 1~MeV and ``back resonances'' (the \gt back resonances are states reached by the strong \gt transitions in the electron capture process built on the ground and excited states, see \cite{Ful80,Auf94}. These states were discussed earlier by Klapdor \cite{Kla76}) from daughter states of energy below 1~MeV. Back resonances from daughter states higher than 1~MeV were not considered. The calculations were done  for the temperature range, T9 = 1--10 (T9 is temperature in units of 10$^{9}$ K) and density range, $\rho_{7}$ = 10--1000 ($\rho_{7}$ is the density in units of 10$^{7}$ g cm$^{-3}$). Shell model calculations of \ec rates were reported in \cite{Lan98,Lan98a}. A comparison of their \ec rates with the QRPA \ec rates is presented below (also see \cite{Nab99z}).
\begin{figure}
\epsfxsize=7.8cm
\epsffile{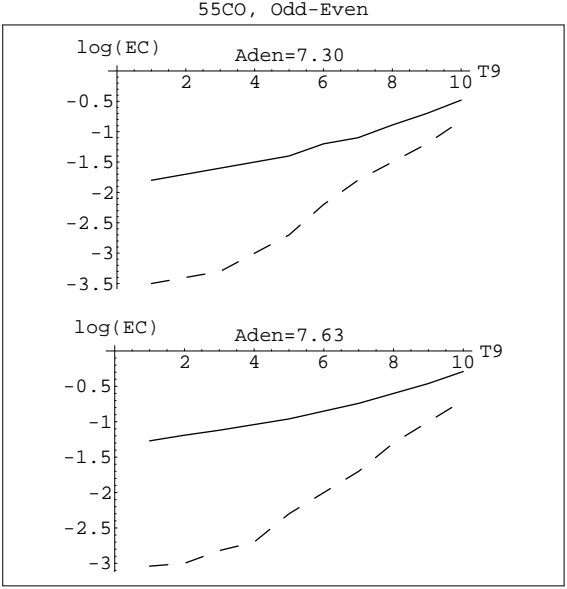}
\caption{ \footnotesize Comparison of the QRPA  \ec (EC) rates (this work) with those of \cite{Lan98}. Solid lines represent the QRPA \ec rates of this work while broken lines represent the \ec rates of \cite{Lan98}. Log(T) is the log of temperature in units of Kelvin and log(EC) represents the log of \ec rates in units of sec$^{-1}$. Aden represents the log of density in units of g cm$^{-3}$.}
\end{figure}
\begin{figure}
\epsfxsize=7.8cm
\epsffile{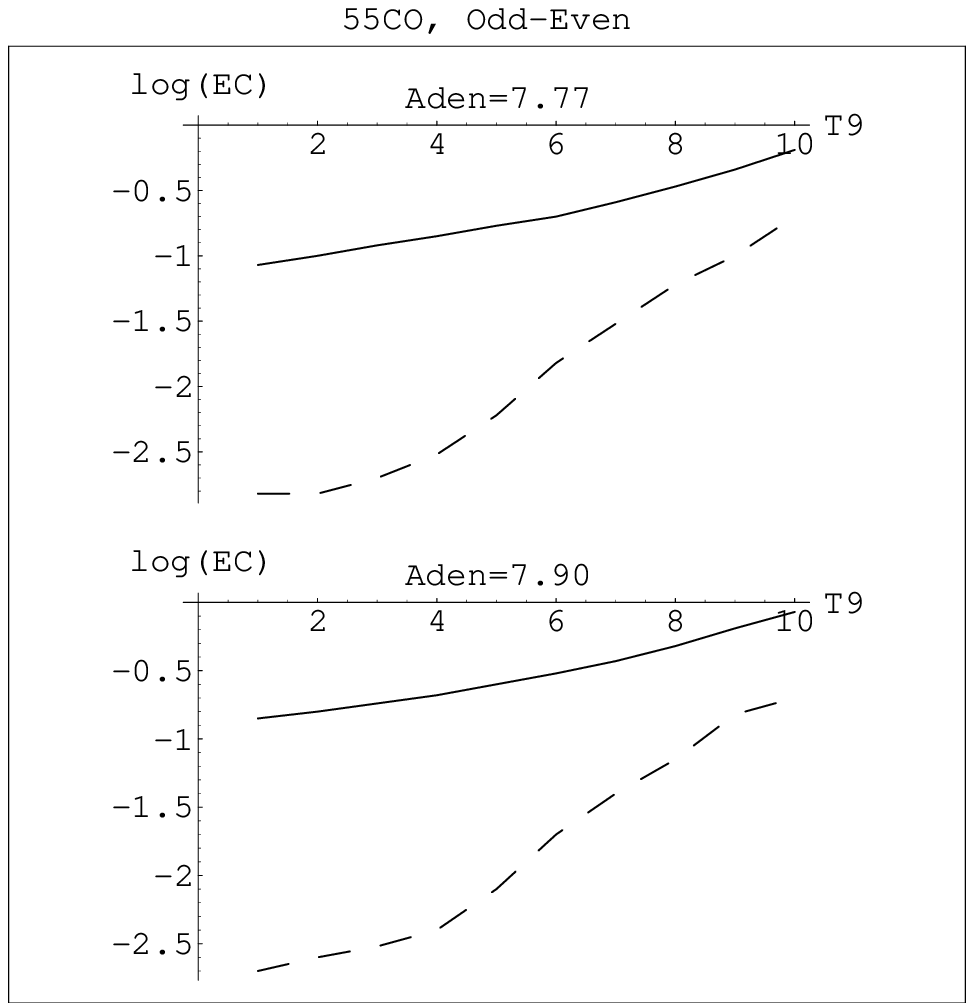}
\caption{ \footnotesize Same as Fig.~21 but for \ec by $^{55}$Co. }
\end{figure}

The authors in \cite{Lan98} calculated the \gt strength distributions for the ground state and first excited states in $^{55}$Co and $^{56}$Ni. The authors quote a noticeably smaller rate than the usually adopted rate for $^{55}$Co. The authors claim that in odd-A nuclei the \gt strength resides at higher energies than assumed by FFN. As seen in Fig.~21 and Fig.~22, the shell model rates are suppressed compared to the pn-QRPA rates  by an average factor of 65 at \lt = 1. At \lt = 10 the rates are suppressed on the average by a factor of 3. Due to the model limitations, the authors restricted themselves to parent excited states of a few MeV. They, for example, went to an $E_{i}$ of 2.565 MeV in the case of $^{55}$Co whereas the QRPA rates considered $E_{i}$ as high as 6.064 MeV (after which protons should be emitted from $^{55}$Co). 
\begin{figure}[h!]
\epsfxsize=7.8cm
\epsffile{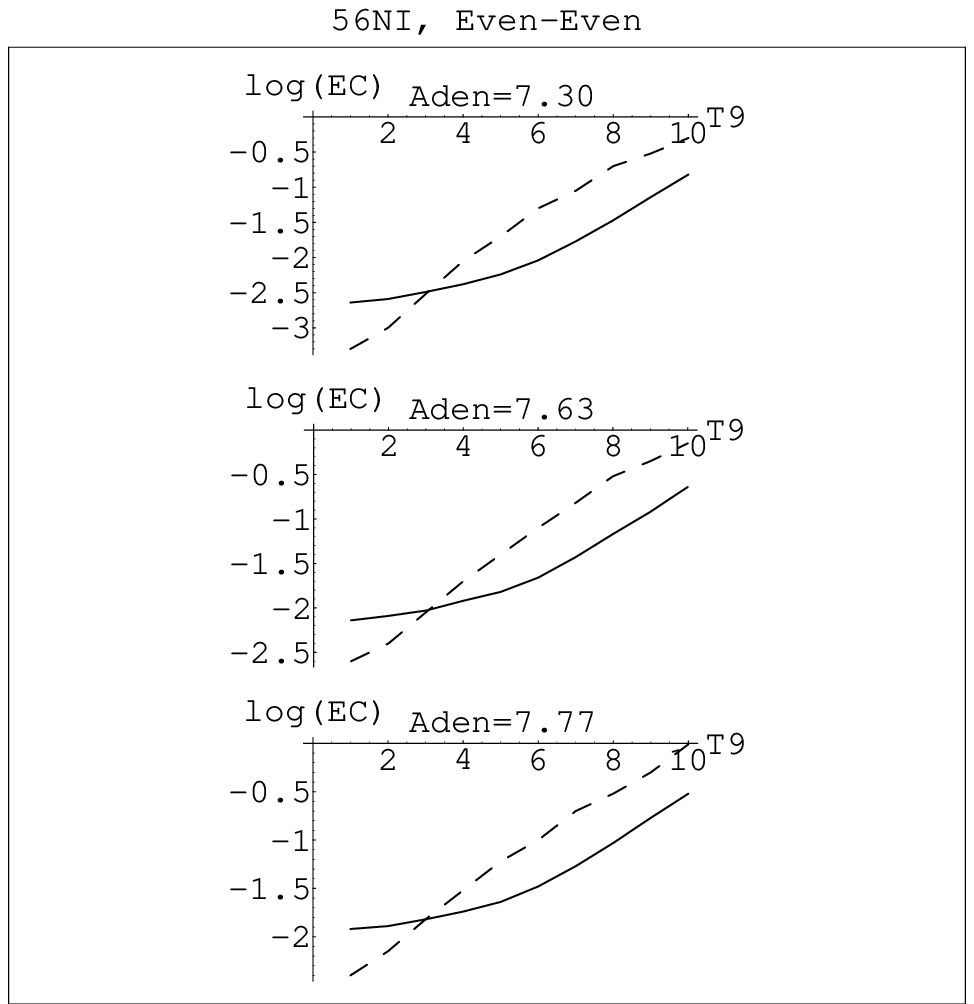}
\caption{ \footnotesize Same as Fig.~21 but for \ec by $^{56}$Ni. } 
\end{figure}

The story is somewhat different for $^{56}$Ni. Fig.~23 depicts the comparison. Till \lt = 3 the shell model rates are suppressed. At \lt = 1, eg., shell model rates are suppressed on the average by a factor of 3. The comparison at \lt =3 is perfect and then shell model rates start increasing. As the authors argue, for the even-even case the centroid of the \gt strength is placed at too high excitation energies in FFN and \cite{Auf94}. Placement of the \gt strength at a lower excitation energy can explain the sudden increase in their calculated rates in comparison with the QRPA rates with increasing temperature.
\begin{figure}[h!]
\epsfxsize=7.8cm
\epsffile{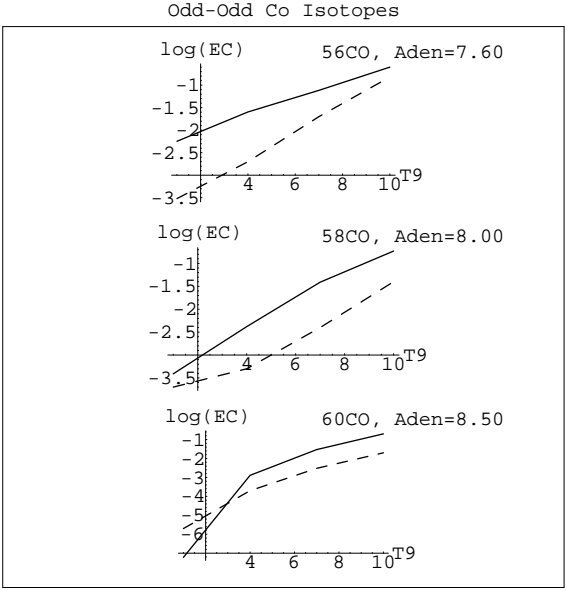}
\caption{ \footnotesize  Same as Fig.~21 but for \ec by odd-odd cobalt isotopes. } 
\end{figure}

Fig.~24 shows the comparison of the \ec rates of the odd-odd cobalt isotopes, $^{56,58,60}$Co.
In the first two graphs the QRPA rates are enhanced because all possible excited states in parent nuclei were considered for calculation in this work which were below the particle decay channel, whereas in the work of \cite{Lan98a}, the first few excited states were considered. In the last case at higher temperatures (\lt $>$ 3) our rates surpass the shell model rates. The probability of occupation of excited states increases with increasing temperature [Eq.~(15)]. Considerable enhancements to the rates come from transitions from these excited states. 
\begin{figure}[h!]
\epsfxsize=7.8cm
\epsffile{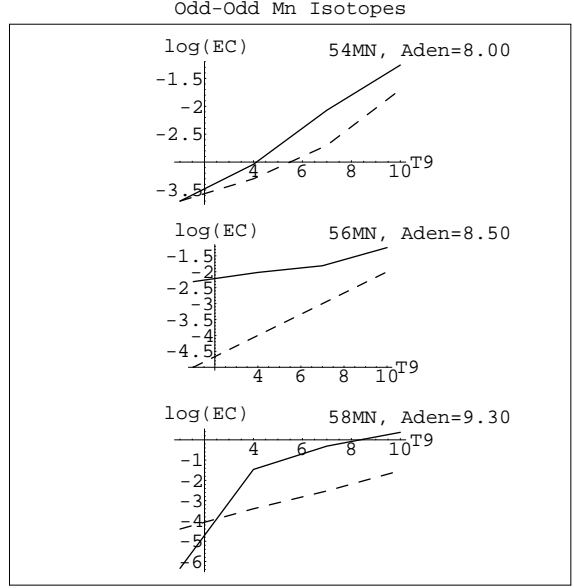}
\caption{ \footnotesize  Same as Fig.~21 but for \ec by odd-odd manganese isotopes. }
\end{figure}

Fig.~25 which shows a comparison of \ec rates for odd-odd manganese isotopes is similar in explanation to Fig.~24. No experimental information exists for the \gt strength for odd-odd nuclei and one has to rely totally on the theoretical model, and the truncations might cause the suppression in their rates. Further for the case of odd-odd nuclei, the shell model gives very weak transition strength to the low lying states in the daughter nucleus \cite{Lan98a}.

The calculation of \bt decay rates using SMDA was discussed in \cite{Mar98}. Here they tried to enhance their rates by using the back resonances from daughter states below 1~MeV energy. However no estimate of the energy was given to which the back resonances reach. Figures~26--28 depict a comparison of the \bt decay rates. 
\begin{figure}[h!]
\epsfxsize=7.8cm
\epsffile{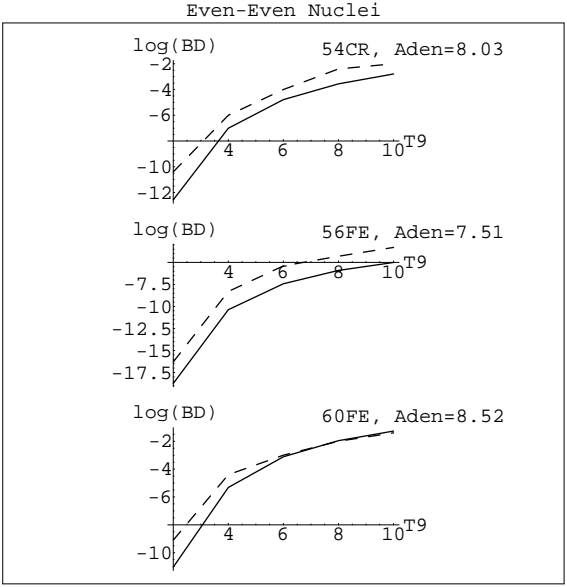}
\caption{ \footnotesize Comparison of the QRPA beta-decay (BD) rates (this work) with those of \cite{Mar98}. Solid lines represent the QRPA \bt rates of this work while broken lines represent the \bt rates of \cite{Mar98}. Log(T) is the log of temperature in units of Kelvin and log(BD) represents the log of \bt rates in units of sec$^{-1}$. Aden represents the log of density in units of g cm$^{-3}$.}
\end{figure}
\begin{figure}
\epsfxsize=7.8cm
\epsffile{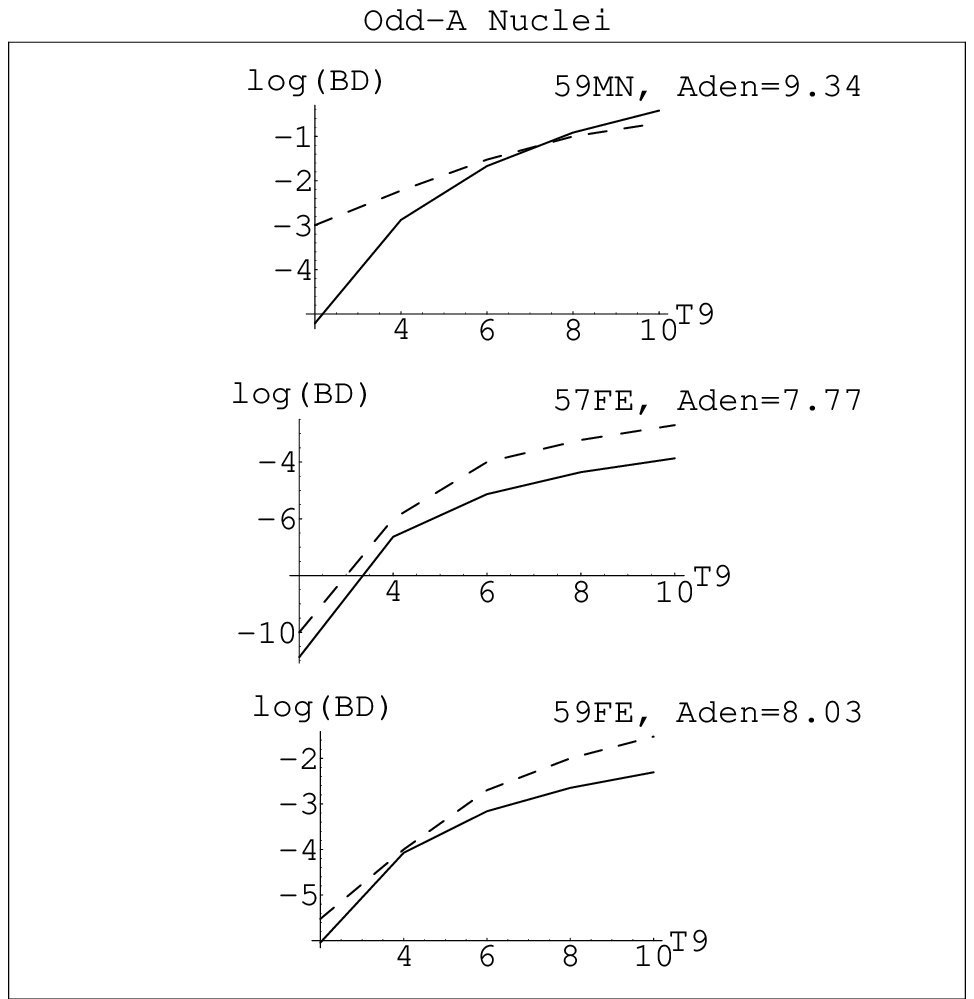}
\caption{ \footnotesize Same as Fig.~26 but for beta decay by odd-A nuclei. }
\end{figure}
\begin{figure}[h!]
\epsfxsize=7.8cm
\epsffile{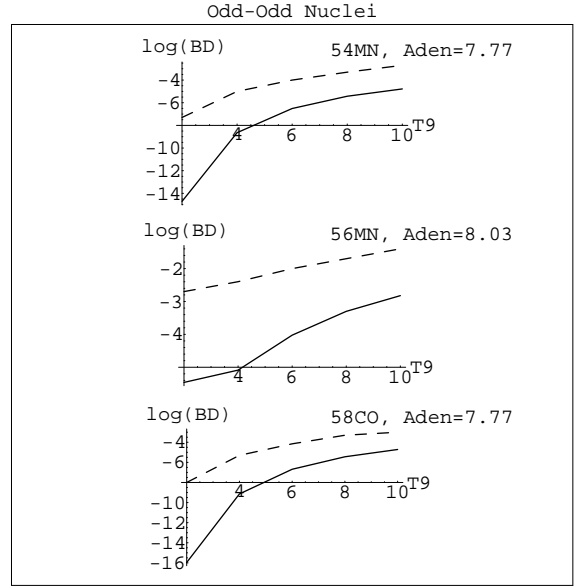}
\caption{ \footnotesize Same as Fig.~26 but for beta decay by odd-odd nuclei. }
\end{figure}
As can be seen (Fig.~26) the QRPA rates are suppressed by about one order of magnitude. The same trend is seen in the comparison of odd-A nuclei (Fig.~27). The results support the statement in \cite{Mar98} that experimental data and shell model studies tend to decrease the contribution of the back-resonance to the \bt decay rates for even-even and odd-A nuclei.
 One should note that at high densities and temperatures the QRPA rates surpass the SMDA rates. This shows the necessity of considering transitions from high-lying states at high temperatures and densities. The QRPA approach, used in the present calculation, allowed the use of a very large model space considering excitation of particles up to 7 major shells (SMDA, in comparison, considered 1 major shell). 

The difference in comparison is greatest for the case of odd-odd nuclei. Here the QRPA rates are suppressed by more than 7 orders of magnitude (Fig.~28). Our results do not support the conclusions drawn in \cite{Mar98} that the \bt decay rate of odd-odd nuclei should be large compared to previous compilations. Table~(2) and Table~(3) show the comparison of the QRPA rates, SMDA rates, FFN rates and the calculation of \cite{Auf94} for \ec and \bt decay rates, respectively. Table~(2) and Table~(3) are similar to Table~I and Table~II of \cite{Mar98}. 

The probabilities of $\beta$-delayed proton (neutron) $P^{p(n)}$ emission at temperature 10$^{7}$ K and density $\rho Y_{e}=$ 10 g cm$^{-3}$ can be compared to terrestrial probabilities. We searched the literature for measured probabilities to check the reliability of the present calculation. Unfortunately measured probabilities are not always given; often it is only stated that particle emission processes are observed. We found that agreement with available experiment is satisfactory and that in all instances of reported particle emission our calculation also gives finite probabilities $P^{p(n)}$. (See \cite{Hir92} for a discussion of comparison of such probabilities with experimental observations.)

\section{RELIABILITY OF THE pn-QRPA CALCULATIONS}
It should be pointed out that the uncertainty in calculations of stellar weak rates can be considerably large. Lack of experimental data deteriorates the situation. The ``\ec direction'' can be explored experimentally by (n,p) experiments, whereas the ``\btm decay'' side can be explored by the (p,n) reactions. There are a handful of other experiments which have been used by many theorists to shape the centroid and width of the \gt strength. However these experimental data are not enough to completely explore the domain of nuclei which are interesting from astrophysical viewpoint. There exists, eg., no experimental information about the \gt strength distribution for odd-odd nuclei in the $fp$-shell. Further there is almost no experimental data concerning the \gt strength distribution from parent excited states. In the stellar environment, at high temperatures and densities, there is a finite probability of occupation of parent excited states and transitions from these excited states are sometimes many orders of magnitude higher than transitions from the ground state. For most of the calculations then one uses the so-called Brink hypothesis which assumes that the \gt strength distribution on excited states is the same as for the ground state, only shifted by the excitation energy of the state. This hypothesis is a very crude approximation and sometimes can lead to misleading results.

The \wi rates are calculated using 
\be
\lambda_{ij}=\frac{ln2}{D}f_{ij}B_{ij}.
\ee
Here $i$ represents the parent excites states and $j$ the daughter's. The first factor is a constant and the second are phase space integrals which can be calculated relatively accurate. In this work the \wi  rates are calculated using a microscopic theory. This theory (pn-QRPA) constructs parent and daughter excited states and also calculates the \gt strength distribution among these states. In other words the Brink hypothesis is not employed in this calculation. However there are still uncertainties present. There are, eg., uncertainties present in the calculation of excited states. Mass defects are used as input parameters to calculate the Q-value of the reaction and $S_{n}$ and $S_{p}$ of parent and daughter nuclei. For certain neutron-rich and proton-rich nuclei, no experimental masses are present and one has to rely on theoretical mass models. The uncertainty in the Q-values must be viewed as the limiting factor for the prediction of \wi rates of unknown isotopes (see \cite{Hir93}). Therefore this work incorporates the latest experimental compilations of excitation energies and $logft$ values wherever present (see Table~(1) and Section~3), to ensure the most reliable calculation of stellar weak rates allowed by the pn-QRPA theory.

There, however, exists a check for the calculations of these \wi rates. At the lowest temperature considered in this work (T9 = 0.01), excited parent states are not appreciably populated, while at the lowest density considered in this work ($\rho Y_{e}$ = 10 gcm$^{-3}$), the continuum electron density is quite low, and except for \ec, the stellar rates should be close to the terrestrial values. The exception arises because only continuum \ec and no bound states capture is calculated. There is a reasonable amount of data of measured half-lives available as compared to the measured \gt strength functions. The idea was to check the calculation of the half-lives using the pn-QRPA theory with the measured ones and to also check the predictive power of the pn-QRPA theory (regarding the half-lives) with the recent measurements. We present such comparisons for thousands of nuclei in this section to give a feeling how the pn-QRPA theory can perform in the domain of high temperature and density for the calculation of stellar rates of 719 nuclei considered in this project. We also present the comparison of our calculated total \gt strength function with those measured from $(n,p)$ and $(p,n)$ reactions at the end of this section.

A comparison between the terrestrial half-lives (calculated by \cite{Sta90a,Hir93}) and the stellar half-lives at temperature (T9 = 0.01) and density ($\rho Y_{e}$ = 10 gcm$^{-3}$) is given in Table~(4) and Table~(5). The terrestrial half-lives are calculated using the mass formula of \cite{Moe81} whereas the stellar rates are presented using the mass formulae of \cite{Moe81,Mye96}. The slight differences are attributed to the fact that in the present calculation a model space of 7 major shells was considered for all nuclei, while the calculation for terrestrial half-lives \cite{Sta90a,Hir93} considered sometimes also smaller model spaces.

\begin{figure}[h!]
\centerline{\epsfig{file=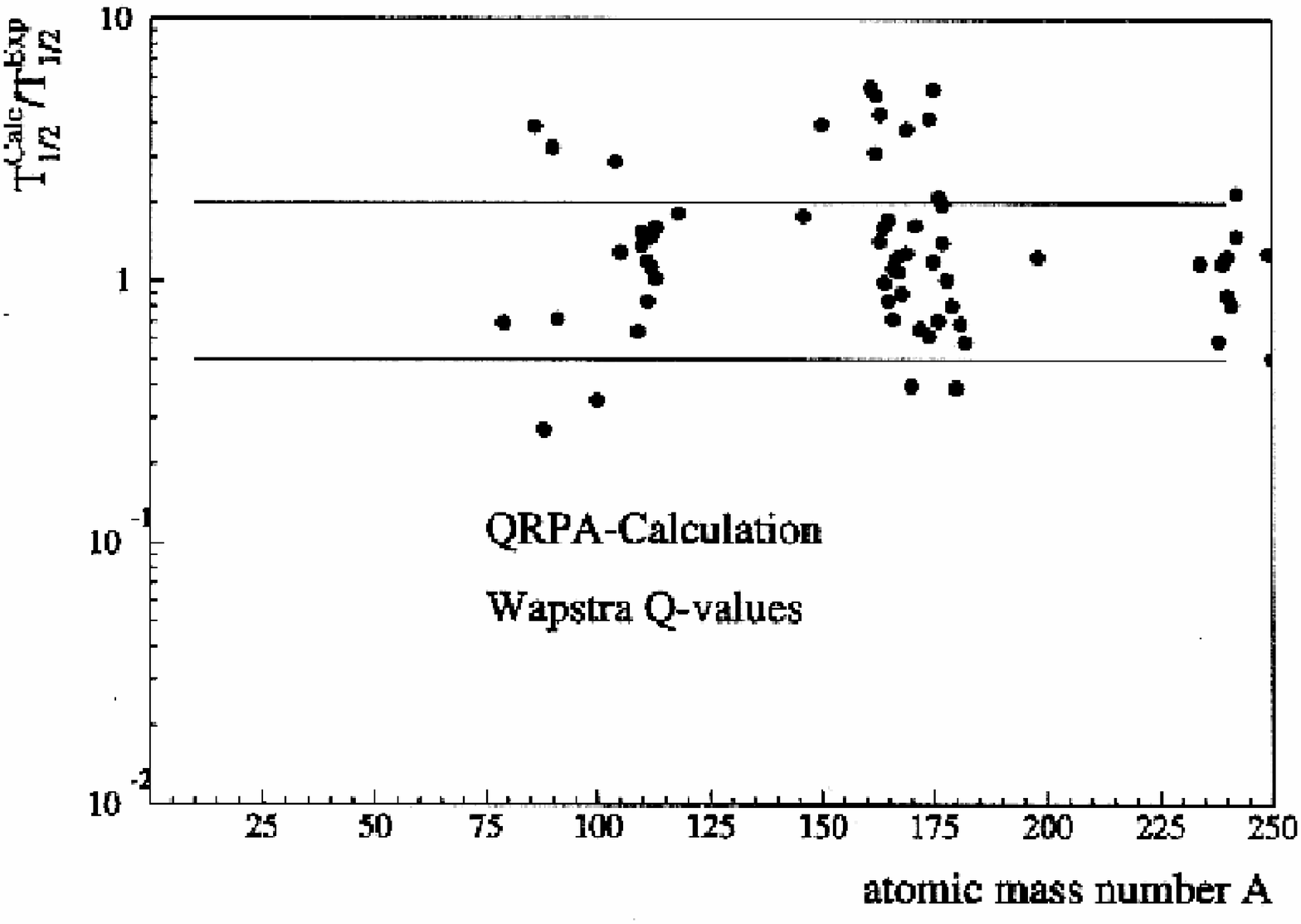,height=3.2in,width=3.0in}}
\caption{Predictive power of the pn-QRPA theory. Here the ratios of the calculated $\beta^{+}/(EC)$ decay half-lives to measured ones are shown as a function of mass numbers. The measured values were obtained \textit{after} the calculations \cite{Hir93}.}
\end{figure}

The terrestrial half-lives are compared to the experimental data in Table~(6) and Table~(7).
In Tables~(6) and (7), $N$ denotes the number of experimentally known half-lives shorter than the limit in the second line, $n$ is the number (and percentage) of isotopes reproduced under the condition given in the first column, and $\bar{x}$ is the average deviation defined by
\be
\bar{x} = \frac{1}{n}\sum_{i=1}^{n}x_{i},
\ee
where\\
\hspace*{1.5cm}$x_{i}=T^{cal}_{1/2}/T^{exp}_{1/2}$ \hspace*{0.2cm} if \hspace*{0.2cm} $T^{cal}_{1/2} \geq T^{exp}_{1/2}$\\
\vspace*{0.1cm}\\
\hspace*{1.5cm}$x_{i}=T^{exp}_{1/2}/T^{cal}_{1/2}$ \hspace*{0.2cm} if \hspace*{0.2cm} $T^{cal}_{1/2} < T^{exp}_{1/2}$.\\
\vspace*{0.2cm}\\
For example, the pn-QRPA reproduces 93$\%$ (75$\%$) of all experimentally known half-lives shorter than 1 minute for $\beta^{+}$/EC within a factor of 10 (2) with an average deviation of $\bar{x}$ = 1.72 (1.31) and 96$\%$ (82$\%$) of all known $\beta^{-}$-decaying nuclei with half-lives less than a minute are reproduced within a factor of 10 (2) with an average deviation of $\bar{x}$ = 1.67 (1.36). 

\begin{figure}[h!]
\centerline{\epsfig{file=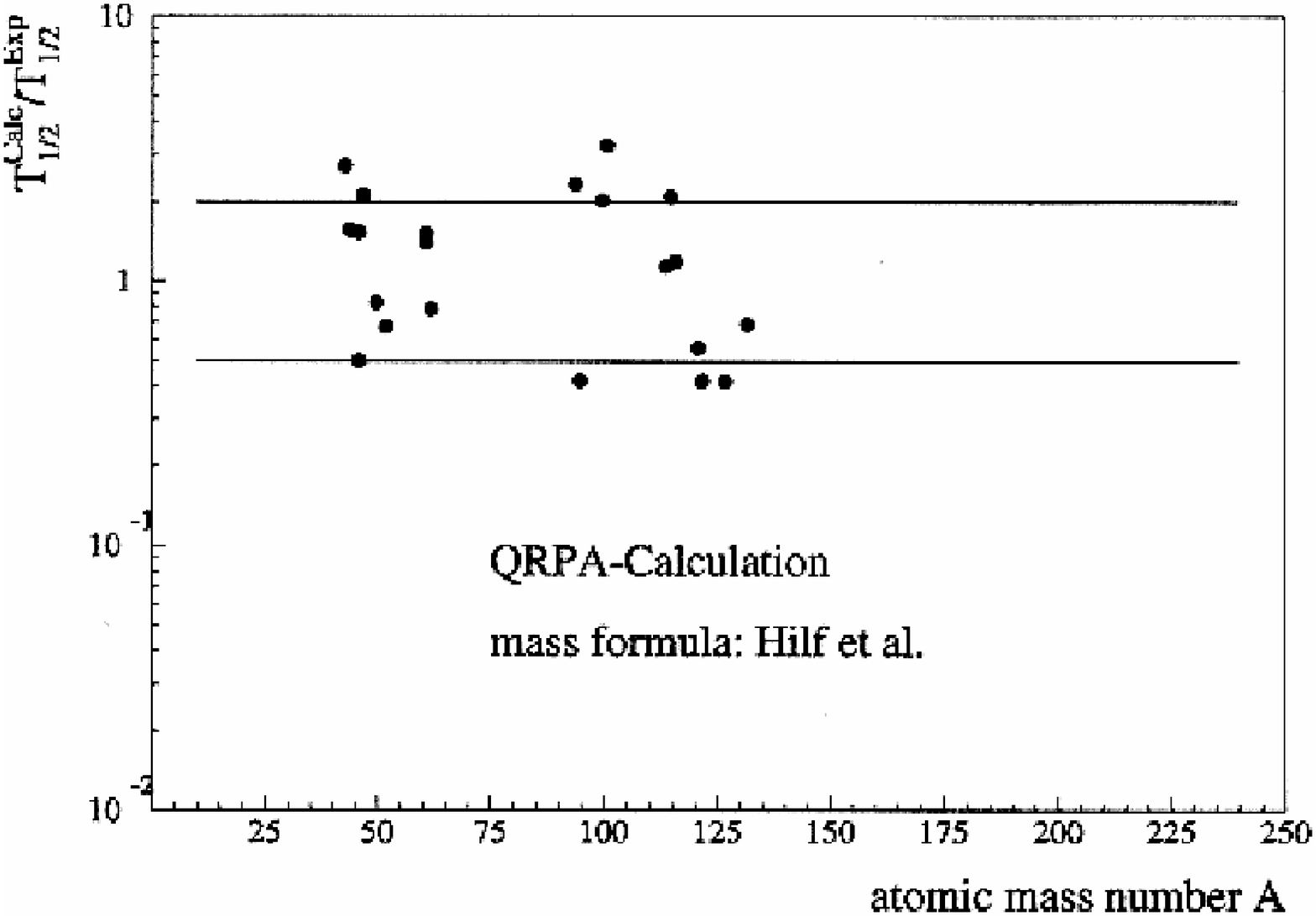,height=3.2in,width=3.0in}}
\caption{Predictive power of the pn-QRPA theory for proton-rich nuclei. Here the ratios of the calculated $\beta^{+}/(EC)$ decay half-lives (using mass formula of \cite{Hil76}) to measured ones are shown as a function of mass numbers. The measured values were obtained \textit{after} the calculations \cite{Hir93}.}
\end{figure}
\begin{figure}
\centerline{\epsfig{file=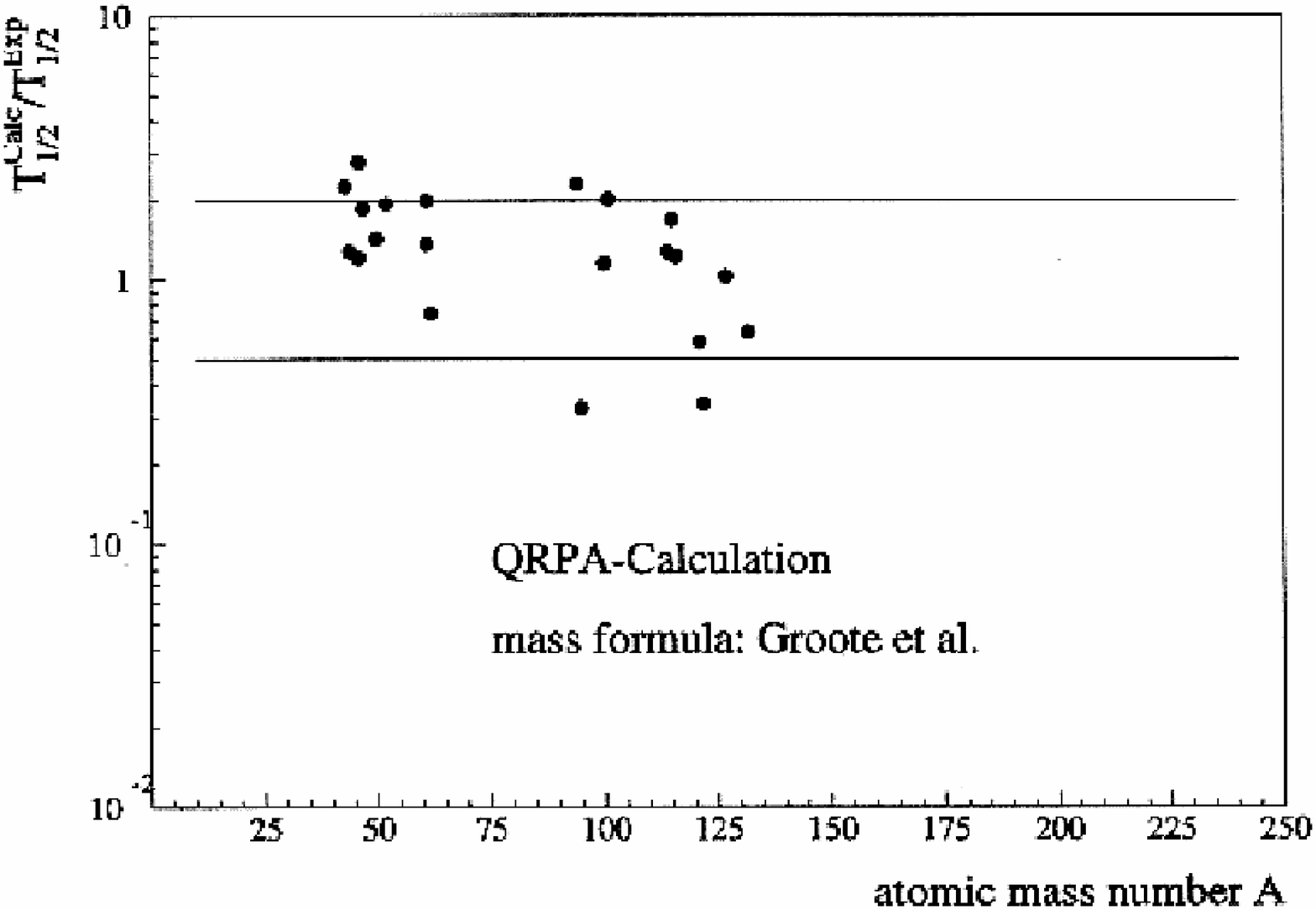,height=3.2in,width=3.0in}}
\caption{Same as Fig.~30 but here the mass formula of \cite{Gro76} was used for the calculated half-lives.}
\end{figure}
\begin{figure}[h!]
\centerline{\epsfig{file=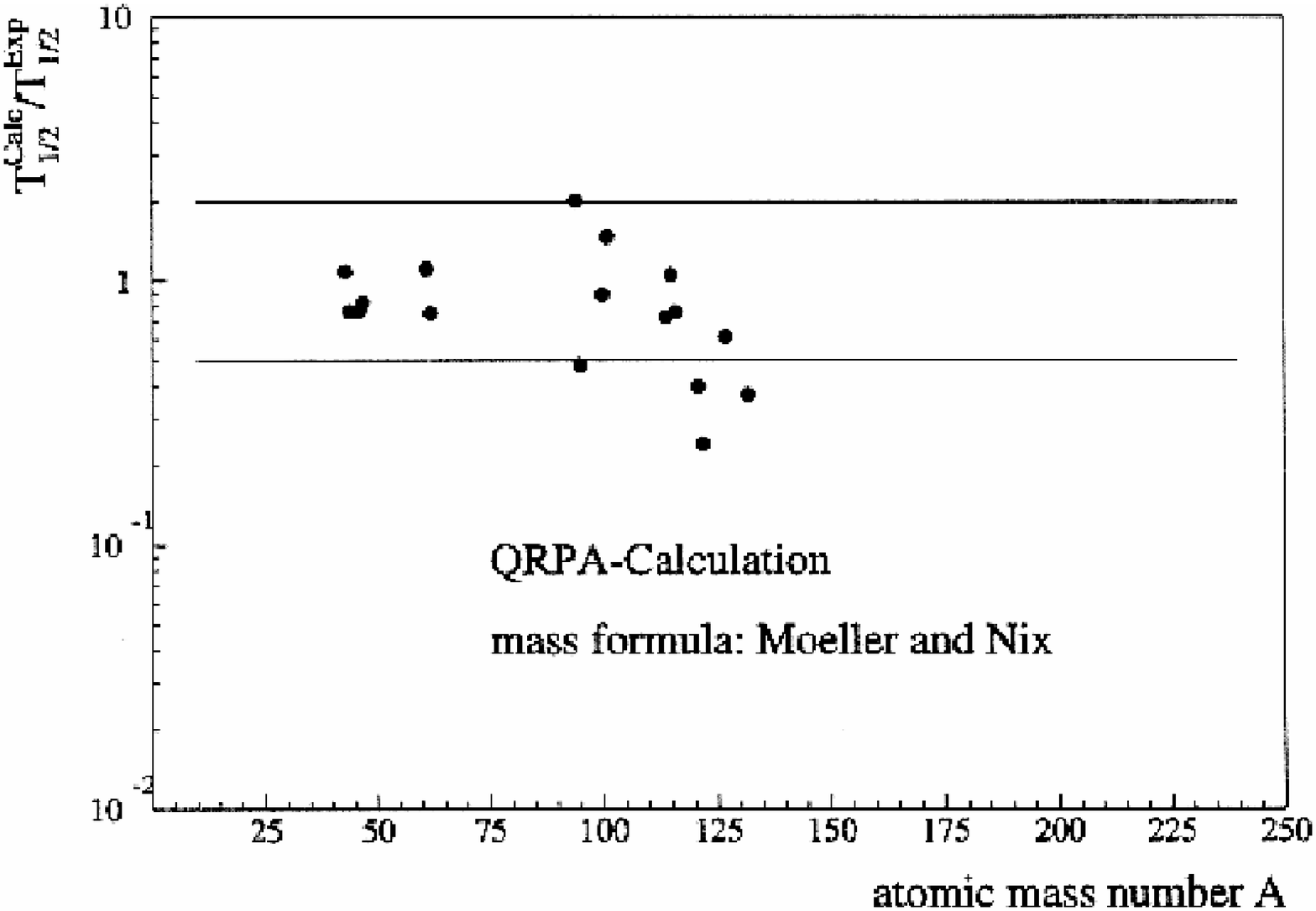,height=3.2in,width=3.0in}}
\caption{Same as Fig.~30 but here the mass formula of \cite{Moe81} was used for the calculated half-lives.}
\end{figure}
\begin{figure}[h!]
\centerline{\epsfig{file=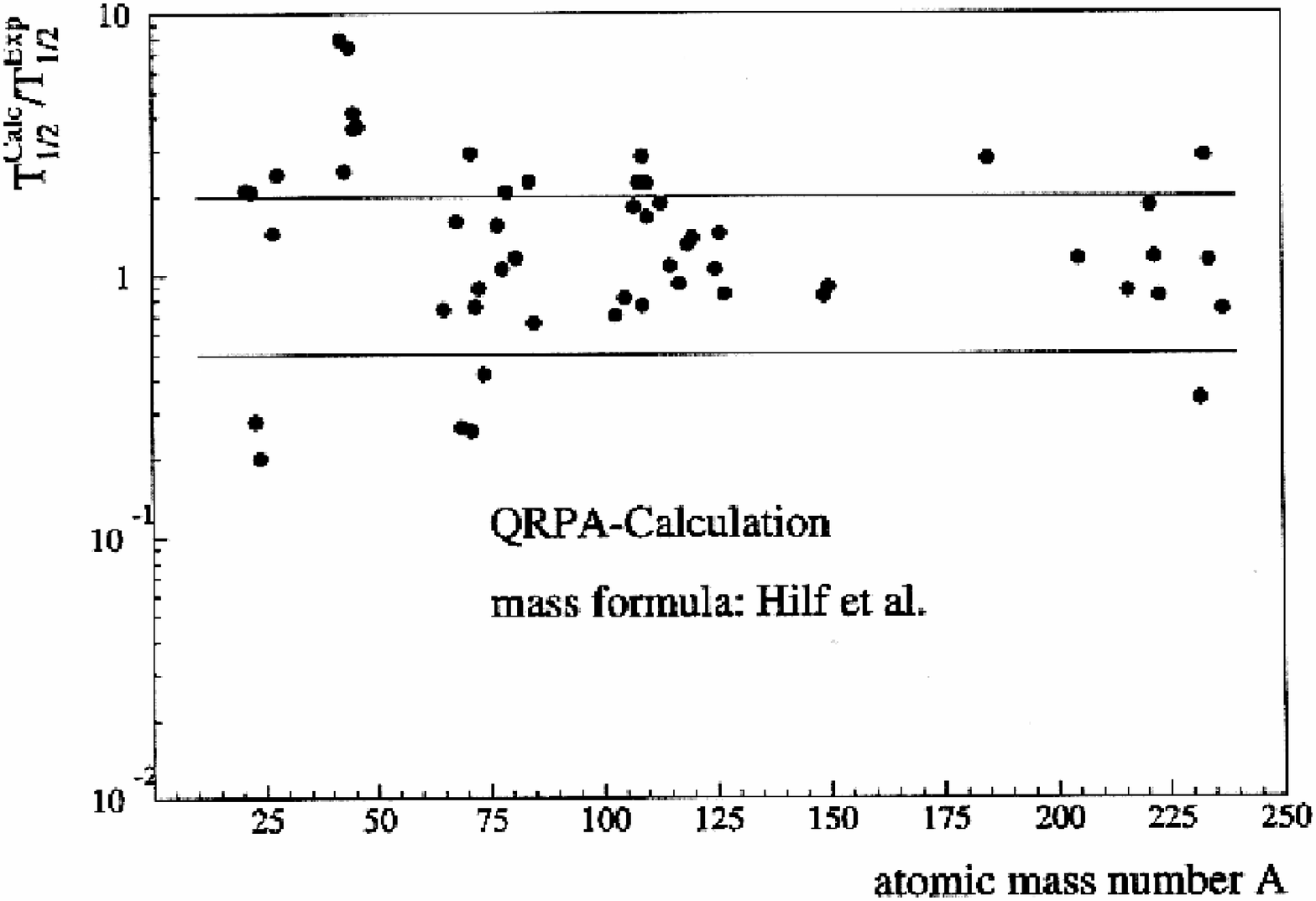,height=3.2in,width=3.0in}}
\caption{Same as Fig.~30 but depicting the predictive power of the pn-QRPA theory for neutron-rich nuclei (\btm decays).} 
\end{figure}
\begin{figure}[h!]
\centerline{\epsfig{file=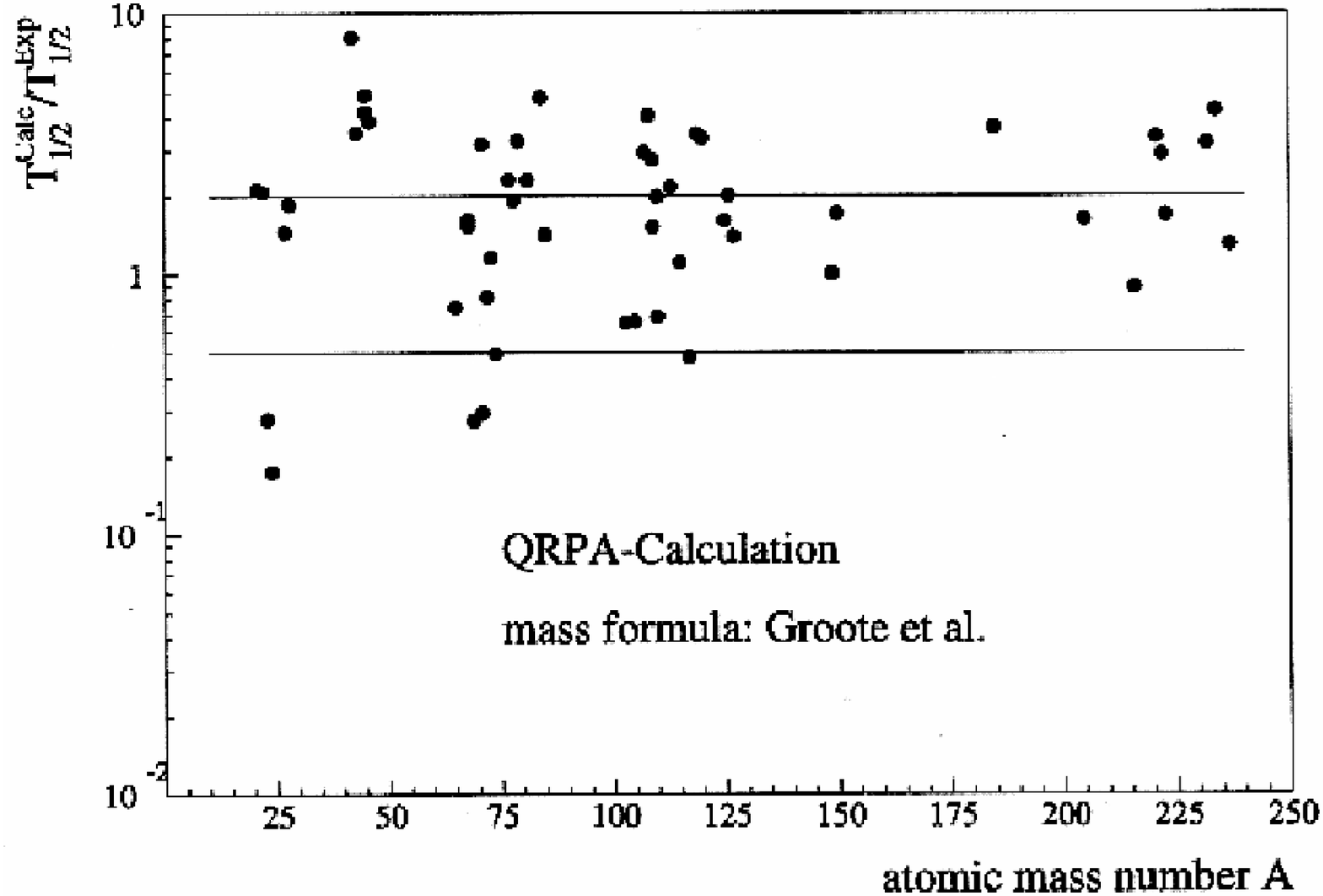,height=3.2in,width=3.0in}}
\caption{Same as Fig.~33 but here the mass formula of \cite{Gro76} was used for the calculated half-lives.}
\end{figure}
\begin{figure}[h!]
\centerline{\epsfig{file=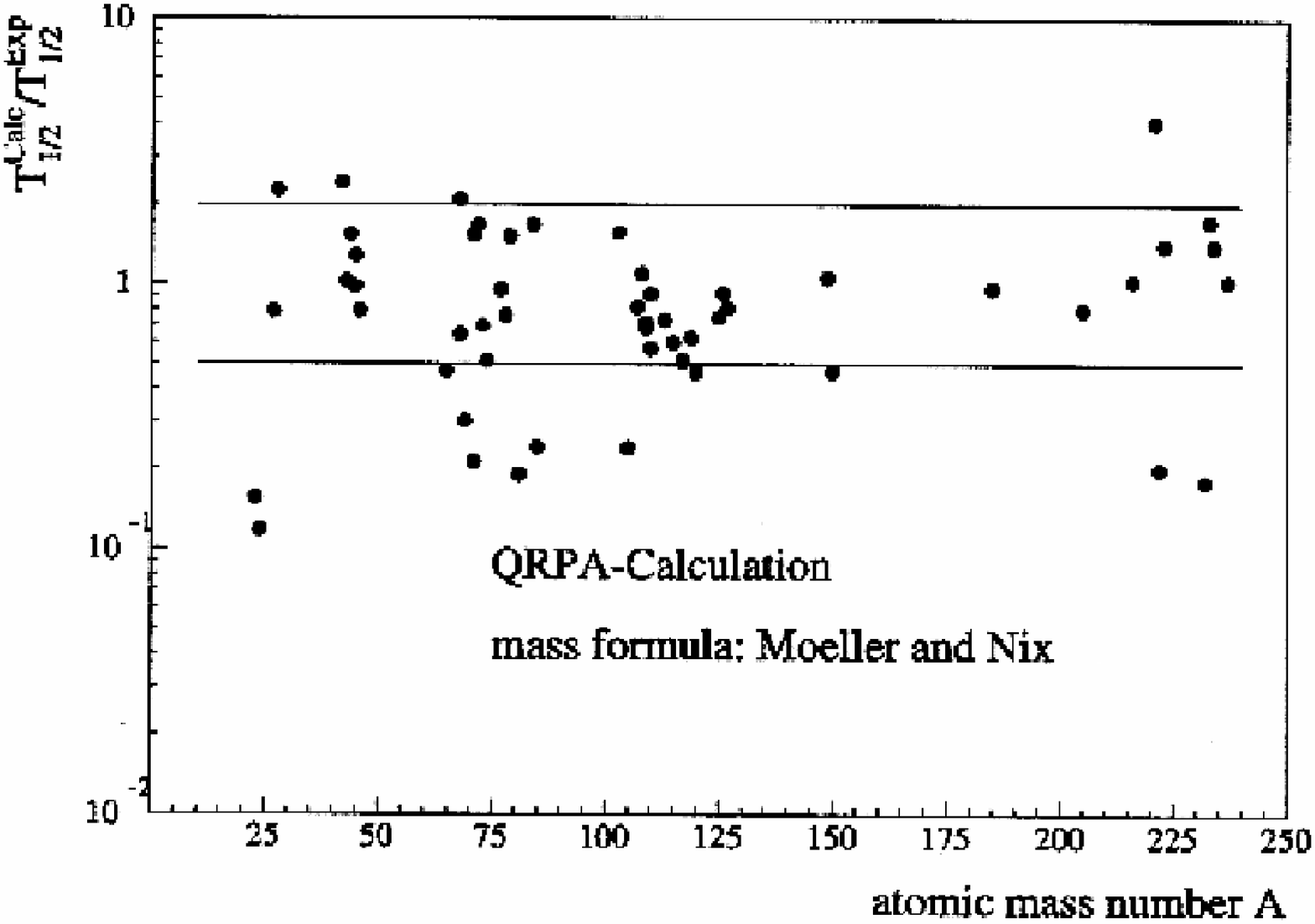,height=3.2in,width=3.0in}}
\caption{Same as Fig.~33 but here the mass formula of \cite{Moe81} was used for the calculated half-lives.}
\end{figure}

It can be seen from the tables that the model works better with increasing neutron excess (corresponding to shorter half-lives), that is, with increasing distance from the line of stability. This is in agreement with the expectation, since forbidden transitions are neglected in the calculation. This is also a promising feature with respect to the prediction of unknown half-lives, implying that the predictions are made on the basis of a realistic physical model (see for example, Table~B and Fig.~6 of \cite{Sta90a}).

The \textit{predictive} power of the pn-QRPA model is shown in Table~(8). Here a comparison of the measured $\beta^{+}/(EC)$ decay half-lives with those calculated in the pn-QRPA theory is given. The Q-values were taken from \cite{Wap88}. All the calculated rates shown were computed \cite{Hir93} \textit{before} their measurements given in the third column. Fig.~29 shows that most of the calculated half-lives are in very good agreement with the measured ones. 

Also the $\beta^{+}/(EC)$ decay half-lives of many proton-rich nuclei were measured after they were calculated in \cite{Hir93}. Since experimental masses were not available for these proton-rich nuclei at that time, Hirsch et al. calculated these half-lives using three different mass formulae from \cite{Hil76}, \cite{Gro76} and \cite{Moe81}. Table~(9) shows the comparison of these calculated half-lives with the measured ones. 
Figures~30-32 show the comparison with these later measured experimental data for the calculations of half-lives using the mass formulae of \cite{Hil76}, \cite{Gro76} and \cite{Moe81}, respectively. Again the comparison is very satisfactory. 

A search of new literature was also made for the measurement of neutron-rich nuclei which were not measured at the time of their calculations in \cite{Sta90a} to check the predictive power of the pn-QRPA for neutron-rich nuclei. Table~(10) shows the measured values and those calculated earlier by \cite{Sta90a} using mass formulae of \cite{Hil76}, \cite{Gro76} and \cite{Moe81}  
whereas the comparison is depicted graphically in Figures~33-35. Once again the comparison is very satisfactory showing the good predictive powers of the pn-QRPA theory and putting weight into the reliability of our calculations.

Finally we also present the comparison of our calculated \gt strength function with those measured. As stated earlier, not much experimental measurements of the \gt strength function are present. $(p,n)$ and $(n,p)$ reactions can be used to measure the total $B(GT)^{-}$ and $B(GT)^{+}$ strength function, respectively. One should be cautioned about the uncertainties present in these sorts of measurements and various energy cutoffs are used as a reasonable upper limit on the energy at which \gt strength could be reliably related to measured $\Delta L = 0$ cross sections. Table~11 and Table~12 show the comparison of our calculated \gt strength function with the measured ones for $(n,p)$ and $(p,n)$ reactions, respectively. In the comparisons we used the same energy cutoff as used in the respective experiments. We also include in these tables the results of the large-scale shell model calculations of the total \gt strength function \cite{Cau99} where ever reported. One notes that our comparison with measured strength for $^{60,62,64}$Ni isotopes is not in good agreement. Mass formula \cite{Moe81} predicts these proton-magic isotopes to be essentially spherical. However this is not the case when one measures the electric quadrupole transition probabilities (see discussions in \cite{Hir93}). This might can cause the enhancement in our calculated strength function (around a factor of 2). Low-lying transitions can contribute considerably in stellar rates due to phase space enhancements. We should like to point here again that in this compilation of stellar rates, experimental informations, derived from measured half-lives, were used, when ever available, and inserted if the pn-QRPA theory was missing them (see Section~3).   
\section{SUMMARY}
In a series of paper beginning with a paper considering the $sd$-shell nuclei \cite{Nab99}, we present extensive microscopic calculations of \wi rates in stellar environment and the first ever one for the $fp$-shell nuclei. The calculations are done for 619 nuclei in the mass range A = 40 to 100. In this first paper of the $fp$-shell series we present the results for nuclei in the mass range A = 40 to 60.
The pn-QRPA theory which gives good results for the terrestrial rates also promises to be a good candidate for calculations of stellar rates.

We compared our results to the FFN calculations and to the recently reported shell model calculations.  We considered for the first time the effect of particle emission processes from parent excited states in these types of calculations. This leads to a general suppression of decay rates, and electron capture rates at high temperatures and densities, in comparison to the FFN rates. Our calculations are in very good agreement with the FFN calculations wherever experimental data are available.  

These calculations, including \wi rates of many important neutron-rich nuclei, are of prime importance for stellar and galactic evolution processes and nucleosynthesis calculations.

In the next papers of this series we plan to present our results for A = 61 to 80 and A = 81 to 100, respectively.

\onecolumn
\noindent
\textbf{Table (1):} References of experimental data incorporated in this calculation (NP $\rightarrow$ Nuclear Physics, NDS $\rightarrow$ Nuclear Data Sheets).\\
\vspace*{0.1cm}\\
\begin{tabular}{cc} \hline\\
Mass Number A & Reference \\ \\ \hline
40 - 44 & NP A521,1 (1990)\\
45      & NDS 65,1 (1992), NDS 40,149 (1983)\\
46      & NDS 68,271 (1993), NDS 49,237 (1986)\\
47      & NDS 74,1 (1995), NDS 48,1 (1986)\\
48      & NDS 68,1 (1993), NDS 45,557 (1985)\\
49      & NDS 76,191 (1995), NDS 48,569 (1986)\\
50      & NDS 75,1 (1995)\\
51      & NDS 81,183 (1997)\\
52      & NDS 71,659 (1994), NDS 58,677 (1989)\\
53      & NDS 61,47 (1990), NDS 43,481 (1984)\\
54      & NDS 68,887 (1993), NDS 50,255 (1987)\\
55      & NDS 64,723 (1991)\\
56      & NDS 67,523 (1992), NDS 51,1 (1987)\\
57      & NDS 67,195 (1992), NDS 47,1 (1986)\\
58      & NDS 80,789 (1997)\\ 
59      & NDS 69,733 (1993), NDS 39,641 (1983)\\ 
60      & NDS 69,1 (1993), NDS 48,251 (1986)\\\hline
\end{tabular}
\newpage
\noindent
\textbf{Table (2):} Comparison of the electron capture rates calculated by the pn-QRPA theory (this work) with previous works. $\rho_{7}$ is the density (in units of 10$^{7}$ g cm$^{-3}$), $T_{9}$ is the temperature (in units of 10$^{9}$ K). QRPA, SM, FFN and AFWH denote the rates calculated in this work, \cite{Mar98}, \cite{Ful82} and \cite{Auf94}, respectively. Exponents are given in parenthesis. All rates are given in units of $s^{-1}$.
\begin{table}[h!]
\begin{tabular}{ccccccc}\hline\\
Nucleus & $\rho_{7}$ & $T_{9}$ & QRPA & SM & FFN & AFWH\\ \\ \hline
$^{56}$Ni & 4.32 & 3.26 & 9.9 (-3) & 1.3 (-2) & 7.4 (-3) & 8.6 (-3)\\
$^{54}$Fe & 5.86 & 3.40 & 1.3 (-5) & 4.2 (-5) & 2.9 (-4) & 3.1 (-4)\\
$^{58}$Ni & 5.86 & 3.40 & 3.7 (-4) & 8.1 (-5) & 3.7 (-4) & 6.3 (-4)\\
$^{56}$Fe & 10.7 & 3.65 & 1.1 (-6) & 2.1 (-6) & 1.0 (-5) & 4.7 (-7)\\ \hline
$^{55}$Co & 4.32 & 3.26 & 8.0 (-2) & 1.6 (-3) & 8.4 (-2) & 5.1 (-2)\\
$^{57}$Co & 5.86 & 3.40 & 1.6 (-3) & 1.3 (-4) & 1.9 (-3) & 3.4 (-3)\\
$^{55}$Fe & 5.86 & 3.40 & 4.8 (-3) & 1.9 (-4) & 5.8 (-3) & 3.8 (-3)\\
$^{59}$Ni & 5.86 & 3.40 & 4.1 (-3) & 4.7 (-4) & 4.4 (-3) & 4.4 (-3)\\ 
$^{59}$Co & 10.7 & 3.65 & 4.9 (-4) & 7.8 (-6) & 2.1 (-4) & 2.1 (-4)\\
$^{53}$Mn & 10.7 & 3.65 & 1.4 (-2) & 3.3 (-4) & 3.8 (-3) & 5.6 (-3)\\ \hline
$^{56}$Co & 5.86 & 3.40 & 3.3 (-2) & 1.7 (-3) & 6.9 (-2) & 5.1 (-2)\\
$^{54}$Mn & 10.7 & 3.65 & 7.5 (-4) & 3.1 (-4) & 4.5 (-3) & 1.1 (-2)\\
$^{58}$Co & 10.7 & 3.65 & 3.4 (-3) & 3.5 (-4) & 9.1 (-3) & 2.1 (-2)\\ 
$^{56}$Mn & 33.0 & 4.24 & 1.1 (-2) & 1.0 (-4) & 4.1 (-4) & 2.0 (-3)\\
$^{60}$Co & 33.0 & 4.24 & 2.0 (-3) & 1.7 (-4) & 1.1 (-1) & 6.1 (-2)\\ \hline 
\end{tabular}
\end{table}\\
\textbf{Table (3):} Comparison of the \bt decay rates calculated by the pn-QRPA theory (this work) with previous works. $\rho_{7}$ is the density (in units of 10$^{7}$ g cm$^{-3}$), $T_{9}$ is the temperature (in units of 10$^{9}$ K). QRPA, SM, FFN and AFWH denote the rates calculated in this work, \cite{Mar98}, \cite{Ful82} and \cite{Auf94}, respectively. Exponents are given in parenthesis. All rates are given in units of $s^{-1}$. FFN did not calculate rates for nuclei with A $>$ 60.\\
\begin{table}[h!]
\begin{tabular}{ccccccc}\hline \\
Nucleus & $\rho_{7}$ & $T_{9}$ & QRPA & SM & FFN & AFWH\\ \\ \hline
$^{56}$Fe & 5.86 & 3.40 & 5.5 (-13) & 3.9 (-11) & 2.3 (-10) & 5.9 (-11)\\
$^{54}$Cr & 5.86 & 3.40 & 1.8 (-8) & 2.2 (-7) & 2.2 (-5) & 1.5 (-7)\\
$^{58}$Fe & 10.7 & 3.65 & 7.3 (-9) & 5.2 (-8) & 2.6 (-6) & 1.5 (-7)\\
$^{60}$Fe & 33.0 & 4.24 & 1.1 (-5) & 1.7 (-4) & 4.6 (-3) & 1.0 (-3)\\ 
$^{52}$Ti & 33.0 & 4.24 & 1.8 (-5) & 1.3 (-3) & 1.1 (-2) & 1.2 (-4)\\ \hline
$^{59}$Fe & 33.0 & 4.24 & 6.2 (-5) & 6.0 (-5) & 6.3 (-3) & 5.3 (-3)\\
$^{61}$Fe & 33.0 & 4.24 & 4.2 (-3) & 1.7 (-3) &  & 6.4 (-2)\\
$^{61}$Co & 33.0 & 4.24 & 2.1 (-5) & 1.6 (-4) &  & 9.3 (-4)\\ 
$^{63}$Co & 33.0 & 4.24 & 3.8 (-2) & 1.6 (-2) &  & 1.4 (-2)\\
$^{59}$Mn & 220  & 5.39 & 1.1 (-2) & 2.2 (-2) & 7.2 (-1) & 1.4 (-1)\\ \hline
$^{58}$Co & 4.32 & 3.26 & 2.7 (-11) & 2.7 (-6) & 1.2 (-6) & 3.8 (-6)\\
$^{54}$Mn & 5.86 & 3.40 & 1.8 (-10) & 2.7 (-6) & 1.6 (-6) & 7.5 (-6)\\
$^{56}$Mn & 10.7 & 3.65 & 5.7 (-6) & 3.4 (-3) & 3.0 (-3) & 9.1 (-3)\\ 
$^{60}$Co & 10.7 & 3.65 & 8.3 (-7) & 6.6 (-4) & 1.4 (-3) & 3.4 (-3)\\
$^{50}$Sc & 33.0 & 4.24 & 6.6 (-4) & 1.2 (-2) & 2.8 (-2) & 1.8 (-1)\\ \hline 
\end{tabular}
\end{table}
\clearpage
\noindent
\begin{center}
\textbf{Comparison of the terrestrial and stellar half-lives at lowest temperature and density}
\end{center}
\vspace*{0.5cm}
\textbf{Table (4):} Comparison of calculated $\beta^{+}/(EC)$ decay half-lives at temperature (T9 = 0.01) and density ($\rho Y_{e}$ = 10 gcm$^{-3}$), with the calculation of terrestrial half-lives using the pn-QRPA theory \cite{Hir93}. The Q-values used in the calculations of rates in columns 3 and 4 are from \cite{Moe81} while the mass formula of \cite{Mye96} was used to calculate the rates in column 5. $^{55}$Zn is an unstable nucleus according to the calculations of \cite{Mye96}. All half-lives are given in seconds. Exponents are given in parenthesis. \\
\vspace*{0.1cm}\\
\begin{tabular}{ccccc} \hline \\
$A$ & Element & $T_{1/2}^{cal}$ (terrestrial) & $T_{1/2}^{cal}$ (stellar) & $T_{1/2}^{cal}$ (stellar) \\  
& &Q-values from \cite{Moe81} & Q-values from \cite{Moe81} & Q-values from \cite{Mye96} \\ \\ \hline
 42 & Cr & 1.16 (-2) & 1.19 (-2) & 1.36 (-2)\\
 43 & Cr & 2.27 (-2) & 2.42 (-2) & 2.97 (-2)\\
 45 & Fe & 7.03 (-3) & 9.37 (-3) & 1.01 (-2)\\
 46 & Fe & 1.53 (-2) & 4.53 (-2) & 5.79 (-2)\\ 
 47 & Fe & 2.23 (-2) & 2.12 (-2) & 2.28 (-2)\\
 50 & Ni & 1.45 (-2) & 7.44 (-3) & 1.59 (-2)\\
 51 & Ni & 1.82 (-2) & 1.47 (-2) & 2.45 (-2)\\
 54 & Zn & 4.39 (-3) & 2.58 (-3) & 4.98 (-3)\\ 
 55 & Zn & 8.16 (-3) & 6.70 (-3) & \\
 58 & Ge & 1.12 (-2) & 1.18 (-2) & 1.36 (-2)\\
 69 & Kr & 1.65 (-2) & 6.74 (-2) & 9.37 (-2)\\
 73 & Sr & 2.26 (-2) & 5.32 (-2) & 7.34 (-2)\\
 76 & Sr & 1.22 (+1) & 4.91 (+1) & 4.68 (+1)\\
 80 & Zr & 4.72 (+0) & 7.24 (+0) & 4.64 (+0)\\ 
 84 & Mo & 3.07 (+0) & 3.40 (+0) & 2.85 (+0)\\
 87 & Ru & 3.50 (-2) & 6.25 (-2) & 8.14 (-2)\\
 91 & Pd & 3.30 (-2) & 4.51 (-2) & 8.69 (-2)\\
 92 & Pd & 1.24 (+0) & 4.32 (+0) & 9.79 (+0)\\ 
 93 & Pd & 2.51 (-1) & 3.15 (-1) & 2.53 (-1)\\
 95 & Ag & 9.62 (-1) & 8.31 (-1) & 9.86 (-1)\\
 96 & Cd & 2.53 (-1) & 6.19 (-1) & 1.00 (+0)\\
 99 & In & 3.25 (+0) & 7.67 (+0) & 3.92 (+0)\\ \hline
\end{tabular}
\clearpage
\noindent
\begin{center}
\textbf{Comparison of the terrestrial and stellar half-lives at lowest temperature and density}
\end{center}
\vspace*{0.5cm}
\textbf{Table (5):} Comparison of calculated $\beta^{-}$ decay half-lives at temperature (T9 = 0.01) and density ($\rho Y_{e}$ = 10 gcm$^{-3}$), with the calculation of terrestrial half-lives using the pn-QRPA theory \cite{Sta90a}. The Q-values used in the calculations of rates in columns 3 and 4 are from \cite{Moe81} while the mass formula of \cite{Mye96} was used to calculate the rates in column 5. All half-lives are given in seconds. Exponents are given in parenthesis. \\
\vspace*{0.1cm}\\
\begin{tabular}{ccccc} \hline \\
$A$ & Element & $T_{1/2}^{cal}$ (terrestrial) & $T_{1/2}^{cal}$ (stellar) & $T_{1/2}^{cal}$ (stellar) \\  
& &Q-values from \cite{Moe81} & Q-values from \cite{Moe81} & Q-values from \cite{Mye96} \\ \\ \hline
 60 & Ti & 2.93 (-3) & 3.43 (-3) & 7.26 (-4)\\
 61 & Ti & 1.43 (-3) & 2.35 (-3) & 2.64 (-3)\\
 62 &  V & 1.89 (-2) & 1.94 (-2) & 1.03 (-1)\\
 63 &  V & 4.99 (-2) & 2.16 (-2) & 3.47 (-2)\\
 64 & Cr & 1.28 (-2) & 5.02 (-2) & 5.09 (-3)\\
 69 & Fe & 4.97 (-2) & 9.48 (-2) & 3.51 (+0)\\ 
 70 & Co & 4.05 (-2) & 1.78 (-2) & 1.45 (-2)\\
 71 & Co & 4.26 (-2) & 2.23 (-2) & 2.02 (-2)\\
 72 & Co & 2.23 (-2) & 1.77 (-2) & 1.70 (-1)\\ 
 73 & Ni & 6.27 (-1) & 2.39 (-1) & 3.95 (-1)\\
 74 & Ni & 5.63 (-1) & 4.31 (-1) & 1.21 (+0)\\
 75 & Ni & 2.11 (-1) & 1.81 (-1) & 2.92 (-1)\\
 76 & Ni & 1.17 (-1) & 1.24 (-1) & 2.12 (-1)\\
 77 & Ni & 8.27 (-2) & 9.77 (-2) & 1.41 (-1)\\
 78 & Ni & 3.96 (-2) & 5.93 (-2) & 7.60 (-2)\\ 
 82 & Zn & 4.45 (-2) & 8.97 (-2) & 8.09 (-2)\\
 85 & Ge & 1.31 (-1) & 1.67 (-1) & 1.30 (-1)\\
 86 & Ge & 5.70 (-2) & 7.04 (-2) & 5.31 (-2)\\ 
 88 & As & 5.70 (-2) & 8.14 (-2) & 6.53 (-2)\\
 89 & As & 4.78 (-2) & 5.08 (-2) & 4.84 (-2)\\
 96 & Kr & 2.67 (-1) & 4.08 (-1) & 3.34 (-1)\\
 97 & Kr & 1.70 (-1) & 3.29 (-1) & 2.27 (-1)\\ \hline
\end{tabular}
\clearpage
\noindent
\begin{center}
\textbf{Accuracy of the pn-QRPA theory}
\end{center}
\vspace*{0.5cm}
\textbf{Table (6):} The accuracy of the pn-QRPA model compared to experimental data ($\beta^{+}$/EC decay). $N$ denotes the number of experimentally known half-lives shorter than the limit in the second column, $n$ is the number (and percentage) of isotopes reproduced under the condition given in the first column, and $\bar{x}$ is the average deviation defined in the text (see \cite{Hir93} for details).\\
\vspace*{0.1cm}\\
\begin{tabular}{cccccc} \hline \\
Conditions & $T_{1/2}^{exp}(s) \leq$ & $N$ & $n$ & $n(\%)$ & $\bar{x}$\\ \\ \hline
$\forall x_{i} \leq$ 10 & 10$^{6}$ & 894 & 706 & 79.0 & 2.057 \\
                        & 60       & 327 & 304 & 93.0 & 1.718 \\
                        &  1       &  81 &  78 & 96.3 & 1.848 \\ \hline
$\forall x_{i} \leq$  2 & 10$^{6}$ & 894 & 489 & 54.7 & 1.363 \\
                        & 60       & 327 & 245 & 74.9 & 1.308 \\
                        &  1       &  81 &  59 & 72.8 & 1.230 \\ \hline
\end{tabular}\\
\vspace*{2.0cm}\\
\textbf{Table (7):} The accuracy of the pn-QRPA model compared to experimental data ($\beta^{-}$ decay). $N$ denotes the number of experimentally known half-lives shorter than the limit in the second column, $n$ is the number (and percentage) of isotopes reproduced under the condition given in the first column, and $\bar{x}$ is the average deviation defined in the text (see \cite{Sta90a} for details).\\
\vspace*{0.1cm}\\
\begin{tabular}{cccccc} \hline \\
Conditions & $T_{1/2}^{exp}(s) \leq$ & $N$ & $n$ & $n(\%)$ & $\bar{x}$\\ \\ \hline
$\forall x_{i} \leq$ 10 & 10$^{6}$ & 654 & 472 & 72.2 & 1.85 $\pm$ 1.21 \\
                        & 60       & 325 & 313 & 96.3 & 1.67 $\pm$ 1.02 \\
                        &  1       & 106 & 105 & 99.1 & 1.44 $\pm$ 0.40 \\ \hline
$\forall x_{i} \leq$  5 & 10$^{6}$ & 654 & 456 & 69.7 & 1.68 $\pm$ 0.76 \\
                        & 60       & 325 & 307 & 94.5 & 1.56 $\pm$ 0.66 \\
                        &  1       & 106 & 105 & 99.1 & 1.44 $\pm$ 0.40 \\ \hline
$\forall x_{i} \leq$  3 & 10$^{6}$ & 654 & 420 & 64.2 & 1.50 $\pm$ 0.46 \\
                        & 60       & 325 & 295 & 90.8 & 1.46 $\pm$ 0.43 \\
                        &  1       & 106 & 105 & 99.1 & 1.44 $\pm$ 0.40 \\ \hline
$\forall x_{i} \leq$  2 & 10$^{6}$ & 654 & 369 & 56.4 & 1.37 $\pm$ 0.29 \\
                        & 60       & 325 & 267 & 82.2 & 1.36 $\pm$ 0.29 \\
                        &  1       & 106 &  96 & 90.6 & 1.35 $\pm$ 0.27 \\ \hline
\end{tabular}\\
\clearpage
\noindent
\begin{center}
\textbf{Predictive power of the pn-QRPA theory}
\end{center}
\vspace*{0.5cm}
\textbf{Table (8):} Comparison of calculated $\beta^{+}/(EC)$ decay half-lives using the pn-QRPA theory and Q-values from \cite{Wap88}, with the measured half-lives reported \textit{after} the calculations done in \cite{Hir93}. All half-lives are given in seconds. Experimental half-lives are taken from [43-75]\\
\vspace*{0.1cm}\\
\begin{tabular}{cccccc} \hline \\
$A$ & Element & $T_{1/2}^{exp}$ & $T_{1/2}^{cal}$ & $T_{1/2}^{exp}/T_{1/2}^{cal}$ & Comments \\ \\ \hline
 79 & Y  & 14.800 $\pm$ 6.000 & 10.264 & 0.69 & \\
 86 & Mo & 19.600 $\pm$ 1.100 & 76.433 & 3.90 & \\
 88 & Tc &  6.400 $\pm$ 0.800 &  1.733 & 0.27 & \\
 90 & Ru &  9.000 $\pm$ 1.000 & 29.122 & 3.24 & \\
 
 91 & Ru &  7.600 $\pm$ 0.800 &  5.459 & 0.72 & \\
100 & In &  6.100 $\pm$ 0.900 &  2.133 & 0.35 & \\
104 & Sb &  0.520 $\pm$ 0.180 &  1.489 & 2.86 & \\
105 & Sb &  1.300 $\pm$ 0.200 &  1.679 & 1.29 & \\

109 & Te &  4.600 $\pm$ 0.300 &  2.957 & 0.64 & \\ 
110 & Te & 18.600 $\pm$ 0.800 & 28.609 & 1.54 & \\
108 &  I &  0.036 $\pm$ 0.006 &  0.402 & 11.1 & partly alpha emission\\
110 &  I &  0.650 $\pm$ 0.020 &  0.896 & 1.38 & \\ 

111 &  I &  2.500 $\pm$ 0.200 &  2.977 & 1.19 & \\ 
112 &  I &  3.420 $\pm$ 0.110 &  5.065 & 1.48 & \\
113 &  I &  6.600 $\pm$ 0.200 & 10.611 & 1.61 & \\
111 & Xe &  0.740 $\pm$ 0.200 &  0.618 & 0.84 & partly alpha emission\\ 

112 & Xe &  2.700 $\pm$ 0.800 &  3.052 & 1.13 & \\ 
113 & Xe &  2.740 $\pm$ 0.080 &  2.809 & 1.03 & \\
118 & Ba &  5.200 $\pm$ 0.200 &  9.447 & 1.82 & \\
146 & Er &  1.700 $\pm$ 0.600 &  3.032 & 1.78 & \\ 

150 & Lu &  0.035 $\pm$ 0.010 &  0.138 & 3.96 & partly proton emission\\ 
161 &  W &  0.410 $\pm$ 0.040 &  2.255 & 5.50 & partly alpha emission\\
162 &  W &  1.390 $\pm$ 0.400 &  4.325 & 3.11 & partly alpha emission\\
163 &  W &  2.750 $\pm$ 0.250 &  3.897 & 1.42 & partly alpha emission\\ 

164 &  W &  6.400 $\pm$ 0.800 & 10.273 & 1.61 & \\ 
165 &  W &  5.100 $\pm$ 0.500 &  8.748 & 1.72 & \\
166 &  W & 18.800 $\pm$ 0.400 & 21.117 & 1.12 & \\
167 &  W & 19.900 $\pm$ 0.500 & 24.663 & 1.24 & partly alpha emission\\ 

162 & Re &  0.100 $\pm$ 0.300 &  0.514 & 5.14 & partly alpha emission\\ 
163 & Re &  0.260 $\pm$ 0.040 &  1.132 & 4.35 & partly alpha emission\\
164 & Re &  0.880 $\pm$ 0.240 &  0.867 & 0.99 & partly alpha emission\\
165 & Re &  2.400 $\pm$ 0.600 &  2.018 & 0.84 & \\ 

166 & Re &  2.800 $\pm$ 0.300 &  1.998 & 0.71 & partly alpha emission\\ 
167 & Re &  3.400 $\pm$ 0.400 &  3.693 & 1.09 & partly alpha emission\\
168 & Re &  4.400 $\pm$ 0.100 &  3.922 & 0.89 & \\
169 & Re &  8.100 $\pm$ 0.500 & 10.369 & 1.28 & \\ 

169 & Ir &  0.400 $\pm$ 0.100 &  1.523 & 3.81 & partly alpha emission\\ 
170 & Ir &  1.050 $\pm$ 0.150 &  0.417 & 0.40 & partly alpha emission\\
171 & Ir &  1.500 $\pm$ 0.100 &  2.436 & 1.62 & partly alpha emission\\
172 & Ir &  4.400 $\pm$ 0.300 &  2.903 & 0.66 & \\ 
\end{tabular}
\clearpage
\noindent
\begin{center}
\textbf{Predictive power of the pn-QRPA theory}
\end{center}
\textbf{Table (8)\textit{(contd.)}:} \\
\vspace*{0.1cm}\\
\begin{tabular}{cccccc} \hline \\
$A$ & Element & $T_{1/2}^{exp}$ & $T_{1/2}^{cal}$ & $T_{1/2}^{exp}/T_{1/2}^{cal}$ & Comments \\ \\ \hline
174 & Ir &  9.000 $\pm$ 2.000 &  5.546 & 0.62 & \\ 
175 & Ir &  9.000 $\pm$ 2.000 & 10.652 & 1.18 & \\
176 & Ir &  8.000 $\pm$ 1.000 & 16.729 & 2.09 & \\
177 & Ir & 30.000 $\pm$ 2.000 & 42.039 & 1.40 & \\ 

174 & Au &  0.120 $\pm$ 0.020 &  0.503 & 4.19 & partly alpha emission\\ 
175 & Au &  0.200 $\pm$ 0.022 &  1.079 & 5.39 & partly alpha emission\\
176 & Au &  1.250 $\pm$ 0.300 &  0.881 & 0.70 & partly alpha emission\\ 
177 & Au &  1.180 $\pm$ 0.070 &  2.302 & 1.95 & partly alpha emission\\ 

178 & Au &  2.600 $\pm$ 0.500 &  2.621 & 1.01 & partly alpha emission\\ 
179 & Au &  7.100 $\pm$ 0.300 &  5.732 & 0.81 & partly alpha emission\\
180 & Tl &  1.900 $\pm$ 0.900 &  0.742 & 0.39 & partly alpha emission\\
181 & Tl &  3.400 $\pm$ 0.600 &  2.331 & 0.69 & partly alpha emission\\ 

182 & Tl &  3.100 $\pm$ 1.000 &  1.810 & 0.58 & \\ 
198 & At &  4.200 $\pm$ 0.300 &  5.172 & 1.23 & partly alpha emission\\
234 & Am &139.200 $\pm$ 4.800 &163.325 & 1.17 & \\
238 & Bk &144.000 $\pm$ 5.000 & 85.205 & 0.59 & \\ 

240 & Bk &288.000 $\pm$48.000 &357.840 & 1.24 & \\ 
239 & Cf & 39.000 $\pm$37.000 & 45.802 & 1.17 & partly alpha emission\\
240 & Cf & 63.600 $\pm$ 9.000 & 56.025 & 0.88 & partly alpha emission\\
241 & Cf &226.800 $\pm$42.000 &184.563 & 0.81 & partly alpha emission\\ 

242 & Cf &209.400 $\pm$ 7.200 &311.928 & 1.49 & \\ 
242 & Es &  7.000 $\pm$ 0.000 & 15.271 & 2.18 & \\
249 & Fm &156.000 $\pm$42.000 &199.917 & 1.28 & \\
250 & Fm &1800.00 $\pm$180.00 &915.150 & 0.51 & partly alpha emission\\ 

253 & No &102.000 $\pm$18.000 & 44.769 & 0.44 & partly alpha emission\\ 
254 & Lr & 13.000 $\pm$ 2.000 & 24.733 & 1.90 & partly alpha emission\\
258 & Ha & 20.000 $\pm$10.000 & 13.857 & 0.69 & \\ \hline
\end{tabular}\\   
\clearpage
\noindent
\begin{center}
\textbf{Predictive power of the pn-QRPA for proton-rich nuclei using different mass formulae}
\end{center}
\vspace*{0.5cm}
\textbf{Table (9):} Comparison of calculated $\beta^{+}/(EC)$ decay half-lives using the pn-QRPA theory, with the measured half-lives reported \textit{after} the calculations done in \cite{Hir93}. Columns 4--6 represent the calculated half-lives using the mass formulae of \cite{Hil76}, \cite{Gro76} and \cite{Moe81}, respectively. The last three columns show the ratio of calculated half-lives (using mass formulae from \cite{Hil76}, \cite{Gro76} and \cite{Moe81}, respectively) to measured ones. All half-lives are given in seconds. Experimental half-lives are taken from [43-75]\\
\vspace*{0.1cm}\\
\begin{tabular}{ccccccccc} \hline \\
$A$ & Element & $T_{1/2}^{exp}$ & $T_{1/2}^{cal}$ & $T_{1/2}^{cal}$  & $T_{1/2}^{cal}$  & $T_{1/2}^{cal}/T_{1/2}^{exp}$ & $T_{1/2}^{cal}/T_{1/2}^{exp}$ & $T_{1/2}^{cal}/T_{1/2}^{exp}$ \\ \\ 
& & & (Hilf) & (Groote) & (M\"oller) & (Hilf) & (Groote) & (M\"oller)\\ \\ \hline
 22 & Si & 0.006 $\pm$ 0.003 & 0.099 & 0.081 & 0.068 & 16.5 & 13.54 & 11.2\\
 43 & Cr & 0.021 $\pm$ 0.004 & 0.057 & 0.047 & 0.023 & 2.73 &  2.26 & 1.08\\
 44 & Cr & 0.053 $\pm$ 0.004 & 0.083 & 0.068 & 0.041 & 1.57 &  1.28 & 0.77\\
 46 & Mn & 0.041 $\pm$ 0.007 & 0.021 & 0.114 & 0.000 & 0.50 &  2.79 & 0.00\\
 46 & Fe & 0.020 $\pm$ 0.020 & 0.031 & 0.024 & 0.015 & 1.53 &  1.21 & 0.77\\
 47 & Fe & 0.027 $\pm$ 0.032 & 0.057 & 0.050 & 0.022 & 2.12 &  1.87 & 0.83\\
 50 & Fe & 0.150 $\pm$ 0.030 & 0.125 & 0.214 & 0.000 & 0.83 &  1.43 & 0.00\\
 52 & Ni & 0.038 $\pm$ 0.005 & 0.026 & 0.073 & 0.000 & 0.67 &  1.93 & 0.00\\
 61 & Ga & 0.150 $\pm$ 0.030 & 0.227 & 0.298 & 0.000 & 1.51 &  1.99 & 0.00\\
 61 & Ge & 0.040 $\pm$ 0.015 & 0.056 & 0.054 & 0.044 & 1.39 &  1.36 & 1.11\\
 62 & Ge & 0.110 $\pm$ 0.060 & 0.086 & 0.082 & 0.083 & 0.78 &  0.75 & 0.76\\
 94 & Ag & 0.010 $\pm$ 0.000 & 0.023 & 0.023 & 0.020 & 2.33 &  2.30 & 2.01\\
 95 & Ag & 2.000 $\pm$ 0.100 & 0.842 & 0.654 & 0.962 & 0.42 &  0.33 & 0.48\\
100 & Sn & 1.000 $\pm$ 0.800 & 2.018 & 1.151 & 0.882 & 2.02 &  1.15 & 0.88\\
101 & Sn & 3.000 $\pm$ 1.000 & 9.765 & 6.049 & 4.417 & 3.26 &  2.02 & 1.47\\
114 & Ba & 0.400 $\pm$ 0.300 & 0.457 & 0.510 & 0.292 & 1.14 &  1.27 & 0.73\\
115 & Ba & 0.400 $\pm$ 0.200 & 0.835 & 0.675 & 0.424 & 2.09 &  1.69 & 1.06\\
116 & Ba & 1.350 $\pm$ 0.150 & 1.613 & 1.649 & 1.033 & 1.19 &  1.22 & 0.76\\
122 & Ce & 8.700 $\pm$ 0.700 & 3.660 & 2.934 & 2.101 & 0.42 &  0.34 & 0.24\\
121 & Pr & 1.400 $\pm$ 0.800 & 0.784 & 0.813 & 0.560 & 0.56 &  0.58 & 0.40\\
127 & Pr &15.100 $\pm$ 0.000 & 6.291 &15.514 & 9.269 & 0.42 &  1.03 & 0.61\\
132 & Sm & 4.000 $\pm$ 0.300 & 2.736 & 2.533 & 1.477 & 0.68 &  0.63 & 0.37\\ \hline
\end{tabular}
\clearpage
\noindent
\begin{center}
\textbf{Predictive power of the pn-QRPA for neutron-rich nuclei using different mass formulae}
\end{center}
\vspace*{0.5cm}
\textbf{Table (10):} Comparison of calculated $\beta^{-}$ decay half-lives using the pn-QRPA theory, with the measured half-lives reported \textit{after} the calculations done in \cite{Sta90a}. Columns 4--6 represent the calculated half-lives using the mass formulae of \cite{Hil76}, \cite{Gro76} and \cite{Moe81}, respectively. The last three columns show the ratio of calculated half-lives (using mass formulae from \cite{Hil76}, \cite{Gro76} and \cite{Moe81}, respectively) to measured ones. All half-lives are given in seconds. Experimental half-lives are taken from [43-75]\\
\vspace*{0.1cm}\\
\begin{tabular}{ccccccccc} \hline \\
$A$ & Element & $T_{1/2}^{exp}$ & $T_{1/2}^{cal}$ & $T_{1/2}^{cal}$  & $T_{1/2}^{cal}$  & $T_{1/2}^{cal}/T_{1/2}^{exp}$ & $T_{1/2}^{cal}/T_{1/2}^{exp}$ & $T_{1/2}^{cal}/T_{1/2}^{exp}$ \\ \\ 
& & & (Hilf) & (Groote) & (M\"oller) & (Hilf) & (Groote) & (M\"oller)\\ \\ \hline
 21 &  N & 0.095 $\pm$ 0.013 & 0.201 & 0.201 & 0.000 & 2.12 &  2.12 & 0.00\\
 22 &  N & 0.024 $\pm$ 0.007 & 0.050 & 0.050 & 0.000 & 2.07 &  2.08 & 0.00\\
 23 &  O & 0.082 $\pm$ 0.037 & 0.023 & 0.023 & 0.013 & 0.28 &  0.28 & 0.16\\
 24 &  O & 0.061 $\pm$ 0.026 & 0.012 & 0.011 & 0.007 & 0.20 &  0.17 & 0.12\\
 27 & Ne & 0.032 $\pm$ 0.002 & 0.046 & 0.046 & 0.025 & 1.45 &  1.45 & 0.79\\
 28 & Ne & 0.014 $\pm$ 0.010 & 0.034 & 0.026 & 0.032 & 2.43 &  1.85 & 2.26\\
 43 &  P & 0.033 $\pm$ 0.003 & 0.083 & 0.115 & 0.034 & 2.50 &  3.48 & 1.02\\
 42 &  S & 0.560 $\pm$ 0.060 & 4.460 & 4.530 & 1.350 & 7.96 &  8.09 & 2.41\\
 45 &  S & 0.082 $\pm$ 0.013 & 0.299 & 0.397 & 0.105 & 3.65 &  4.84 & 1.28\\
 44 & Cl & 0.430 $\pm$ 0.060 & 3.190 & 5.020 & 0.660 & 7.42 & 11.67 & 1.53\\
 45 & Cl & 0.400 $\pm$ 0.043 & 1.670 & 1.670 & 0.391 & 4.17 &  4.17 & 0.98\\
 46 & Cl & 0.220 $\pm$ 0.040 & 0.818 & 0.848 & 0.174 & 3.72 &  3.85 & 0.79\\
 65 & Fe & 0.450 $\pm$ 0.150 & 0.335 & 0.335 & 0.210 & 0.74 &  0.74 & 0.47\\
 68 & Fe & 0.100 $\pm$ 0.060 & 0.160 & 0.152 & 0.209 & 1.60 &  1.52 & 2.09\\
 68 & Co & 0.180 $\pm$ 0.100 & 0.289 & 0.289 & 0.116 & 1.61 &  1.61 & 0.64\\
 69 & Co & 0.270 $\pm$ 0.050 & 0.071 & 0.074 & 0.082 & 0.26 &  0.27 & 0.30\\
 71 & Co & 0.200 $\pm$ 0.050 & 0.051 & 0.059 & 0.043 & 0.25 &  0.29 & 0.21\\
 71 & Ni & 1.860 $\pm$ 0.350 & 5.440 & 5.860 & 2.850 & 2.92 &  3.15 & 1.53\\
 72 & Ni & 2.060 $\pm$ 0.300 & 1.560 & 1.680 & 3.450 & 0.76 &  0.82 & 1.67\\
 73 & Ni & 0.900 $\pm$ 0.150 & 0.806 & 1.040 & 0.627 & 0.90 &  1.16 & 0.70\\
 74 & Ni & 1.100 $\pm$ 0.500 & 0.462 & 0.543 & 0.563 & 0.42 &  0.49 & 0.51\\
 77 & Cu & 0.469 $\pm$ 0.008 & 0.730 & 1.080 & 0.448 & 1.56 &  2.30 & 0.96\\
 78 & Cu & 0.342 $\pm$ 0.008 & 0.361 & 0.653 & 0.261 & 1.06 &  1.91 & 0.76\\
 79 & Cu & 0.188 $\pm$ 0.011 & 0.390 & 0.608 & 0.284 & 2.07 &  3.23 & 1.51\\
 81 & Zn & 0.290 $\pm$ 0.050 & 0.336 & 0.665 & 0.055 & 1.16 &  2.29 & 0.19\\
 84 & Ga & 0.085 $\pm$ 0.010 & 0.194 & 0.406 & 0.143 & 2.28 &  4.78 & 1.68\\
 85 & Ge & 0.540 $\pm$ 0.050 & 0.355 & 0.761 & 0.131 & 0.66 &  1.41 & 0.24\\
103 &  Y & 0.230 $\pm$ 0.030 & 0.162 & 0.149 & 0.362 & 0.70 &  0.65 & 1.57\\
105 & Zr & 1.000 $\pm$ 0.000 & 0.820 & 0.654 & 0.241 & 0.82 &  0.65 & 0.24\\
107 & Nb & 0.330 $\pm$ 0.050 & 0.602 & 0.971 & 0.271 & 1.82 &  2.94 & 0.82\\
\end{tabular}
\clearpage
\begin{center}
\textbf{Predictive power of the pn-QRPA for neutron-rich nuclei using different mass formulae}
\end{center}
\textbf{Table (10)\textit{(contd.)}:} \\
\vspace*{0.1cm}\\
\begin{tabular}{ccccccccc} \hline \\
$A$ & Element & $T_{1/2}^{exp}$ & $T_{1/2}^{cal}$ & $T_{1/2}^{cal}$  & $T_{1/2}^{cal}$  & $T_{1/2}^{cal}/T_{1/2}^{exp}$ & $T_{1/2}^{cal}/T_{1/2}^{exp}$ & $T_{1/2}^{cal}/T_{1/2}^{exp}$ \\ \\ 
& & & (Hilf) & (Groote) & (M\"oller) & (Hilf) & (Groote) & (M\"oller)\\ \\ \hline
108 & Nb & 0.170 $\pm$ 0.020 & 0.382 & 0.688 & 0.187 & 2.25 &  4.05 & 1.10\\
109 & Nb & 0.190 $\pm$ 0.030 & 0.542 & 0.520 & 0.129 & 2.85 &  2.74 & 0.68\\
110 & Nb & 0.170 $\pm$ 0.020 & 0.284 & 0.116 & 0.097 & 1.67 &  0.68 & 0.57\\
109 & Mo & 0.530 $\pm$ 0.060 & 0.407 & 0.802 & 0.379 & 0.77 &  1.51 & 0.72\\
110 & Mo & 0.300 $\pm$ 0.040 & 0.675 & 0.593 & 0.276 & 2.25 &  1.98 & 0.92\\
113 & Tc & 0.130 $\pm$ 0.050 & 0.244 & 0.281 & 0.095 & 1.88 &  2.16 & 0.73\\  
115 & Ru & 0.400 $\pm$ 0.100 & 0.434 & 0.443 & 0.240 & 1.08 &  1.11 & 0.60\\
117 & Rh & 0.440 $\pm$ 0.040 & 0.408 & 0.210 & 0.226 & 0.93 &  0.48 & 0.51\\
119 & Pd & 0.920 $\pm$ 0.130 & 1.200 & 3.160 & 0.577 & 1.30 &  3.43 & 0.63\\
120 & Pd & 0.500 $\pm$ 0.100 & 0.695 & 1.660 & 0.232 & 1.39 &  3.32 & 0.46\\  
125 & Ag & 0.156 $\pm$ 0.007 & 0.164 & 0.249 & 0.117 & 1.05 &  1.60 & 0.75\\
126 & Ag & 0.097 $\pm$ 0.008 & 0.140 & 0.194 & 0.090 & 1.44 &  2.00 & 0.93\\
127 & Ag & 0.109 $\pm$ 0.015 & 0.092 & 0.151 & 0.089 & 0.84 &  1.39 & 0.81\\
149 & Ba & 0.344 $\pm$ 0.007 & 0.287 & 0.345 & 0.365 & 0.83 &  1.00 & 1.06\\  
150 & La & 0.860 $\pm$ 0.050 & 0.771 & 1.460 & 0.405 & 0.90 &  1.70 & 0.47\\
185 & Hf &210.000 $\pm$36.000 &584.000 &766.000 &204.000 & 2.78 &  3.65 & 0.97\\
205 & Au &31.000 $\pm$ 2.000 & 35.700 & 50.000 &25.100 & 1.15 &  1.61 & 0.81\\
216 & Bi &216.000 $\pm$24.000 &189.000 & 191.000 &223.000 & 0.88 &  0.88 & 1.03\\  
221 & At & 138.000 $\pm$ 12.000 & 256.000 & 462.000 & 564.000 & 1.86 &  3.35 & 4.09\\
222 & At &54.000 $\pm$ 10.000 &63.300 &155.000 &10.800 & 1.17 &  2.87 & 0.20\\
223 & At &50.000 $\pm$  7.000 &41.600 & 84.100 &70.700 & 0.83 &  1.68 & 1.41\\
232 & Fr & 5.000 $\pm$ 1.000 & 1.690 & 15.900 & 0.903 & 0.34 &  3.18 & 0.18\\  
233 & Ra &30.000 $\pm$  5.000 & 86.000 & 622.000 & 52.000 & 2.87 &  20.73 & 1.73\\
234 & Ra &30.000 $\pm$10.000 & 34.300 & 127.000 & 42.000 & 1.14 &  4.23 & 1.40\\
237 & Th & 300.000 $\pm$ 54.000 & 221.000 & 389.000 & 311.000 & 0.74 &  1.30 & 1.04\\ \hline
\end{tabular}
\clearpage
\noindent
\textbf{Table (11):} Comparison of calculated $B(GT)^{+}$ strength function using the pn-QRPA theory (this work), with the measured ones from $(n,p)$ reactions. QRPA and SM represent calculations using the pn-QRPA theory and those from the recent large-scale shell model calculations \cite{Cau99}, respectively. The experimental data are taken from [79-83].
\vspace*{0.1cm}\\
\begin{tabular}{cccc} \hline \\
Nucleus & Measured Strength & QRPA & SM \\ \\ \hline
 $^{51}$V  & 1.2 $\pm$ 0.1 & 1.4 & 1.4\\
 $^{54}$Fe & 3.5 $\pm$ 0.3 & 3.8 & 3.6\\
 $^{55}$Mn & 1.7 $\pm$ 0.2 & 2.4 & 2.2\\
 $^{56}$Fe & 2.9 $\pm$ 0.3 & 2.7 & 2.7\\
 $^{58}$Ni & 3.8 $\pm$ 0.4 & 3.3 & 4.4\\
 $^{59}$Co & 1.9 $\pm$ 0.1 & 0.8 & 2.5\\
 $^{60}$Ni & 3.1 $\pm$ 0.1 & 7.0 & 3.4\\
 $^{62}$Ni & 2.5 $\pm$ 0.1 & 4.4 & 2.1\\
 $^{64}$Ni & 1.7 $\pm$ 0.1 & 3.6 & 1.3\\
 $^{70}$Ge & 0.7 $\pm$ 0.1 & 0.3 & -\\
 $^{72}$Ge & 0.2 $\pm$ 0.1 & 0.1 & -\\ \hline
\end{tabular}
\vspace*{2.3cm}\\
\textbf{Table (12):} Comparison of calculated $B(GT)^{-}$ strength function using the pn-QRPA theory (this work), with the measured ones from $(p,n)$ reactions.  QRPA and SM represent calculations using the pn-QRPA theory and those from the recent large-scale shell model calculations \cite{Cau99}, respectively. The experimental data are taken from [84-87]. See \cite{And87} regarding the uncertainty in measured \gt strength value of $^{32}$S.
\vspace*{0.1cm}\\
\begin{tabular}{cccc} \hline \\
Nucleus & Measured Strength & QRPA & SM \\ \\ \hline
 $^{32}$S  & 2.21  & 1.1 & - \\
 $^{51}$V  & 12.6 $\pm$ 2.5 & 16.6 & - \\
 $^{54}$Fe & 7.8 $\pm$ 1.9 & 5.9 & 6.9 \\
 $^{58}$Ni & 7.4 $\pm$ 1.8 & 7.5 & 7.7 \\
 $^{60}$Ni & 7.2 $\pm$ 1.8 & 7.7 & 10.0 \\
 $^{71}$Ga & 4.3 $\pm$ 0.7 & 5.6 & - \\ \hline
\end{tabular}
\clearpage
\normalsize
\begin{center}
\textbf{\Large EXPLANATION OF TABLE}
\end{center}
\vspace{0.7in}
\textbf{\large TABLE A. $\mathbf{fp}$-Shell Nuclei Weak Rates in Stellar Matter}
\vspace{0.4in}
\noindent
\newline
The calculated weak interaction rates [Eqs. (16) and (19)-(21)] are all tabulated in log$_{10} \lambda$. The probabilities of $\beta$-delayed proton (neutron) emission [Eq.~(22)] are also tabulated in logarithmic scale. These probabilities are calculated only to one significant figure and are given up to three places of decimal only for designing purposes. All rates listed for a particular direction concern the parent nucleus except for the last two columns which concern the daughter nucleus. For each daughter nucleus, either the proton energy rate and probability of $\beta$-delayed proton emission is stated (if $S_{p} < S_{n}$) or the neutron energy rate and probability of $\beta$-delayed neutron emission is stated (if $S_{n} < S_{p}$). In the table -100 means that the rate (or the probability) is smaller than $10^{-100}$. \\
\begin {tabbing}
\textsf{Q} \hspace{0.7in}        \= Mass of parent minus mass of daughter nucleus\\
S$_{\mathsf{nP}}$                       \> Separation energy of neutron for parent nucleus\\
S$_{\mathsf{pP}}$                       \> Separation energy of proton for parent nucleus\\
S$_{\mathsf{nD}}$                       \> Separation energy of neutron for daughter nucleus\\
S$_{\mathsf{pD}}$                     \> Separation energy of neutron for daughter nucleus\\
\textsf{ADen}                    \> log $(\rho Y_{e})$ (g.cm$^{-3}$), where $\rho$ is the density of the baryon, and $Y_{e}$ is the ratio of the\\
                        \> electron number to the baryon number\\
\textsf{T9}                      \> Temperature in units of $10^{9}$ K\\
\textsf{EFermi}                  \> Total Fermi energy of electron and positron, including the rest mass (MeV)\\
\textsf{E+Cap}                   \> Positron capture rate (s$^{-1}$)\\
\textsf{E-Dec}                   \> Electron decay rate (s$^{-1}$)\\ 
\textsf{ANuEn}                   \> Anti-Neutrino energy loss rate (MeV.s$^{-1}$)\\
\textsf{GamEn}                 \> Gamma ray heating rate (MeV.s$^{-1}$), tabulated separately for $\beta^{+}$ direction \\
                        \> and $\beta^{-}$ direction \\
\textsf{E-Cap}                   \> Electron capture rate (s$^{-1}$)\\
\textsf{E+Dec}                   \> Positron decay rate (s$^{-1}$)\\
\textsf{NuEn}                 \> Neutrino energy loss rate (MeV.s$^{-1}$)\\     
\textsf{ProEn}                  \> Energy rate of $\beta$-delayed proton (MeV.s$^{-1}$)\\
\textsf{NeuEn}                  \> Energy rate of $\beta$-delayed neutron (MeV.s$^{-1}$)\\
\textsf{PPEm}                  \> Probability of $\beta$-delayed proton emission\\
\textsf{PNEm}                  \> Probability of $\beta$-delayed neutron emission\\             
\end{tabbing}
\clearpage
\sffamily \normalsize
\noindent
\begin{tabular}{rr|rrrrrr}
\multicolumn{7}{c}{\normalsize{\hspace*{3.5cm}$^{  40}_{20}$ CA $_{ 20} \rightarrow \hspace*{0.2cm}^{  40}_{21}$ SC $_{ 19}$}}\\
\hline
Aden& T9&  E+Cap &   E-Dec &  ANuEn &  GamEn &  ProEn &  PPEm \\
\hline
     3.0&  1.0 & -76.256 &    -100 & -76.205 & -80.721 & -85.961 & -9.000 \\
     3.0&  3.0 & -25.504 &    -100 & -25.348 & -26.538 & -28.464 & -3.000 \\
     3.0& 30.0 &   0.278 &    -100 &   1.277 &  -0.468 &   0.766 & -0.100 \\
     7.0&  1.0 & -82.076 &    -100 & -82.025 & -86.541 & -91.781 & -9.000 \\
     7.0&  3.0 & -27.220 &    -100 & -27.064 & -28.254 & -30.180 & -3.000 \\
     7.0& 30.0 &   0.275 &    -100 &   1.274 &  -0.471 &   0.763 & -0.100 \\
    11.0&  1.0 &    -100 &    -100 &    -100 &    -100 &    -100 &-100\\
    11.0&  3.0 & -65.697 &    -100 & -65.541 & -66.731 & -68.657 & -3.000 \\
    11.0& 30.0 &  -3.588 &    -100 &  -2.589 &  -4.334 &  -3.100 & -0.100 \\
\hline
\end{tabular}\\
\vspace{0.20cm}\\
\begin{tabular}{rr|rrrrrr}
\multicolumn{7}{c}{\normalsize{\hspace*{3.5cm}$^{  40}_{20}$ CA $_{ 20} \rightarrow \hspace*{0.2cm}^{  40}_{19}$ K  $_{ 21}$}}\\
\hline
Aden& T9&  E-Cap &   E+Dec &  NuEn &  GamEn &  ProEn &  PPEm \\
\hline
     3.0&  1.0 & -25.483 & -25.143 & -25.002 & -25.103 & -56.523 &-30.000 \\
     3.0&  3.0 &  -9.824 &  -9.645 &  -9.280 &  -9.484 & -19.330 & -9.000 \\
     3.0& 30.0 &   1.905 &  -2.098 &   3.004 &   2.485 &   1.260 & -0.800 \\
     7.0&  1.0 & -19.663 & -25.143 & -20.234 & -19.283 & -50.703 &-30.000 \\
     7.0&  3.0 &  -8.162 &  -9.640 &  -8.021 &  -7.840 & -17.615 & -9.000 \\
     7.0& 30.0 &   1.909 &  -2.097 &   3.007 &   2.489 &   1.263 & -0.800 \\
    11.0&  1.0 &   3.650 & -25.143 &   4.781 &   4.182 &   2.745 & -0.700 \\
    11.0&  3.0 &   3.651 &  -9.640 &   4.783 &   4.183 &   2.747 & -0.700 \\
    11.0& 30.0 &   4.793 &  -1.905 &   6.112 &   5.373 &   4.458 & -0.600 \\
\hline
\end{tabular}\\
\vspace{0.20cm}\\
\noindent
\begin{tabular}{rr|rrrrrr}
\multicolumn{7}{c}{\normalsize{\hspace*{3.5cm}$^{  40}_{21}$ SC $_{ 19} \rightarrow \hspace*{0.2cm}^{  40}_{22}$ TI $_{ 18}$}}\\
\hline
Aden& T9&  E+Cap &   E-Dec &  ANuEn &  GamEn &  ProEn &  PPEm \\
\hline
     3.0&  1.0 & -81.245 &    -100 & -81.194 &    -100 & -81.298 &  0.000 \\
     3.0&  3.0 & -26.642 &    -100 & -26.490 &    -100 & -26.661 &  0.000 \\
     3.0& 30.0 &   0.015 &    -100 &   0.994 &    -100 &   0.390 &  0.000 \\
     7.0&  1.0 & -87.065 &    -100 & -87.014 &    -100 & -87.118 &  0.000 \\
     7.0&  3.0 & -28.357 &    -100 & -28.206 &    -100 & -28.377 &  0.000 \\
     7.0& 30.0 &   0.011 &    -100 &   0.991 &    -100 &   0.387 &  0.000 \\
    11.0&  1.0 &    -100 &    -100 &    -100 &    -100 &    -100 &-100\\
    11.0&  3.0 & -66.834 &    -100 & -66.683 &    -100 & -66.854 &  0.000 \\
    11.0& 30.0 &  -3.852 &    -100 &  -2.872 &    -100 &  -3.476 &  0.000 \\
\hline
\end{tabular}\\
\vspace{0.20cm}\\
\begin{tabular}{rr|rrrrrr}
\multicolumn{7}{c}{\normalsize{\hspace*{3.5cm}$^{  40}_{21}$ SC $_{ 19} \rightarrow \hspace*{0.2cm}^{  40}_{20}$ CA $_{ 20}$}}\\
\hline
Aden& T9&  E-Cap &   E+Dec &  NuEn &  GamEn &  ProEn &  PPEm \\
\hline
     3.0&  1.0 &  -3.236 &   1.696 &   2.374 &   2.320 &  -2.292 & -3.000 \\
     3.0&  3.0 &  -0.961 &   1.664 &   2.344 &   2.307 &  -1.816 & -3.000 \\
     3.0& 30.0 &   2.812 &   1.673 &   4.094 &   3.516 &   2.549 & -0.900 \\
     7.0&  1.0 &   0.474 &   1.696 &   2.427 &   2.347 &  -1.189 & -3.000 \\
     7.0&  3.0 &   0.455 &   1.665 &   2.397 &   2.333 &  -0.957 & -3.000 \\
     7.0& 30.0 &   2.816 &   1.673 &   4.097 &   3.520 &   2.552 & -0.900 \\
    11.0&  1.0 &   5.285 &   1.696 &   6.714 &   5.895 &   5.196 & -0.800 \\
    11.0&  3.0 &   5.262 &   1.665 &   6.690 &   5.883 &   5.221 & -0.800 \\
    11.0& 30.0 &   5.397 &   1.746 &   6.858 &   6.030 &   5.451 & -0.700 \\
\hline
\end{tabular}\\
\vspace{0.20cm}\\
\noindent
\begin{tabular}{rr|rrrrrr}
\multicolumn{7}{c}{\normalsize{\hspace*{3.5cm}$^{  40}_{22}$ TI $_{ 18} \rightarrow \hspace*{0.2cm}^{  40}_{23}$ V  $_{ 17}$}}\\
\hline
Aden& T9&  E+Cap &   E-Dec &  ANuEn &  GamEn &  ProEn &  PPEm \\
\hline
     3.0&  1.0 &    -100 &    -100 &    -100 &    -100 &    -100 &-100\\
     3.0&  3.0 & -36.419 &    -100 & -36.269 &    -100 & -36.314 &  0.000 \\
     3.0& 30.0 &  -1.620 &    -100 &  -0.661 &    -100 &  -1.005 &  0.000 \\
     7.0&  1.0 &    -100 &    -100 &    -100 &    -100 &    -100 &-100\\
     7.0&  3.0 & -38.135 &    -100 & -37.985 &    -100 & -38.029 &  0.000 \\
     7.0& 30.0 &  -1.624 &    -100 &  -0.665 &    -100 &  -1.008 &  0.000 \\
    11.0&  1.0 &    -100 &    -100 &    -100 &    -100 &    -100 &-100\\
    11.0&  3.0 & -76.612 &    -100 & -76.462 &    -100 & -76.507 &  0.000 \\
    11.0& 30.0 &  -5.487 &    -100 &  -4.528 &    -100 &  -4.871 &  0.000 \\
\hline
\end{tabular}\\
\vspace{0.20cm}\\
\begin{tabular}{rr|rrrrrr}
\multicolumn{7}{c}{\normalsize{\hspace*{3.5cm}$^{  40}_{22}$ TI $_{ 18} \rightarrow \hspace*{0.2cm}^{  40}_{21}$ SC $_{ 19}$}}\\
\hline
Aden& T9&  E-Cap &   E+Dec &  NuEn &  GamEn &  ProEn &  PPEm \\
\hline
     3.0&  1.0 &  -3.260 &   1.452 &   2.100 &   1.257 &   1.338 & -0.300 \\
     3.0&  3.0 &  -0.948 &   1.452 &   2.103 &   1.257 &   1.343 & -0.300 \\
     3.0& 30.0 &   2.851 &   1.373 &   4.087 &   2.226 &   3.420 & -0.090 \\
     7.0&  1.0 &   0.456 &   1.452 &   2.176 &   1.279 &   1.452 & -0.300 \\
     7.0&  3.0 &   0.473 &   1.452 &   2.181 &   1.279 &   1.459 & -0.300 \\
     7.0& 30.0 &   2.855 &   1.373 &   4.090 &   2.229 &   3.423 & -0.090 \\
    11.0&  1.0 &   5.435 &   1.452 &   6.827 &   4.676 &   6.087 & -0.060 \\
    11.0&  3.0 &   5.436 &   1.452 &   6.828 &   4.676 &   6.087 & -0.060 \\
    11.0& 30.0 &   5.484 &   1.452 &   6.915 &   4.711 &   6.149 & -0.060 \\
\hline
\end{tabular}\\
\vspace{0.20cm}\\
\noindent
\begin{tabular}{rr|rrrrrr}
\multicolumn{7}{c}{\normalsize{\hspace*{3.5cm}$^{  41}_{20}$ CA $_{ 21} \rightarrow \hspace*{0.2cm}^{  41}_{21}$ SC $_{ 20}$}}\\
\hline
Aden& T9&  E+Cap &   E-Dec &  ANuEn &  GamEn &  ProEn &  PPEm \\
\hline
     3.0&  1.0 & -37.297 & -43.398 & -37.245 & -41.481 & -53.133 &-20.000 \\
     3.0&  3.0 & -12.955 & -16.868 & -12.798 & -14.400 & -18.304 & -5.000 \\
     3.0& 30.0 &   0.894 &  -5.663 &   1.920 &  -0.086 &   1.667 & -0.080 \\
     7.0&  1.0 & -43.116 & -45.778 & -43.065 & -45.784 & -58.953 &-20.000 \\
     7.0&  3.0 & -14.671 & -17.505 & -14.513 & -16.100 & -20.019 & -5.000 \\
     7.0& 30.0 &   0.891 &  -5.665 &   1.916 &  -0.090 &   1.664 & -0.080 \\
    11.0&  1.0 &    -100 &    -100 &    -100 &    -100 &    -100 &-100\\
    11.0&  3.0 & -53.148 & -55.860 & -52.990 & -54.572 & -58.497 & -5.000 \\
    11.0& 30.0 &  -2.970 &  -9.168 &  -1.945 &  -3.945 &  -2.198 & -0.080 \\
\hline
\end{tabular}\\
\vspace{0.20cm}\\
\begin{tabular}{rr|rrrrrr}
\multicolumn{7}{c}{\normalsize{\hspace*{3.5cm}$^{  41}_{20}$ CA $_{ 21} \rightarrow \hspace*{0.2cm}^{  41}_{19}$ K  $_{ 22}$}}\\
\hline
Aden& T9&  E-Cap &   E+Dec &  NuEn &  GamEn &  ProEn &  PPEm \\
\hline
     3.0&  1.0 &  -7.534 & -20.309 &  -7.712 & -12.560 & -49.416 &-40.000 \\
     3.0&  3.0 &  -4.649 &  -8.688 &  -4.515 &  -6.374 & -17.128 &-10.000 \\
     3.0& 30.0 &   1.867 &  -1.706 &   2.934 &   2.325 &   2.072 & -0.300 \\
     7.0&  1.0 &  -3.438 & -20.309 &  -3.405 &  -8.473 & -43.596 &-40.000 \\
     7.0&  3.0 &  -3.097 &  -8.687 &  -2.907 &  -4.822 & -15.415 &-10.000 \\
     7.0& 30.0 &   1.870 &  -1.705 &   2.937 &   2.328 &   2.075 & -0.300 \\
    11.0&  1.0 &   4.374 & -20.309 &   5.557 &   5.034 &   3.455 & -1.000 \\
    11.0&  3.0 &   4.378 &  -8.687 &   5.563 &   5.033 &   3.478 & -1.000 \\
    11.0& 30.0 &   4.873 &  -1.569 &   6.150 &   5.221 &   5.239 & -0.200 \\
\hline
\end{tabular}\\
\vspace{0.20cm}\\
\noindent
\begin{tabular}{rr|rrrrrr}
\multicolumn{7}{c}{\normalsize{\hspace*{3.5cm}$^{  41}_{21}$ SC $_{ 20} \rightarrow \hspace*{0.2cm}^{  41}_{22}$ TI $_{ 19}$}}\\
\hline
Aden& T9&  E+Cap &   E-Dec &  ANuEn &  GamEn &  ProEn &  PPEm \\
\hline
     3.0&  1.0 & -69.473 &    -100 & -69.423 & -75.817 & -93.550 &-20.000 \\
     3.0&  3.0 & -23.545 &    -100 & -23.393 & -26.513 & -30.873 & -8.000 \\
     3.0& 30.0 &  -0.755 &    -100 &   0.226 &  -2.498 &  -0.581 & -0.200 \\
     7.0&  1.0 & -75.293 &    -100 & -75.242 & -81.637 & -99.369 &-20.000 \\
     7.0&  3.0 & -25.261 &    -100 & -25.109 & -28.228 & -32.588 & -8.000 \\
     7.0& 30.0 &  -0.758 &    -100 &   0.223 &  -2.502 &  -0.584 & -0.200 \\
    11.0&  1.0 &    -100 &    -100 &    -100 &    -100 &    -100 &-100\\
    11.0&  3.0 & -63.738 &    -100 & -63.586 & -66.706 & -71.066 & -8.000 \\
    11.0& 30.0 &  -4.621 &    -100 &  -3.640 &  -6.364 &  -4.447 & -0.200 \\
\hline
\end{tabular}\\
\vspace{0.20cm}\\
\begin{tabular}{rr|rrrrrr}
\multicolumn{7}{c}{\normalsize{\hspace*{3.5cm}$^{  41}_{21}$ SC $_{ 20} \rightarrow \hspace*{0.2cm}^{  41}_{20}$ CA $_{ 21}$}}\\
\hline
Aden & T9& E-Cap &   E+Dec &  NuEn &  GamEn &  NeuEn &  PNEm \\
\hline
     3.0&  1.0 &  -4.416 &  -0.112 &   0.355 &  -3.080 & -22.675 &-20.000 \\
     3.0&  3.0 &  -2.163 &  -0.181 &   0.296 &  -1.415 &  -9.162 & -8.000 \\
     3.0& 30.0 &   1.938 &  -0.192 &   3.116 &   2.300 &   1.416 & -0.900 \\
     7.0&  1.0 &  -0.691 &  -0.112 &   0.566 &  -2.071 & -16.855 &-20.000 \\
     7.0&  3.0 &  -0.740 &  -0.180 &   0.512 &  -1.241 &  -7.447 & -7.000 \\
     7.0& 30.0 &   1.942 &  -0.191 &   3.120 &   2.303 &   1.419 & -0.900 \\
    11.0&  1.0 &   4.445 &  -0.112 &   5.786 &   3.768 &   4.316 & -0.500 \\
    11.0&  3.0 &   4.416 &  -0.180 &   5.748 &   4.097 &   4.329 & -0.500 \\
    11.0& 30.0 &   4.694 &  -0.071 &   6.065 &   5.101 &   4.578 & -0.600 \\
\hline
\end{tabular}\\
\vspace{0.20cm}\\
\noindent
\begin{tabular}{rr|rrrrrr}
\multicolumn{7}{c}{\normalsize{\hspace*{3.5cm}$^{  41}_{22}$ TI $_{ 19} \rightarrow \hspace*{0.2cm}^{  41}_{23}$ V  $_{ 18}$}}\\
\hline
Aden& T9&  E+Cap &   E-Dec &  ANuEn &  GamEn &  ProEn &  PPEm \\
\hline
     3.0&  1.0 & -82.127 &    -100 & -82.076 &    -100 & -83.571 &  0.000 \\
     3.0&  3.0 & -27.435 &    -100 & -27.283 &    -100 & -28.464 &  0.000 \\
     3.0& 30.0 &  -0.805 &    -100 &   0.179 &    -100 &  -0.401 &  0.000 \\
     7.0&  1.0 & -87.947 &    -100 & -87.896 &    -100 & -89.391 &  0.000 \\
     7.0&  3.0 & -29.150 &    -100 & -28.998 &    -100 & -30.180 &  0.000 \\
     7.0& 30.0 &  -0.809 &    -100 &   0.176 &    -100 &  -0.404 &  0.000 \\
    11.0&  1.0 &    -100 &    -100 &    -100 &    -100 &    -100 &-100\\
    11.0&  3.0 & -67.628 &    -100 & -67.476 &    -100 & -68.657 &  0.000 \\
    11.0& 30.0 &  -4.672 &    -100 &  -3.687 &    -100 &  -4.267 &  0.000 \\
\hline
\end{tabular}\\
\vspace{0.20cm}\\
\begin{tabular}{rr|rrrrrr}
\multicolumn{7}{c}{\normalsize{\hspace*{3.5cm}$^{  41}_{22}$ TI $_{ 19} \rightarrow \hspace*{0.2cm}^{  41}_{21}$ SC $_{ 20}$}}\\
\hline
Aden& T9&  E-Cap &   E+Dec &  NuEn &  GamEn &  ProEn &  PPEm \\
\hline
     3.0&  1.0 &  -3.591 &   0.944 &   1.581 &   0.759 &   1.241 & -0.300 \\
     3.0&  3.0 &  -1.275 &   0.940 &   1.581 &   0.754 &   1.244 & -0.300 \\
     3.0& 30.0 &   2.686 &   0.972 &   3.887 &   1.321 &   3.547 & -0.020 \\
     7.0&  1.0 &   0.132 &   0.944 &   1.681 &   0.774 &   1.362 & -0.300 \\
     7.0&  3.0 &   0.149 &   0.941 &   1.685 &   0.769 &   1.367 & -0.300 \\
     7.0& 30.0 &   2.689 &   0.973 &   3.891 &   1.324 &   3.551 & -0.020 \\
    11.0&  1.0 &   5.286 &   0.944 &   6.645 &   3.959 &   6.139 & -0.020 \\
    11.0&  3.0 &   5.282 &   0.941 &   6.642 &   3.952 &   6.137 & -0.020 \\
    11.0& 30.0 &   5.370 &   1.054 &   6.774 &   3.756 &   6.292 & -0.008 \\
\hline
\end{tabular}\\
\vspace{0.20cm}\\
\noindent
\begin{tabular}{rr|rrrrrr}
\multicolumn{7}{c}{\normalsize{\hspace*{3.5cm}$^{  42}_{20}$ CA $_{ 22} \rightarrow \hspace*{0.2cm}^{  42}_{21}$ SC $_{ 21}$}}\\
\hline
Aden& T9&  E+Cap &   E-Dec &  ANuEn &  GamEn &  ProEn &  PPEm \\
\hline
     3.0&  1.0 & -35.853 & -47.080 & -36.198 & -38.323 & -63.872 &-30.000 \\
     3.0&  3.0 & -12.594 & -16.777 & -12.443 & -14.215 & -21.477 & -8.000 \\
     3.0& 30.0 &   1.266 &  -1.940 &   2.323 &   1.319 &   1.544 & -0.200 \\
     7.0&  1.0 & -41.672 & -48.803 & -42.017 & -44.143 & -69.692 &-30.000 \\
     7.0&  3.0 & -14.309 & -16.933 & -14.158 & -15.882 & -23.192 & -8.000 \\
     7.0& 30.0 &   1.263 &  -1.941 &   2.319 &   1.315 &   1.540 & -0.200 \\
    11.0&  1.0 &    -100 &    -100 &    -100 &    -100 &    -100 &-100\\
    11.0&  3.0 & -52.787 & -52.818 & -52.478 & -52.986 & -61.669 & -8.000 \\
    11.0& 30.0 &  -2.595 &  -5.293 &  -1.541 &  -2.542 &  -2.321 & -0.200 \\
\hline
\end{tabular}\\
\vspace{0.20cm}\\
\begin{tabular}{rr|rrrrrr}
\multicolumn{7}{c}{\normalsize{\hspace*{3.5cm}$^{  42}_{20}$ CA $_{ 22} \rightarrow \hspace*{0.2cm}^{  42}_{19}$ K  $_{ 23}$}}\\
\hline
Aden & T9& E-Cap &   E+Dec &  NuEn &  GamEn &  NeuEn &  PNEm \\
\hline
     3.0&  1.0 & -26.477 & -30.847 & -26.972 & -28.640 & -67.776 &-40.000 \\
     3.0&  3.0 & -10.901 & -12.362 & -10.818 & -11.313 & -22.919 &-10.000 \\
     3.0& 30.0 &   1.482 &  -1.062 &   2.568 &   2.032 &   1.138 & -0.800 \\
     7.0&  1.0 & -21.464 & -30.847 & -21.772 & -22.838 & -61.956 &-40.000 \\
     7.0&  3.0 &  -9.256 & -12.347 &  -9.174 &  -9.639 & -21.203 &-10.000 \\
     7.0& 30.0 &   1.485 &  -1.061 &   2.572 &   2.035 &   1.142 & -0.800 \\
    11.0&  1.0 &   3.832 & -30.847 &   4.962 &   4.320 &   1.484 & -3.000 \\
    11.0&  3.0 &   3.831 & -12.346 &   4.963 &   4.319 &   1.487 & -3.000 \\
    11.0& 30.0 &   4.481 &  -0.955 &   5.760 &   5.082 &   4.332 & -0.600 \\
\hline
\end{tabular}\\
\vspace{0.20cm}\\
\noindent
\begin{tabular}{rr|rrrrrr}
\multicolumn{7}{c}{\normalsize{\hspace*{3.5cm}$^{  42}_{21}$ SC $_{ 21} \rightarrow \hspace*{0.2cm}^{  42}_{22}$ TI $_{ 20}$}}\\
\hline
Aden& T9&  E+Cap &   E-Dec &  ANuEn &  GamEn &  ProEn &  PPEm \\
\hline
     3.0&  1.0 & -39.232 &    -100 & -39.181 & -60.032 & -64.586 &-20.000 \\
     3.0&  3.0 & -13.288 &    -100 & -13.131 & -19.841 & -21.648 & -8.000 \\
     3.0& 30.0 &   0.307 &    -100 &   1.316 &  -0.267 &   0.664 & -0.200 \\
     7.0&  1.0 & -45.052 &    -100 & -45.001 & -65.852 & -70.406 &-20.000 \\
     7.0&  3.0 & -15.003 &    -100 & -14.846 & -21.556 & -23.364 & -8.000 \\
     7.0& 30.0 &   0.303 &    -100 &   1.313 &  -0.270 &   0.661 & -0.200 \\
    11.0&  1.0 &    -100 &    -100 &    -100 &    -100 &    -100 &-100\\
    11.0&  3.0 & -53.481 &    -100 & -53.323 & -60.034 & -61.841 & -8.000 \\
    11.0& 30.0 &  -3.559 &    -100 &  -2.550 &  -4.132 &  -3.202 & -0.200 \\
\hline
\end{tabular}\\
\vspace{0.20cm}\\
\begin{tabular}{rr|rrrrrr}
\multicolumn{7}{c}{\normalsize{\hspace*{3.5cm}$^{  42}_{21}$ SC $_{ 21} \rightarrow \hspace*{0.2cm}^{  42}_{20}$ CA $_{ 22}$}}\\
\hline
Aden& T9&  E-Cap &   E+Dec &  NuEn &  GamEn &  ProEn &  PPEm \\
\hline
     3.0&  1.0 &  -4.288 &  -0.008 &   0.453 &  -2.936 & -31.563 &-30.000 \\
     3.0&  3.0 &  -2.116 &  -0.182 &   0.290 &  -2.155 & -11.505 &-10.000 \\
     3.0& 30.0 &   1.918 &  -1.118 &   3.024 &   2.590 &   1.845 & -0.500 \\
     7.0&  1.0 &  -0.561 &  -0.008 &   0.673 &  -1.421 & -25.743 &-20.000 \\
     7.0&  3.0 &  -0.690 &  -0.181 &   0.519 &  -1.078 &  -9.790 & -9.000 \\
     7.0& 30.0 &   1.921 &  -1.117 &   3.027 &   2.594 &   1.848 & -0.500 \\
    11.0&  1.0 &   4.733 &  -0.008 &   6.045 &   5.256 &   4.270 & -0.800 \\
    11.0&  3.0 &   4.722 &  -0.181 &   6.009 &   5.322 &   4.413 & -0.700 \\
    11.0& 30.0 &   4.829 &  -0.996 &   6.136 &   5.403 &   5.027 & -0.300 \\
\hline
\end{tabular}\\
\vspace{0.20cm}\\
\noindent
\begin{tabular}{rr|rrrrrr}
\multicolumn{7}{c}{\normalsize{\hspace*{3.5cm}$^{  42}_{22}$ TI $_{ 20} \rightarrow \hspace*{0.2cm}^{  42}_{23}$ V  $_{ 19}$}}\\
\hline
Aden& T9&  E+Cap &   E-Dec &  ANuEn &  GamEn &  ProEn &  PPEm \\
\hline
     3.0&  1.0 & -92.376 &    -100 & -92.325 &    -100 & -92.590 &  0.000 \\
     3.0&  3.0 & -30.760 &    -100 & -30.609 &    -100 & -30.951 &  0.000 \\
     3.0& 30.0 &  -0.720 &    -100 &   0.251 &    -100 &  -0.083 &  0.000 \\
     7.0&  1.0 & -98.196 &    -100 & -98.145 &    -100 & -98.410 &  0.000 \\
     7.0&  3.0 & -32.475 &    -100 & -32.325 &    -100 & -32.666 &  0.000 \\
     7.0& 30.0 &  -0.723 &    -100 &   0.248 &    -100 &  -0.087 &  0.000 \\
    11.0&  1.0 &    -100 &    -100 &    -100 &    -100 &    -100 &-100\\
    11.0&  3.0 & -70.953 &    -100 & -70.802 &    -100 & -71.144 &  0.000 \\
    11.0& 30.0 &  -4.587 &    -100 &  -3.615 &    -100 &  -3.950 &  0.000 \\
\hline
\end{tabular}\\
\vspace{0.20cm}\\
\begin{tabular}{rr|rrrrrr}
\multicolumn{7}{c}{\normalsize{\hspace*{3.5cm}$^{  42}_{22}$ TI $_{ 20} \rightarrow \hspace*{0.2cm}^{  42}_{21}$ SC $_{ 21}$}}\\
\hline
Aden& T9&  E-Cap &   E+Dec &  NuEn &  GamEn &  ProEn &  PPEm \\
\hline
     3.0&  1.0 &  -3.720 &   0.569 &   1.043 &   0.167 &  -5.834 & -6.000 \\
     3.0&  3.0 &  -1.400 &   0.568 &   1.053 &   0.172 &  -3.081 & -4.000 \\
     3.0& 30.0 &   2.843 &   1.217 &   4.050 &   2.780 &   2.985 & -0.400 \\
     7.0&  1.0 &   0.002 &   0.569 &   1.256 &   0.290 &  -1.774 & -2.000 \\
     7.0&  3.0 &   0.024 &   0.569 &   1.270 &   0.295 &  -1.517 & -2.000 \\
     7.0& 30.0 &   2.847 &   1.218 &   4.054 &   2.783 &   2.988 & -0.400 \\
    11.0&  1.0 &   5.201 &   0.569 &   6.552 &   4.670 &   5.230 & -0.400 \\
    11.0&  3.0 &   5.202 &   0.569 &   6.553 &   4.671 &   5.231 & -0.400 \\
    11.0& 30.0 &   5.526 &   1.298 &   6.931 &   5.362 &   5.848 & -0.300 \\
\hline
\end{tabular}\\
\vspace{0.20cm}\\
\noindent
\begin{tabular}{rr|rrrrrr}
\multicolumn{7}{c}{\normalsize{\hspace*{3.5cm}$^{  43}_{20}$ CA $_{ 23} \rightarrow \hspace*{0.2cm}^{  43}_{21}$ SC $_{ 22}$}}\\
\hline
Aden& T9&  E+Cap &   E-Dec &  ANuEn &  GamEn &  ProEn &  PPEm \\
\hline
     3.0&  1.0 & -17.683 & -25.442 & -17.639 & -22.136 & -47.013 &-30.000 \\
     3.0&  3.0 &  -7.685 & -11.028 &  -7.510 &  -8.827 & -16.248 & -8.000 \\
     3.0& 30.0 &   1.515 &  -2.279 &   2.559 &   1.456 &   2.085 & -0.100 \\
     7.0&  1.0 & -23.502 & -27.651 & -23.459 & -27.870 & -52.831 &-30.000 \\
     7.0&  3.0 &  -9.400 & -11.091 &  -9.215 & -10.444 & -17.963 & -8.000 \\
     7.0& 30.0 &   1.512 &  -2.280 &   2.556 &   1.453 &   2.082 & -0.100 \\
    11.0&  1.0 &    -100 &    -100 &    -100 &    -100 &    -100 &-100\\
    11.0&  3.0 & -47.877 & -46.559 & -46.689 & -46.796 & -56.441 &-10.000 \\
    11.0& 30.0 &  -2.348 &  -5.632 &  -1.305 &  -2.405 &  -1.779 & -0.100 \\
\hline
\end{tabular}\\
\vspace{0.20cm}\\
\begin{tabular}{rr|rrrrrr}
\multicolumn{7}{c}{\normalsize{\hspace*{3.5cm}$^{  43}_{20}$ CA $_{ 23} \rightarrow \hspace*{0.2cm}^{  43}_{19}$ K  $_{ 24}$}}\\
\hline
Aden& T9&  E-Cap &   E+Dec &  NuEn &  GamEn &  ProEn &  PPEm \\
\hline
     3.0&  1.0 & -17.961 & -26.918 & -18.499 & -20.815 & -67.252 &-50.000 \\
     3.0&  3.0 &  -8.399 & -11.348 &  -8.431 &  -8.644 & -22.313 &-10.000 \\
     3.0& 30.0 &   1.533 &  -2.313 &   2.569 &   2.128 &   1.279 & -0.400 \\
     7.0&  1.0 & -12.181 & -26.918 & -12.704 & -14.995 & -61.432 &-50.000 \\
     7.0&  3.0 &  -6.689 & -11.345 &  -6.720 &  -6.928 & -20.597 &-10.000 \\
     7.0& 30.0 &   1.537 &  -2.312 &   2.572 &   2.132 &   1.283 & -0.400 \\
    11.0&  1.0 &   4.039 & -26.918 &   5.181 &   4.650 &   2.678 & -2.000 \\
    11.0&  3.0 &   4.114 & -11.345 &   5.260 &   4.730 &   2.769 & -2.000 \\
    11.0& 30.0 &   4.659 &  -2.160 &   5.899 &   5.240 &   4.523 & -0.300 \\
\hline
\end{tabular}\\
\vspace{0.20cm}\\
\noindent
\begin{tabular}{rr|rrrrrr}
\multicolumn{7}{c}{\normalsize{\hspace*{3.5cm}$^{  43}_{21}$ SC $_{ 22} \rightarrow \hspace*{0.2cm}^{  43}_{22}$ TI $_{ 21}$}}\\
\hline
Aden& T9&  E+Cap &   E-Dec &  ANuEn &  GamEn &  ProEn &  PPEm \\
\hline
     3.0&  1.0 & -39.083 &    -100 & -39.031 & -42.382 & -67.354 &-30.000 \\
     3.0&  3.0 & -13.552 &    -100 & -13.395 & -15.379 & -22.324 & -8.000 \\
     3.0& 30.0 &   0.656 &    -100 &   1.676 &   0.532 &   0.968 & -0.200 \\
     7.0&  1.0 & -44.903 &    -100 & -44.851 & -48.202 & -73.174 &-30.000 \\
     7.0&  3.0 & -15.268 &    -100 & -15.111 & -17.094 & -24.040 & -8.000 \\
     7.0& 30.0 &   0.652 &    -100 &   1.672 &   0.529 &   0.964 & -0.200 \\
    11.0&  1.0 &    -100 &    -100 &    -100 &    -100 &    -100 &-100\\
    11.0&  3.0 & -53.745 &    -100 & -53.588 & -55.572 & -62.517 & -8.000 \\
    11.0& 30.0 &  -3.209 &    -100 &  -2.190 &  -3.332 &  -2.898 & -0.200 \\
\hline
\end{tabular}\\
\vspace{0.20cm}\\
\begin{tabular}{rr|rrrrrr}
\multicolumn{7}{c}{\normalsize{\hspace*{3.5cm}$^{  43}_{21}$ SC $_{ 22} \rightarrow \hspace*{0.2cm}^{  43}_{20}$ CA $_{ 23}$}}\\
\hline
Aden & T9& E-Cap &   E+Dec &  NuEn &  GamEn &  NeuEn &  PNEm \\
\hline
     3.0&  1.0 &  -6.480 &  -4.382 &  -4.553 &  -5.516 & -40.994 &-40.000 \\
     3.0&  3.0 &  -3.775 &  -3.301 &  -2.949 &  -4.230 & -14.461 &-10.000 \\
     3.0& 30.0 &   1.680 &  -2.050 &   2.787 &   2.042 &   1.575 & -0.400 \\
     7.0&  1.0 &  -2.658 &  -4.381 &  -2.254 &  -3.279 & -35.174 &-30.000 \\
     7.0&  3.0 &  -2.319 &  -3.296 &  -1.761 &  -3.117 & -12.746 &-10.000 \\
     7.0& 30.0 &   1.683 &  -2.049 &   2.791 &   2.046 &   1.578 & -0.400 \\
    11.0&  1.0 &   3.822 &  -4.381 &   4.987 &   4.489 &   3.656 & -0.600 \\
    11.0&  3.0 &   3.886 &  -3.296 &   5.067 &   4.458 &   3.679 & -0.500 \\
    11.0& 30.0 &   4.573 &  -1.877 &   5.887 &   4.873 &   4.731 & -0.300 \\
\hline
\end{tabular}\\
\vspace{0.20cm}\\
\noindent
\begin{tabular}{rr|rrrrrr}
\multicolumn{7}{c}{\normalsize{\hspace*{3.5cm}$^{  43}_{22}$ TI $_{ 21} \rightarrow \hspace*{0.2cm}^{  43}_{23}$ V  $_{ 20}$}}\\
\hline
Aden& T9&  E+Cap &   E-Dec &  ANuEn &  GamEn &  ProEn &  PPEm \\
\hline
     3.0&  1.0 & -60.958 &    -100 & -60.907 & -62.694 & -70.918 & -9.000 \\
     3.0&  3.0 & -20.423 &    -100 & -20.270 & -21.521 & -25.563 & -5.000 \\
     3.0& 30.0 &  -0.072 &    -100 &   0.922 &  -1.417 &   0.590 & -0.100 \\
     7.0&  1.0 & -66.778 &    -100 & -66.727 & -68.514 & -76.738 & -9.000 \\
     7.0&  3.0 & -22.139 &    -100 & -21.985 & -23.236 & -27.278 & -5.000 \\
     7.0& 30.0 &  -0.075 &    -100 &   0.919 &  -1.420 &   0.587 & -0.100 \\
    11.0&  1.0 &    -100 &    -100 &    -100 &    -100 &    -100 &-100\\
    11.0&  3.0 & -60.616 &    -100 & -60.463 & -61.713 & -65.756 & -5.000 \\
    11.0& 30.0 &  -3.938 &    -100 &  -2.944 &  -5.283 &  -3.276 & -0.100 \\
\hline
\end{tabular}\\
\vspace{0.20cm}\\
\begin{tabular}{rr|rrrrrr}
\multicolumn{7}{c}{\normalsize{\hspace*{3.5cm}$^{  43}_{22}$ TI $_{ 21} \rightarrow \hspace*{0.2cm}^{  43}_{21}$ SC $_{ 22}$}}\\
\hline
Aden& T9&  E-Cap &   E+Dec &  NuEn &  GamEn &  ProEn &  PPEm \\
\hline
     3.0&  1.0 &  -4.213 &   0.076 &   0.567 &  -1.249 &  -6.527 & -6.000 \\
     3.0&  3.0 &  -1.955 &  -0.043 &   0.454 &  -0.970 &  -3.640 & -3.000 \\
     3.0& 30.0 &   2.439 &   0.110 &   3.575 &   2.397 &   2.983 & -0.100 \\
     7.0&  1.0 &  -0.472 &   0.076 &   0.766 &  -0.479 &  -2.409 & -2.000 \\
     7.0&  3.0 &  -0.518 &  -0.043 &   0.669 &  -0.358 &  -2.089 & -2.000 \\
     7.0& 30.0 &   2.443 &   0.111 &   3.579 &   2.400 &   2.987 & -0.100 \\
    11.0&  1.0 &   5.046 &   0.076 &   6.345 &   5.187 &   5.315 & -0.300 \\
    11.0&  3.0 &   5.030 &  -0.043 &   6.322 &   5.182 &   5.335 & -0.200 \\
    11.0& 30.0 &   5.258 &   0.203 &   6.601 &   5.059 &   5.937 & -0.070 \\
\hline
\end{tabular}\\
\vspace{0.20cm}\\
\noindent
\begin{tabular}{rr|rrrrrr}
\multicolumn{7}{c}{\normalsize{\hspace*{3.5cm}$^{  43}_{23}$ V  $_{ 20} \rightarrow \hspace*{0.2cm}^{  43}_{24}$ CR $_{ 19}$}}\\
\hline
Aden& T9&  E+Cap &   E-Dec &  ANuEn &  GamEn &  ProEn &  PPEm \\
\hline
     3.0&  1.0 & -90.162 &    -100 & -90.111 & -90.160 &    -100 &-20.000 \\
     3.0&  3.0 & -30.872 &    -100 & -30.721 & -30.870 & -38.039 & -8.000 \\
     3.0& 30.0 &  -1.803 &    -100 &  -0.838 &  -2.240 &  -1.210 & -0.200 \\
     7.0&  1.0 & -95.982 &    -100 & -95.931 & -95.980 &    -100 &-20.000 \\
     7.0&  3.0 & -32.587 &    -100 & -32.436 & -32.586 & -39.754 & -8.000 \\
     7.0& 30.0 &  -1.806 &    -100 &  -0.841 &  -2.243 &  -1.213 & -0.200 \\
    11.0&  1.0 &    -100 &    -100 &    -100 &    -100 &    -100 &-100\\
    11.0&  3.0 & -71.065 &    -100 & -70.914 & -71.063 & -78.231 & -8.000 \\
    11.0& 30.0 &  -5.669 &    -100 &  -4.704 &  -6.106 &  -5.076 & -0.200 \\
\hline
\end{tabular}\\
\vspace{0.20cm}\\
\begin{tabular}{rr|rrrrrr}
\multicolumn{7}{c}{\normalsize{\hspace*{3.5cm}$^{  43}_{23}$ V  $_{ 20} \rightarrow \hspace*{0.2cm}^{  43}_{22}$ TI $_{ 21}$}}\\
\hline
Aden& T9&  E-Cap &   E+Dec &  NuEn &  GamEn &  ProEn &  PPEm \\
\hline
     3.0&  1.0 &  -3.665 &   1.186 &   1.870 &   1.183 &  -0.728 & -2.000 \\
     3.0&  3.0 &  -1.334 &   1.212 &   1.899 &   1.211 &  -0.663 & -2.000 \\
     3.0& 30.0 &   2.530 &   1.153 &   3.751 &   2.218 &   2.907 & -0.300 \\
     7.0&  1.0 &   0.043 &   1.186 &   1.927 &   1.208 &  -0.195 & -2.000 \\
     7.0&  3.0 &   0.084 &   1.212 &   1.957 &   1.235 &  -0.127 & -2.000 \\
     7.0& 30.0 &   2.533 &   1.154 &   3.755 &   2.221 &   2.910 & -0.300 \\
    11.0&  1.0 &   5.134 &   1.186 &   6.487 &   4.631 &   5.715 & -0.200 \\
    11.0&  3.0 &   5.161 &   1.212 &   6.515 &   4.657 &   5.742 & -0.200 \\
    11.0& 30.0 &   5.230 &   1.224 &   6.626 &   4.704 &   5.839 & -0.200 \\
\hline
\end{tabular}\\
\vspace{0.20cm}\\
\noindent
\begin{tabular}{rr|rrrrrr}
\multicolumn{7}{c}{\normalsize{\hspace*{3.5cm}$^{  44}_{20}$ CA $_{ 24} \rightarrow \hspace*{0.2cm}^{  44}_{21}$ SC $_{ 23}$}}\\
\hline
Aden& T9&  E+Cap &   E-Dec &  ANuEn &  GamEn &  ProEn &  PPEm \\
\hline
     3.0&  1.0 & -23.429 & -29.193 & -23.850 & -26.210 & -62.297 &-40.000 \\
     3.0&  3.0 &  -9.852 & -11.225 &  -9.735 & -10.124 & -20.513 &-10.000 \\
     3.0& 30.0 &   1.603 &  -0.854 &   2.671 &   2.047 &   1.835 & -0.300 \\
     7.0&  1.0 & -29.249 & -29.513 & -29.363 & -31.949 & -67.996 &-40.000 \\
     7.0&  3.0 & -11.564 & -11.428 & -11.174 & -11.800 & -22.228 &-10.000 \\
     7.0& 30.0 &   1.600 &  -0.855 &   2.668 &   2.043 &   1.832 & -0.300 \\
    11.0&  1.0 &    -100 &    -100 &    -100 &    -100 &    -100 &-100\\
    11.0&  3.0 & -50.042 & -46.815 & -46.955 & -48.255 & -60.706 &-10.000 \\
    11.0& 30.0 &  -2.258 &  -3.967 &  -1.190 &  -1.811 &  -2.029 & -0.300 \\
\hline
\end{tabular}\\
\vspace{0.20cm}\\
\begin{tabular}{rr|rrrrrr}
\multicolumn{7}{c}{\normalsize{\hspace*{3.5cm}$^{  44}_{20}$ CA $_{ 24} \rightarrow \hspace*{0.2cm}^{  44}_{19}$ K  $_{ 25}$}}\\
\hline
Aden & T9& E-Cap &   E+Dec &  NuEn &  GamEn &  NeuEn &  PNEm \\
\hline
     3.0&  1.0 & -36.662 & -48.869 & -37.075 & -39.520 & -76.691 &-40.000 \\
     3.0&  3.0 & -13.810 & -17.353 & -13.758 & -14.377 & -25.676 &-10.000 \\
     3.0& 30.0 &   1.447 &  -1.434 &   2.522 &   1.937 &   1.135 & -0.700 \\
     7.0&  1.0 & -31.221 & -48.869 & -31.703 & -33.776 & -70.871 &-40.000 \\
     7.0&  3.0 & -12.179 & -17.347 & -12.102 & -12.666 & -23.961 &-10.000 \\
     7.0& 30.0 &   1.450 &  -1.433 &   2.525 &   1.940 &   1.138 & -0.700 \\
    11.0&  1.0 &   2.666 & -48.869 &   3.674 &   3.178 &   1.969 & -0.800 \\
    11.0&  3.0 &   2.663 & -17.347 &   3.673 &   3.175 &   1.968 & -0.800 \\
    11.0& 30.0 &   4.501 &  -1.301 &   5.761 &   5.050 &   4.332 & -0.600 \\
\hline
\end{tabular}\\
\vspace{0.20cm}\\
\noindent
\begin{tabular}{rr|rrrrrr}
\multicolumn{7}{c}{\normalsize{\hspace*{3.5cm}$^{  44}_{21}$ SC $_{ 23} \rightarrow \hspace*{0.2cm}^{  44}_{22}$ TI $_{ 22}$}}\\
\hline
Aden& T9&  E+Cap &   E-Dec &  ANuEn &  GamEn &  ProEn &  PPEm \\
\hline
     3.0&  1.0 & -11.109 & -31.751 & -11.059 & -25.595 & -54.963 &-40.000 \\
     3.0&  3.0 &  -7.599 & -12.841 &  -7.358 &  -9.348 & -18.270 &-10.000 \\
     3.0& 30.0 &   1.381 &  -3.550 &   2.420 &   1.827 &   1.722 & -0.200 \\
     7.0&  1.0 & -16.928 & -32.988 & -16.879 & -31.371 & -60.783 &-40.000 \\
     7.0&  3.0 &  -9.308 & -13.151 &  -9.068 & -11.048 & -19.985 &-10.000 \\
     7.0& 30.0 &   1.377 &  -3.552 &   2.417 &   1.824 &   1.718 & -0.200 \\
    11.0&  1.0 &    -100 &    -100 &    -100 &    -100 &    -100 &-100\\
    11.0&  3.0 & -47.785 & -49.846 & -47.544 & -49.070 & -58.463 &-10.000 \\
    11.0& 30.0 &  -2.483 &  -6.970 &  -1.444 &  -2.035 &  -2.144 & -0.200 \\
\hline
\end{tabular}\\
\vspace{0.20cm}\\
\begin{tabular}{rr|rrrrrr}
\multicolumn{7}{c}{\normalsize{\hspace*{3.5cm}$^{  44}_{21}$ SC $_{ 23} \rightarrow \hspace*{0.2cm}^{  44}_{20}$ CA $_{ 24}$}}\\
\hline
Aden & T9& E-Cap &   E+Dec &  NuEn &  GamEn &  NeuEn &  PNEm \\
\hline
     3.0&  1.0 &  -7.015 &  -4.558 &  -4.629 &  -4.491 & -49.294 &-40.000 \\
     3.0&  3.0 &  -4.740 &  -4.845 &  -4.253 &  -4.247 & -17.073 &-10.000 \\
     3.0& 30.0 &   1.807 &  -1.466 &   2.874 &   2.530 &   1.880 & -0.400 \\
     7.0&  1.0 &  -3.190 &  -4.557 &  -2.765 &  -2.980 & -43.474 &-40.000 \\
     7.0&  3.0 &  -3.243 &  -4.836 &  -2.838 &  -2.866 & -15.358 &-10.000 \\
     7.0& 30.0 &   1.811 &  -1.465 &   2.878 &   2.534 &   1.883 & -0.400 \\
    11.0&  1.0 &   4.145 &  -4.557 &   5.330 &   5.016 &   3.280 & -1.000 \\
    11.0&  3.0 &   4.167 &  -4.836 &   5.349 &   5.058 &   3.333 & -1.000 \\
    11.0& 30.0 &   4.817 &  -1.346 &   6.091 &   5.481 &   5.036 & -0.300 \\
\hline
\end{tabular}\\
\vspace{0.20cm}\\
\noindent
\begin{tabular}{rr|rrrrrr}
\multicolumn{7}{c}{\normalsize{\hspace*{3.5cm}$^{  44}_{22}$ TI $_{ 22} \rightarrow \hspace*{0.2cm}^{  44}_{23}$ V  $_{ 21}$}}\\
\hline
Aden& T9&  E+Cap &   E-Dec &  ANuEn &  GamEn &  ProEn &  PPEm \\
\hline
     3.0&  1.0 & -72.653 &    -100 & -72.602 & -73.935 & -89.393 &-20.000 \\
     3.0&  3.0 & -23.797 &    -100 & -23.643 & -24.824 & -29.582 & -6.000 \\
     3.0& 30.0 &   0.447 &    -100 &   1.444 &  -0.229 &   0.869 & -0.200 \\
     7.0&  1.0 & -78.473 &    -100 & -78.422 & -79.755 & -95.213 &-20.000 \\
     7.0&  3.0 & -25.512 &    -100 & -25.358 & -26.539 & -31.298 & -6.000 \\
     7.0& 30.0 &   0.443 &    -100 &   1.441 &  -0.233 &   0.865 & -0.200 \\
    11.0&  1.0 &    -100 &    -100 &    -100 &    -100 &    -100 &-100\\
    11.0&  3.0 & -63.990 &    -100 & -63.836 & -65.017 & -69.775 & -6.000 \\
    11.0& 30.0 &  -3.419 &    -100 &  -2.422 &  -4.095 &  -2.998 & -0.200 \\
\hline
\end{tabular}\\
\vspace{0.20cm}\\
\begin{tabular}{rr|rrrrrr}
\multicolumn{7}{c}{\normalsize{\hspace*{3.5cm}$^{  44}_{22}$ TI $_{ 22} \rightarrow \hspace*{0.2cm}^{  44}_{21}$ SC $_{ 23}$}}\\
\hline
Aden& T9&  E-Cap &   E+Dec &  NuEn &  GamEn &  ProEn &  PPEm \\
\hline
     3.0&  1.0 & -10.093 & -21.094 & -10.528 & -10.918 & -44.475 &-30.000 \\
     3.0&  3.0 &  -6.882 &  -7.120 &  -6.519 &  -7.254 & -15.807 & -8.000 \\
     3.0& 30.0 &   2.212 &  -1.026 &   3.322 &   2.674 &   2.188 & -0.500 \\
     7.0&  1.0 &  -5.652 & -21.094 &  -5.773 &  -6.476 & -38.659 &-30.000 \\
     7.0&  3.0 &  -5.271 &  -7.118 &  -5.139 &  -5.631 & -14.110 & -8.000 \\
     7.0& 30.0 &   2.216 &  -1.025 &   3.325 &   2.678 &   2.192 & -0.500 \\
    11.0&  1.0 &   4.419 & -21.094 &   5.565 &   5.194 &   3.222 & -1.000 \\
    11.0&  3.0 &   4.420 &  -7.118 &   5.567 &   5.195 &   3.225 & -1.000 \\
    11.0& 30.0 &   5.105 &  -0.868 &   6.418 &   5.519 &   5.330 & -0.400 \\
\hline
\end{tabular}\\
\vspace{0.20cm}\\
\noindent
\begin{tabular}{rr|rrrrrr}
\multicolumn{7}{c}{\normalsize{\hspace*{3.5cm}$^{  44}_{23}$ V  $_{ 21} \rightarrow \hspace*{0.2cm}^{  44}_{24}$ CR $_{ 20}$}}\\
\hline
Aden& T9&  E+Cap &   E-Dec &  ANuEn &  GamEn &  ProEn &  PPEm \\
\hline
     3.0&  1.0 & -74.801 &    -100 & -74.751 & -74.434 & -76.468 & -0.400 \\
     3.0&  3.0 & -24.949 &    -100 & -24.797 & -24.911 & -25.694 & -0.200 \\
     3.0& 30.0 &  -0.443 &    -100 &   0.542 &  -1.253 &  -0.016 & -0.020 \\
     7.0&  1.0 & -80.621 &    -100 & -80.570 & -80.254 & -82.288 & -0.400 \\
     7.0&  3.0 & -26.665 &    -100 & -26.513 & -26.626 & -27.410 & -0.200 \\
     7.0& 30.0 &  -0.447 &    -100 &   0.539 &  -1.257 &  -0.019 & -0.020 \\
    11.0&  1.0 &    -100 &    -100 &    -100 &    -100 &    -100 &-100\\
    11.0&  3.0 & -65.142 &    -100 & -64.990 & -65.104 & -65.887 & -0.200 \\
    11.0& 30.0 &  -4.310 &    -100 &  -3.324 &  -5.119 &  -3.882 & -0.020 \\
\hline
\end{tabular}\\
\vspace{0.20cm}\\
\begin{tabular}{rr|rrrrrr}
\multicolumn{7}{c}{\normalsize{\hspace*{3.5cm}$^{  44}_{23}$ V  $_{ 21} \rightarrow \hspace*{0.2cm}^{  44}_{22}$ TI $_{ 22}$}}\\
\hline
Aden& T9&  E-Cap &   E+Dec &  NuEn &  GamEn &  ProEn &  PPEm \\
\hline
     3.0&  1.0 &  -3.876 &   0.692 &   1.302 &   1.377 &  -1.677 & -2.000 \\
     3.0&  3.0 &  -1.549 &   0.713 &   1.328 &   1.415 &  -1.460 & -2.000 \\
     3.0& 30.0 &   2.307 &   0.581 &   3.500 &   2.699 &   2.662 & -0.200 \\
     7.0&  1.0 &  -0.158 &   0.692 &   1.404 &   1.419 &  -0.843 & -1.000 \\
     7.0&  3.0 &  -0.127 &   0.713 &   1.431 &   1.457 &  -0.684 & -1.000 \\
     7.0& 30.0 &   2.311 &   0.581 &   3.504 &   2.702 &   2.665 & -0.200 \\
    11.0&  1.0 &   4.976 &   0.692 &   6.333 &   5.193 &   5.250 & -0.200 \\
    11.0&  3.0 &   4.994 &   0.713 &   6.349 &   5.230 &   5.321 & -0.200 \\
    11.0& 30.0 &   5.029 &   0.666 &   6.413 &   5.237 &   5.593 & -0.100 \\
\hline
\end{tabular}\\
\vspace{0.20cm}\\
\noindent
\begin{tabular}{rr|rrrrrr}
\multicolumn{7}{c}{\normalsize{\hspace*{3.5cm}$^{  44}_{24}$ CR $_{ 20} \rightarrow \hspace*{0.2cm}^{  44}_{25}$ MN $_{ 19}$}}\\
\hline
Aden& T9&  E+Cap &   E-Dec &  ANuEn &  GamEn &  ProEn &  PPEm \\
\hline
     3.0&  1.0 &    -100 &    -100 &    -100 &    -100 &    -100 &-100\\
     3.0&  3.0 & -36.287 &    -100 & -36.136 &    -100 & -36.282 &  0.000 \\
     3.0& 30.0 &  -1.353 &    -100 &  -0.388 &    -100 &  -0.758 &  0.000 \\
     7.0&  1.0 &    -100 &    -100 &    -100 &    -100 &    -100 &-100\\
     7.0&  3.0 & -38.002 &    -100 & -37.851 &    -100 & -37.997 &  0.000 \\
     7.0& 30.0 &  -1.357 &    -100 &  -0.391 &    -100 &  -0.762 &  0.000 \\
    11.0&  1.0 &    -100 &    -100 &    -100 &    -100 &    -100 &-100\\
    11.0&  3.0 & -76.480 &    -100 & -76.329 &    -100 & -76.475 &  0.000 \\
    11.0& 30.0 &  -5.220 &    -100 &  -4.254 &    -100 &  -4.625 &  0.000 \\
\hline
\end{tabular}\\
\vspace{0.20cm}\\
\begin{tabular}{rr|rrrrrr}
\multicolumn{7}{c}{\normalsize{\hspace*{3.5cm}$^{  44}_{24}$ CR $_{ 20} \rightarrow \hspace*{0.2cm}^{  44}_{23}$ V  $_{ 21}$}}\\
\hline
Aden& T9&  E-Cap &   E+Dec &  NuEn &  GamEn &  ProEn &  PPEm \\
\hline
     3.0&  1.0 &  -3.573 &   1.053 &   1.660 &   1.264 &  -2.374 & -4.000 \\
     3.0&  3.0 &  -1.261 &   1.053 &   1.663 &   1.266 &  -1.382 & -3.000 \\
     3.0& 30.0 &   2.985 &   1.787 &   4.227 &   2.753 &   3.569 & -0.200 \\
     7.0&  1.0 &   0.140 &   1.053 &   1.758 &   1.310 &  -0.042 & -2.000 \\
     7.0&  3.0 &   0.162 &   1.053 &   1.763 &   1.311 &   0.065 & -2.000 \\
     7.0& 30.0 &   2.988 &   1.788 &   4.230 &   2.756 &   3.572 & -0.200 \\
    11.0&  1.0 &   5.337 &   1.053 &   6.686 &   5.092 &   6.009 & -0.200 \\
    11.0&  3.0 &   5.338 &   1.053 &   6.687 &   5.092 &   6.010 & -0.200 \\
    11.0& 30.0 &   5.638 &   1.846 &   7.056 &   5.230 &   6.356 & -0.100 \\
\hline
\end{tabular}\\
\vspace{0.20cm}\\
\noindent
\begin{tabular}{rr|rrrrrr}
\multicolumn{7}{c}{\normalsize{\hspace*{3.5cm}$^{  45}_{20}$ CA $_{ 25} \rightarrow \hspace*{0.2cm}^{  45}_{21}$ SC $_{ 24}$}}\\
\hline
Aden& T9&  E+Cap &   E-Dec &  ANuEn &  GamEn &  ProEn &  PPEm \\
\hline
     3.0&  1.0 &  -8.632 &  -7.364 &  -8.016 &  -8.747 & -44.102 &-40.000 \\
     3.0&  3.0 &  -5.040 &  -7.020 &  -4.884 &  -5.070 & -14.859 &-10.000 \\
     3.0& 30.0 &   1.980 &  -1.400 &   3.043 &   2.120 &   2.420 & -0.100 \\
     7.0&  1.0 & -14.452 & -10.356 & -11.205 & -14.567 & -49.922 &-40.000 \\
     7.0&  3.0 &  -6.750 &  -7.238 &  -6.491 &  -6.779 & -16.574 &-10.000 \\
     7.0& 30.0 &   1.977 &  -1.401 &   3.040 &   2.116 &   2.417 & -0.100 \\
    11.0&  1.0 &    -100 &    -100 &    -100 &    -100 &    -100 &-100\\
    11.0&  3.0 & -45.227 & -41.766 & -41.908 & -42.540 & -55.052 &-10.000 \\
    11.0& 30.0 &  -1.881 &  -4.583 &  -0.821 &  -1.739 &  -1.443 & -0.100 \\
\hline
\end{tabular}\\
\vspace{0.20cm}\\
\begin{tabular}{rr|rrrrrr}
\multicolumn{7}{c}{\normalsize{\hspace*{3.5cm}$^{  45}_{20}$ CA $_{ 25} \rightarrow \hspace*{0.2cm}^{  45}_{19}$ K  $_{ 26}$}}\\
\hline
Aden & T9& E-Cap &   E+Dec &  NuEn &  GamEn &  NeuEn &  PNEm \\
\hline
     3.0&  1.0 & -29.412 & -35.186 & -29.973 & -32.733 & -76.442 &-50.000 \\
     3.0&  3.0 & -11.748 & -14.084 & -11.791 & -11.680 & -25.487 &-10.000 \\
     3.0& 30.0 &   1.621 &  -3.275 &   2.648 &   2.359 &   0.697 & -1.000 \\
     7.0&  1.0 & -23.592 & -35.186 & -24.153 & -26.919 & -70.622 &-50.000 \\
     7.0&  3.0 & -10.043 & -14.081 & -10.092 &  -9.973 & -23.771 &-10.000 \\
     7.0& 30.0 &   1.624 &  -3.274 &   2.652 &   2.362 &   0.700 & -1.000 \\
    11.0&  1.0 &   4.263 & -35.186 &   5.406 &   4.550 &   2.257 & -2.000 \\
    11.0&  3.0 &   4.338 & -14.081 &   5.483 &   4.622 &   2.364 & -2.000 \\
    11.0& 30.0 &   4.761 &  -3.089 &   5.997 &   5.499 &   4.114 & -0.800 \\
\hline
\end{tabular}\\
\vspace{0.20cm}\\
\noindent
\begin{tabular}{rr|rrrrrr}
\multicolumn{7}{c}{\normalsize{\hspace*{3.5cm}$^{  45}_{21}$ SC $_{ 24} \rightarrow \hspace*{0.2cm}^{  45}_{22}$ TI $_{ 23}$}}\\
\hline
Aden& T9&  E+Cap &   E-Dec &  ANuEn &  GamEn &  ProEn &  PPEm \\
\hline
     3.0&  1.0 & -15.739 & -27.226 & -16.025 & -17.927 & -62.767 &-50.000 \\
     3.0&  3.0 &  -7.152 & -11.257 &  -6.984 &  -8.200 & -20.321 &-10.000 \\
     3.0& 30.0 &   1.744 &  -2.819 &   2.791 &   2.280 &   1.761 & -0.400 \\
     7.0&  1.0 & -21.558 & -29.117 & -21.845 & -23.746 & -68.587 &-50.000 \\
     7.0&  3.0 &  -8.866 & -11.693 &  -8.699 &  -9.902 & -22.037 &-10.000 \\
     7.0& 30.0 &   1.741 &  -2.820 &   2.787 &   2.276 &   1.757 & -0.400 \\
    11.0&  1.0 &    -100 &    -100 &    -100 &    -100 &    -100 &-100\\
    11.0&  3.0 & -47.344 & -45.826 & -45.954 & -48.366 & -60.514 &-10.000 \\
    11.0& 30.0 &  -2.119 &  -6.043 &  -1.074 &  -1.582 &  -2.105 & -0.400 \\
\hline
\end{tabular}\\
\vspace{0.20cm}\\
\begin{tabular}{rr|rrrrrr}
\multicolumn{7}{c}{\normalsize{\hspace*{3.5cm}$^{  45}_{21}$ SC $_{ 24} \rightarrow \hspace*{0.2cm}^{  45}_{20}$ CA $_{ 25}$}}\\
\hline
Aden & T9& E-Cap &   E+Dec &  NuEn &  GamEn &  NeuEn &  PNEm \\
\hline
     3.0&  1.0 & -10.493 & -13.408 & -10.732 & -14.428 & -50.333 &-40.000 \\
     3.0&  3.0 &  -5.170 &  -8.162 &  -4.971 &  -7.596 & -17.102 &-10.000 \\
     3.0& 30.0 &   1.808 &  -1.613 &   2.870 &   2.112 &   1.960 & -0.300 \\
     7.0&  1.0 &  -5.684 & -13.408 &  -5.934 & -10.492 & -44.513 &-40.000 \\
     7.0&  3.0 &  -3.641 &  -8.128 &  -3.395 &  -6.066 & -15.387 &-10.000 \\
     7.0& 30.0 &   1.812 &  -1.612 &   2.874 &   2.116 &   1.963 & -0.300 \\
    11.0&  1.0 &   4.011 & -13.408 &   5.142 &   4.772 &   3.201 & -1.000 \\
    11.0&  3.0 &   4.042 &  -8.127 &   5.181 &   4.799 &   3.217 & -1.000 \\
    11.0& 30.0 &   4.831 &  -1.499 &   6.101 &   5.058 &   5.108 & -0.200 \\
\hline
\end{tabular}\\
\vspace{0.20cm}\\
\noindent
\begin{tabular}{rr|rrrrrr}
\multicolumn{7}{c}{\normalsize{\hspace*{3.5cm}$^{  45}_{22}$ TI $_{ 23} \rightarrow \hspace*{0.2cm}^{  45}_{23}$ V  $_{ 22}$}}\\
\hline
Aden& T9&  E+Cap &   E-Dec &  ANuEn &  GamEn &  ProEn &  PPEm \\
\hline
     3.0&  1.0 & -40.236 &    -100 & -40.320 & -41.587 & -59.564 &-20.000 \\
     3.0&  3.0 & -14.008 &    -100 & -13.852 & -15.102 & -19.521 & -6.000 \\
     3.0& 30.0 &   0.973 &    -100 &   1.988 &  -1.019 &   1.743 & -0.030 \\
     7.0&  1.0 & -46.056 &    -100 & -46.140 & -47.407 & -65.384 &-20.000 \\
     7.0&  3.0 & -15.723 &    -100 & -15.567 & -16.817 & -21.236 & -6.000 \\
     7.0& 30.0 &   0.969 &    -100 &   1.985 &  -1.022 &   1.740 & -0.030 \\
    11.0&  1.0 &    -100 &    -100 &    -100 &    -100 &    -100 &-100\\
    11.0&  3.0 & -54.200 &    -100 & -54.045 & -55.294 & -59.714 & -6.000 \\
    11.0& 30.0 &  -2.893 &    -100 &  -1.877 &  -4.879 &  -2.123 & -0.030 \\
\hline
\end{tabular}\\
\vspace{0.20cm}\\
\begin{tabular}{rr|rrrrrr}
\multicolumn{7}{c}{\normalsize{\hspace*{3.5cm}$^{  45}_{22}$ TI $_{ 23} \rightarrow \hspace*{0.2cm}^{  45}_{21}$ SC $_{ 24}$}}\\
\hline
Aden& T9&  E-Cap &   E+Dec &  NuEn &  GamEn &  ProEn &  PPEm \\
\hline
     3.0&  1.0 &  -6.318 &  -4.267 &  -4.470 &  -5.944 & -36.495 &-30.000 \\
     3.0&  3.0 &  -3.922 &  -4.286 &  -3.448 &  -4.808 & -12.983 & -9.000 \\
     3.0& 30.0 &   2.208 &  -1.394 &   3.289 &   2.478 &   2.591 & -0.200 \\
     7.0&  1.0 &  -2.509 &  -4.267 &  -2.093 &  -4.064 & -30.675 &-30.000 \\
     7.0&  3.0 &  -2.457 &  -4.272 &  -1.989 &  -3.439 & -11.268 & -8.000 \\
     7.0& 30.0 &   2.211 &  -1.393 &   3.293 &   2.482 &   2.595 & -0.200 \\
    11.0&  1.0 &   4.528 &  -4.267 &   5.695 &   4.791 &   4.630 & -0.300 \\
    11.0&  3.0 &   4.531 &  -4.272 &   5.700 &   4.803 &   4.641 & -0.300 \\
    11.0& 30.0 &   5.171 &  -1.283 &   6.458 &   5.260 &   5.728 & -0.100 \\
\hline
\end{tabular}\\
\vspace{0.20cm}\\
\noindent
\begin{tabular}{rr|rrrrrr}
\multicolumn{7}{c}{\normalsize{\hspace*{3.5cm}$^{  45}_{23}$ V  $_{ 22} \rightarrow \hspace*{0.2cm}^{  45}_{24}$ CR $_{ 21}$}}\\
\hline
Aden& T9&  E+Cap &   E-Dec &  ANuEn &  GamEn &  ProEn &  PPEm \\
\hline
     3.0&  1.0 & -67.839 &    -100 & -67.789 & -69.679 & -89.515 &-20.000 \\
     3.0&  3.0 & -23.479 &    -100 & -23.326 & -24.734 & -30.918 & -7.000 \\
     3.0& 30.0 &  -0.875 &    -100 &   0.106 &  -1.397 &  -0.511 & -0.200 \\
     7.0&  1.0 & -73.659 &    -100 & -73.609 & -75.499 & -95.335 &-20.000 \\
     7.0&  3.0 & -25.194 &    -100 & -25.042 & -26.450 & -32.634 & -7.000 \\
     7.0& 30.0 &  -0.878 &    -100 &   0.102 &  -1.400 &  -0.515 & -0.200 \\
    11.0&  1.0 &    -100 &    -100 &    -100 &    -100 &    -100 &-100\\
    11.0&  3.0 & -63.671 &    -100 & -63.519 & -64.927 & -71.111 & -7.000 \\
    11.0& 30.0 &  -4.741 &    -100 &  -3.761 &  -5.263 &  -4.378 & -0.200 \\
\hline
\end{tabular}\\
\vspace{0.20cm}\\
\begin{tabular}{rr|rrrrrr}
\multicolumn{7}{c}{\normalsize{\hspace*{3.5cm}$^{  45}_{23}$ V  $_{ 22} \rightarrow \hspace*{0.2cm}^{  45}_{22}$ TI $_{ 23}$}}\\
\hline
Aden& T9&  E-Cap &   E+Dec &  NuEn &  GamEn &  ProEn &  PPEm \\
\hline
     3.0&  1.0 &  -4.322 &   0.053 &   0.565 &  -1.476 & -21.208 &-20.000 \\
     3.0&  3.0 &  -2.024 &   0.032 &   0.553 &  -1.127 &  -8.508 & -8.000 \\
     3.0& 30.0 &   1.937 &  -0.150 &   3.106 &   2.299 &   1.513 & -0.800 \\
     7.0&  1.0 &  -0.603 &   0.053 &   0.739 &  -0.871 & -15.388 &-20.000 \\
     7.0&  3.0 &  -0.600 &   0.032 &   0.730 &  -0.671 &  -6.793 & -7.000 \\
     7.0& 30.0 &   1.940 &  -0.149 &   3.110 &   2.303 &   1.516 & -0.800 \\
    11.0&  1.0 &   4.668 &   0.053 &   5.992 &   5.004 &   4.478 & -0.600 \\
    11.0&  3.0 &   4.669 &   0.032 &   5.991 &   5.029 &   4.491 & -0.600 \\
    11.0& 30.0 &   4.703 &  -0.040 &   6.066 &   5.072 &   4.647 & -0.500 \\
\hline
\end{tabular}\\
\vspace{0.20cm}\\
\noindent
\begin{tabular}{rr|rrrrrr}
\multicolumn{7}{c}{\normalsize{\hspace*{3.5cm}$^{  45}_{24}$ CR $_{ 21} \rightarrow \hspace*{0.2cm}^{  45}_{25}$ MN $_{ 20}$}}\\
\hline
Aden& T9&  E+Cap &   E-Dec &  ANuEn &  GamEn &  ProEn &  PPEm \\
\hline
     3.0&  1.0 & -76.031 &    -100 & -75.980 &    -100 & -78.136 &  0.000 \\
     3.0&  3.0 & -25.465 &    -100 & -25.313 &    -100 & -26.894 &  0.000 \\
     3.0& 30.0 &  -0.865 &    -100 &   0.116 &    -100 &  -0.339 &  0.000 \\
     7.0&  1.0 & -81.851 &    -100 & -81.800 &    -100 & -83.956 &  0.000 \\
     7.0&  3.0 & -27.181 &    -100 & -27.028 &    -100 & -28.610 &  0.000 \\
     7.0& 30.0 &  -0.869 &    -100 &   0.113 &    -100 &  -0.342 &  0.000 \\
    11.0&  1.0 &    -100 &    -100 &    -100 &    -100 &    -100 &-100\\
    11.0&  3.0 & -65.658 &    -100 & -65.506 &    -100 & -67.087 &  0.000 \\
    11.0& 30.0 &  -4.732 &    -100 &  -3.750 &    -100 &  -4.205 &  0.000 \\
\hline
\end{tabular}\\
\vspace{0.20cm}\\
\begin{tabular}{rr|rrrrrr}
\multicolumn{7}{c}{\normalsize{\hspace*{3.5cm}$^{  45}_{24}$ CR $_{ 21} \rightarrow \hspace*{0.2cm}^{  45}_{23}$ V  $_{ 22}$}}\\
\hline
Aden& T9&  E-Cap &   E+Dec &  NuEn &  GamEn &  ProEn &  PPEm \\
\hline
     3.0&  1.0 &  -3.615 &   0.799 &   1.387 &   0.725 &   1.047 & -0.200 \\
     3.0&  3.0 &  -1.325 &   0.794 &   1.399 &   0.698 &   1.048 & -0.300 \\
     3.0& 30.0 &   2.638 &   0.853 &   3.820 &   1.362 &   3.510 & -0.020 \\
     7.0&  1.0 &   0.096 &   0.799 &   1.526 &   0.749 &   1.176 & -0.200 \\
     7.0&  3.0 &   0.096 &   0.794 &   1.534 &   0.722 &   1.179 & -0.200 \\
     7.0& 30.0 &   2.642 &   0.853 &   3.824 &   1.365 &   3.513 & -0.020 \\
    11.0&  1.0 &   5.318 &   0.799 &   6.657 &   4.159 &   6.174 & -0.020 \\
    11.0&  3.0 &   5.312 &   0.794 &   6.650 &   4.126 &   6.182 & -0.010 \\
    11.0& 30.0 &   5.367 &   0.932 &   6.747 &   3.827 &   6.318 & -0.009 \\
\hline
\end{tabular}\\
\vspace{0.20cm}\\
\noindent
\begin{tabular}{rr|rrrrrr}
\multicolumn{7}{c}{\normalsize{\hspace*{3.5cm}$^{  46}_{20}$ CA $_{ 26} \rightarrow \hspace*{0.2cm}^{  46}_{21}$ SC $_{ 25}$}}\\
\hline
Aden& T9&  E+Cap &   E-Dec &  ANuEn &  GamEn &  ProEn &  PPEm \\
\hline
     3.0&  1.0 & -14.016 & -21.653 & -13.963 & -14.387 & -58.534 &-40.000 \\
     3.0&  3.0 &  -5.296 &  -9.225 &  -5.118 &  -5.665 & -19.541 &-10.000 \\
     3.0& 30.0 &   1.711 &  -0.612 &   2.788 &   2.214 &   1.910 & -0.400 \\
     7.0&  1.0 & -19.836 & -22.188 & -19.782 & -20.203 & -64.354 &-40.000 \\
     7.0&  3.0 &  -7.011 &  -9.438 &  -6.832 &  -7.378 & -21.256 &-10.000 \\
     7.0& 30.0 &   1.708 &  -0.612 &   2.785 &   2.210 &   1.906 & -0.400 \\
    11.0&  1.0 &    -100 &    -100 &    -100 &    -100 &    -100 &-100\\
    11.0&  3.0 & -45.489 & -44.276 & -44.375 & -44.382 & -59.734 &-10.000 \\
    11.0& 30.0 &  -2.148 &  -3.761 &  -1.072 &  -1.641 &  -1.955 & -0.400 \\
\hline
\end{tabular}\\
\vspace{0.20cm}\\
\begin{tabular}{rr|rrrrrr}
\multicolumn{7}{c}{\normalsize{\hspace*{3.5cm}$^{  46}_{20}$ CA $_{ 26} \rightarrow \hspace*{0.2cm}^{  46}_{19}$ K  $_{ 27}$}}\\
\hline
Aden & T9& E-Cap &   E+Dec &  NuEn &  GamEn &  NeuEn &  PNEm \\
\hline
     3.0&  1.0 & -45.744 & -53.621 & -46.295 & -47.173 & -84.473 &-40.000 \\
     3.0&  3.0 & -16.513 & -20.862 & -16.544 & -17.195 & -28.006 &-10.000 \\
     3.0& 30.0 &   1.412 &  -6.894 &   2.434 &   1.959 &   0.552 & -0.900 \\
     7.0&  1.0 & -39.925 & -53.621 & -40.476 & -41.354 & -78.653 &-40.000 \\
     7.0&  3.0 & -14.802 & -20.847 & -14.831 & -15.484 & -26.291 &-10.000 \\
     7.0& 30.0 &   1.415 &  -6.892 &   2.437 &   1.962 &   0.555 & -0.900 \\
    11.0&  1.0 &   2.306 & -53.621 &   3.253 &   2.839 &   1.538 & -0.900 \\
    11.0&  3.0 &   2.307 & -20.847 &   3.256 &   2.839 &   1.546 & -0.900 \\
    11.0& 30.0 &   4.550 &  -6.665 &   5.788 &   5.100 &   3.785 & -0.800 \\
\hline
\end{tabular}\\
\vspace{0.20cm}\\
\noindent
\begin{tabular}{rr|rrrrrr}
\multicolumn{7}{c}{\normalsize{\hspace*{3.5cm}$^{  46}_{21}$ SC $_{ 25} \rightarrow \hspace*{0.2cm}^{  46}_{22}$ TI $_{ 24}$}}\\
\hline
Aden& T9&  E+Cap &   E-Dec &  ANuEn &  GamEn &  ProEn &  PPEm \\
\hline
     3.0&  1.0 &  -9.855 &  -7.398 &  -7.999 &  -7.093 & -50.163 &-40.000 \\
     3.0&  3.0 &  -5.955 &  -7.555 &  -5.766 &  -5.409 & -16.599 &-10.000 \\
     3.0& 30.0 &   1.939 &  -1.542 &   3.001 &   2.514 &   2.241 & -0.300 \\
     7.0&  1.0 & -15.675 & -10.122 & -10.881 &  -9.820 & -55.966 &-40.000 \\
     7.0&  3.0 &  -7.665 &  -8.217 &  -7.454 &  -7.046 & -18.314 &-10.000 \\
     7.0& 30.0 &   1.936 &  -1.543 &   2.997 &   2.510 &   2.238 & -0.300 \\
    11.0&  1.0 &    -100 &    -100 &    -100 &    -100 &    -100 &-100\\
    11.0&  3.0 & -46.142 & -43.835 & -43.981 & -43.271 & -56.792 &-10.000 \\
    11.0& 30.0 &  -1.922 &  -4.810 &  -0.863 &  -1.346 &  -1.623 & -0.300 \\
\hline
\end{tabular}\\
\vspace{0.20cm}\\
\begin{tabular}{rr|rrrrrr}
\multicolumn{7}{c}{\normalsize{\hspace*{3.5cm}$^{  46}_{21}$ SC $_{ 25} \rightarrow \hspace*{0.2cm}^{  46}_{20}$ CA $_{ 26}$}}\\
\hline
Aden & T9& E-Cap &   E+Dec &  NuEn &  GamEn &  NeuEn &  PNEm \\
\hline
     3.0&  1.0 & -18.554 & -28.432 & -19.100 & -17.997 & -56.687 &-40.000 \\
     3.0&  3.0 &  -7.972 & -11.350 &  -7.990 &  -7.399 & -19.219 &-10.000 \\
     3.0& 30.0 &   1.832 &  -1.801 &   2.882 &   2.449 &   2.007 & -0.300 \\
     7.0&  1.0 & -12.754 & -28.432 & -13.296 & -12.198 & -50.867 &-40.000 \\
     7.0&  3.0 &  -6.266 & -11.345 &  -6.279 &  -5.693 & -17.503 &-10.000 \\
     7.0& 30.0 &   1.836 &  -1.801 &   2.885 &   2.453 &   2.011 & -0.300 \\
    11.0&  1.0 &   4.198 & -28.432 &   5.338 &   5.065 &   3.007 & -2.000 \\
    11.0&  3.0 &   4.257 & -11.345 &   5.397 &   5.134 &   3.087 & -2.000 \\
    11.0& 30.0 &   4.916 &  -1.655 &   6.167 &   5.472 &   5.194 & -0.200 \\
\hline
\end{tabular}\\
\vspace{0.20cm}\\
\noindent
\begin{tabular}{rr|rrrrrr}
\multicolumn{7}{c}{\normalsize{\hspace*{3.5cm}$^{  46}_{22}$ TI $_{ 24} \rightarrow \hspace*{0.2cm}^{  46}_{23}$ V  $_{ 23}$}}\\
\hline
Aden& T9&  E+Cap &   E-Dec &  ANuEn &  GamEn &  ProEn &  PPEm \\
\hline
     3.0&  1.0 & -39.628 & -46.141 & -39.582 & -41.449 & -72.664 &-30.000 \\
     3.0&  3.0 & -13.258 & -15.731 & -13.096 & -14.585 & -24.094 &-10.000 \\
     3.0& 30.0 &   1.330 &  -2.571 &   2.378 &   1.631 &   1.494 & -0.400 \\
     7.0&  1.0 & -45.448 & -46.437 & -45.369 & -47.251 & -78.484 &-30.000 \\
     7.0&  3.0 & -14.974 & -15.875 & -14.773 & -16.249 & -25.810 &-10.000 \\
     7.0& 30.0 &   1.326 &  -2.572 &   2.375 &   1.628 &   1.491 & -0.400 \\
    11.0&  1.0 &    -100 &    -100 &    -100 &    -100 &    -100 &-100\\
    11.0&  3.0 & -53.451 & -52.937 & -52.917 & -54.097 & -64.287 &-10.000 \\
    11.0& 30.0 &  -2.533 &  -5.993 &  -1.486 &  -2.232 &  -2.372 & -0.400 \\
\hline
\end{tabular}%
\\
\vspace{0.20cm}\\

\begin{center}
\textbf{\Large EXPLANATION OF TABLE}
\end{center}
\vspace{0.7in}
\textbf{\large TABLE B. $\mathbf{fp}$-Shell Nuclei Weak Rates of Neutron-Rich and Proton-Rich Nuclei in Stellar Matter}
\vspace{0.4in}
\noindent
\newline
The weak interaction rates for neutron-rich and proton-rich nuclei are calculated using the mass formula of Myers and Swiatecki \cite{Mye96}. The explanation of the table is similar to that of Table~A. The superscript \textit{(MS)} on each parent nucleus indicates that the mass formula \cite{Mye96} is used for the calculation of Q-values and separation energies. $^{55}$Zn is an unstable nucleus according to the calculations of \cite{Mye96}. Hence for the case of $^{55}$Zn the calculations of the \wi rates are presented using the mass formula of \cite{Moe81} which is also indicated by the superscript \textit{(MN)} on this nucleus.
\clearpage
\sffamily \normalsize
\noindent
\\
\vspace{0.20cm}\\

\end{document}